\documentclass[aip,pop,preprint,showpacs]{revtex4-1}

\draft % marks overfull lines with a black rule on the right
\usepackage[utf8]{inputenc}
\usepackage{amsmath,amsthm,amssymb}
\usepackage{float}
\usepackage{graphicx}
\usepackage{dcolumn}% Align table columns on decimal point
\usepackage{bm}% bold math
\usepackage{color}
\usepackage[caption=false]{subfig}
\usepackage{hyperref}   % use for hypertext links, including those to external documents and URLs
\usepackage[all]{hypcap}
\hypersetup{
    colorlinks=true,
    linkcolor=blue,
    filecolor=magenta,      
    urlcolor=cyan,
    citecolor=red
     }
\pdfoutput=1
\DeclareUnicodeCharacter{2212}{-}
\begin{document}
\title{Stationary Langmuir structures in a relativistic current carrying cold plasma}
 \author{Roopendra Singh Rajawat}
 \email{rupn999@gmail.com}
 \altaffiliation{Present address :- School of Applied and Engineering Physics, Cornell University, Ithaca, NY 14850, United States of America}
\affiliation{Institute for Plasma Research, Bhat , Gandhinagar - 382428, India }
\affiliation{Homi Bhaba National Institute, Training School Complex, Anushakti Nagar, Mumbai 400085, India}

 \author{Sudip Sengupta}
\affiliation{Institute for Plasma Research, Bhat , Gandhinagar - 382428, India }
\affiliation{Homi Bhaba National Institute, Training School Complex, Anushakti Nagar, Mumbai 400085, India}
 \author{Nikhil Chakrabarti}
\affiliation{Saha Institute of Nuclear Physics, 1/AF, Bidhanagar, Kolkata - 70064, India}
\affiliation{Homi Bhaba National Institute, Training School Complex, Anushakti Nagar, Mumbai 400085, India}
\date{\today}
\begin{abstract}
Nonlinear stationary structures formed in a cold plasma with immobile ions in the presence of a relativistic electron current beam have been investigated analytically in the collisionless limit. 
%These are cold plasma version of the relativistic Berstein-Greene-Kruskal (BGK) waves. 
The structure profile is governed by the ratio of maximum electrostatic field energy density to the kinetic energy density of the electron beam, {\it i.e.}, $\kappa = E_{m}/ (4 \pi n_{0} m_{0} v^{2}_{0})^{1/2}$, where $E_{m}$ is the maximum electric field associated with the nonlinear structure and $v_{0}$ is the electron beam velocity. It is found that, in the linear limit, {\it i.e.}, $\kappa \ll \sqrt{2 \gamma_{0}/(1+ \gamma_{0})}$, the fluid variables, {\it viz}, density, electric field, and velocity vary harmonically in space, where $\gamma_{0}$ is the Lorentz factor associated with beam velocity ($v_{0}$). In the range $0 < \kappa \leq \kappa_{c} (= \sqrt{2 \gamma_{0}/(1+ \gamma_{0})})$, the fluid variables exhibit an-harmonic behavior. For values of $ \kappa_{c} < \kappa < + \infty$, the electric field shows finite discontinuities at specific spatial locations indicating the formation of negatively charged planes at these locations.
\end{abstract}

\maketitle

\section{Introduction} \label{stat:introduction}
 { Bernstein-Greene-Kruskal \cite{BGK_pr_1957, Schamel_pr_1986,Ng_prl_2005,Eliasson_pr_2006} (BGK) waves are well known nonlinear potential structures supported by background distribution function of the plasma. These are typically formed during nonlinear evolution of streaming instabilities \cite{Roberts_prl_1967,Rajawat_pop_2017}, breaking of the cold plasma waves \cite{Verma_pre_2012}, etc. BGK modes accelerates charge particle up to relativistic energies \cite{Dieckmann_aj_2009}, forms double layer\cite{Nsingh_grl_1982,Nsingh_jgr_1982}, excites ion acoustic wakes\cite{Dawson_pop_1995} etc. Non-relativistic BGK modes are often found in low temperature laboratory plasmas\cite{Saeki_prl_1979,Fox_prl_2008}, magnetic re-connection \cite{Drake_science_2003,Goldman_prl_2014}, in Earth's magnetosphere \cite{Ergun_prl_1998, Muschietti_grl_1999,Goldman_prl_2007,Holmes_jgr_2018} and foreshock region \cite{Shimada_pop_2003,Shimada_pop_2004}, whereas relativistic BGK modes are observed in high energy laser-plasma interaction \cite{Shukla_pr_1986,Montgomery_prl_2001} and strong relativistic astrophysical scenarios, viz. supernova remnant and gamma ray burst\cite{Piran_pr_1999}. Theory of non-relativistic BGK structures \cite{Schamel_pp_1972,Ng_prl_2005,Eliasson_pr_2006} have been studied for a long time and well understood, while theory of relativistic BGK structures is still a hot topic of research; our work contributes to the theory of relativistic BGK structures.

In this paper, we present a very special class\cite{Nicholson} of stationary BGK structures when all the plasma electrons are moving with a single velocity as a beam and thermal effects are neglected, hereinafter called Langmuir structures \cite{Akhiezer_dans_1951, Psimopoulos_pop_1997, Psimopoulos_pop_1997a}. In the non-relativistic regime, and in the absence of a beam, propagating Langmuir waves in a cold plasma have been derived by Albritton et. al. \cite{Albritton_nf_1975}. The Langmuir mode in this case was obtained from the exact space-time dependent solution \cite{Davidson_nf_1968} of the full nonlinear non-relativistic fluid-Maxwell set of equations. Similarly propagating Langmuir waves in a cold relativistic plasma in the absence of a relativistic electron beam is obtained by transforming the governing equations in such a frame, where the wave is at rest, the so-called wave frame \cite{Akhiezer_spj_1956, Verma_prl_2012,Verma_pop_2012}. Verma et. al. \cite{Verma_prl_2012,Verma_pop_2012} constructed such a solution for propagating Langmuir waves (Akhiezer-Polovin wave \cite{Akhiezer_spj_1956}) from exact space-time dependent solution \cite{Infeld_prl_1989} of the full relativistic fluid-Maxwell set of equations by choosing special initial conditions. In the presence of a beam Psimopolous et. al. \cite{Psimopoulos_pop_1997,Psimopoulos_pop_1997a} obtained the solutions for stationary Langmuir waves (stationary in lab frame) in current carrying non-relativistic cold plasmas for a wide range of parameter ($\kappa = E_{m}/(4\pi n_{0}mv^{2}_{0})^{1/2}$), where $E_{m}$ is the maximum amplitude of the electric field, $v_{0}$ is electron beam velocity and other symbols have their usual meanings. Here we present exact solution of stationary Langmuir structures in a relativistic current carrying cold plasma. These solutions are physically significant in the sense that they allow one to estimate the limiting amplitude of large amplitude stationary waves in relativistic current carrying plasmas. Apart from the above physical significance, exact solutions are also useful for benchmarking of simulation codes.

As mentioned above, in this paper, we study stationary Langmuir structures (stationary in lab frame) in the presence of a relativistic electron beam which is propagating through homogeneous positive background of immobile ions. Under the influence of applied harmonic perturbation, periodic compression and rarefaction occurs in density, so according to equation of continuity electrons accelerate and retard periodically in space, to maintain the constant flux throughout the system. These periodic departures from charge neutrality in turn induce a longitudinal electric field which produces the necessary force on the electrons so that the whole system is kept in stationary state. It is found that the basic parameter that controls the nonlinearity in the system, is a ratio of maximum electrostatic field energy density to relativistic kinetic energy density, {\it i.e.}, $\kappa = E_{m}/(4 \pi n_{0} m_{0} v^{2}_{0})^{1/2}$, where $E_{m}$ is the maximum amplitude of the electric field, $v_{0}$ is the electron beam velocity, $n_{0}$ and $m_{0}$ are respectively the equilibrium number density and rest mass of the electron.

In the non-relativistic limit \cite{Psimopoulos_pop_1997}, it is found that if $\kappa \rightarrow \kappa_{c} = E_{m}/(4 \pi n_{0} m v^{2}_{0})^{1/2} = 1$, electric field becomes discontinuous \cite{Sen_pr_1955} at specific spatial locations indicating formation of negative charge planes at these locations. In the case of a relativistic beam, the critical parameter $\kappa_{c}$ is modified and is found to depend on the beam velocity $v_{0}$ as $\kappa_{c} = \sqrt{2 \gamma_{0}/(1+ \gamma_{0})}$. If $\kappa \ll \kappa_{c}$, the fluid variables,  {\it viz.}, electron fluid velocity $v_{e}(x)$, electron number density $n_{e}(x)$, electrostatic potential $\phi(x)$, and electric field $E(x)$ vary harmonically in space in accordance with linear theory. As $\kappa$ increases, and approaches $\kappa_{c}$ within the interval $0 \ll \kappa < \kappa_{c}$, the above variables gradually become anharmonic in space. In the case of $\kappa \geq \kappa_{c}$ it is shown that gradient of electric field becomes infinitely steep periodically at certain singular points which in turn implies discontinuity in electric field and explosive behavior of electron number density. This discontinuous \cite{Sen_pr_1955} electric field implies formation of negatively charged perfectly conducting planes, infinitely extended in the  transverse direction. At these locations electron density becomes singular and the electron fluid velocity becomes vanishingly small. In the limit $\kappa \rightarrow \infty$ ($v_{0}=0, \, \gamma_{0}=1$), the Langmuir structure collapses to a 1-D crystal.

In this paper, we derive exact stationary solutions of Langmuir structures in current carrying cold relativistic fluid-Maxwell system. Some attempts to describe stationary Langmuir solutions in a relativistic current carrying cold plasma were also made by earlier authors \cite{Farokhi2006}. Here we present exact expressions for electrostatic potential, electric field, electron density and electron velocity as a function of position which describe the nonlinear Langmuir structures. It is also shown that, in an appropriate limit, results of relativistic theory coincide with the non-relativistic results \cite{Psimopoulos_pop_1997}. In section \ref{stat:linear}, we present the governing equations and derive the linear results and in section \ref{stat:non linear} nonlinear theory is derived and results are described. We conclude the paper with a brief discussion in section \ref{stat:conclusion}.

\section{Linear Theory} \label{stat:linear}
Let us consider an one dimensional system, where a relativistic electron beam of density $n_{0}$ and velocity $v_{0}$ is propagating through a homogeneous positive background of immobile ions of density $n_{0}$. The basic set of governing equations required to study nonlinear stationary Langmuir structures are  

\begin{equation} \label{stat:eq1}
  \frac{\partial n_{e}{v}_{e}}{\partial x} = 0,    
  \end{equation}

  \begin{equation}	\label{stat:eq2}
  {v}_{e} \frac{\partial {p}_{e}}{\partial x} = {- e E} = e \frac{\partial \phi}{\partial x},
  \end{equation}

	 \begin{equation} \label{stat:eq3}
	  \frac{\partial E} {\partial x} = 4{\pi}e(n_{0}-n_{e}).
	 \end{equation}
	  where $p_{e} = \gamma_{e} m_{0} v_{e}$ is momentum of electrons and other symbols have their usual meaning.
	  
In the linear limit and in the spirit of weakly relativistic flow $v_{0} \ll c$, fluid variables describing the spatial profile can be obtained using linearized set of steady state fluid equations. The continuity equation is,
\begin{equation} \label{stat:eq4}
n_{0} \frac{\partial v_{e}}{\partial x} + v_{0} \frac{\partial n_{e}}{\partial x} = 0,
\end{equation}

the momentum equation is, 
\begin{equation} \label{stat:eq5}
m_{0}v_{0}\gamma^{3}_{0} \frac{\partial v_{e}}{\partial x} = e \frac{\partial \phi}{\partial x},
\end{equation}

and the Poisson equation is,
\begin{equation} \label{stat:eq6}
\frac{\partial E}{\partial x} =  4 \pi e (n_{0} - n_{e}).
\end{equation}
Using Eqs. \eqref{stat:eq4}, \eqref{stat:eq5} and \eqref{stat:eq6}, solution of stationary equations in the linear limit can be obtained straightforwardly as  
\begin{equation} \label{stat:eq6a}
E(X) = \kappa \sin (X/\gamma^{3/2}_{0}),
\end{equation}
\begin{equation} \label{stat:eq7a}
\Phi(X) = 2 \kappa \gamma^{3/2}_{0} \cos (X/\gamma^{3/2}_{0}),
\end{equation}
\begin{equation}   \label{stat:eq8a}
v_{e}(X) = 1 + \kappa \gamma^{-3/2}_{0} \cos (X/\gamma^{3/2}_{0}),
\end{equation}
and
\begin{equation}	\label{stat:eq9a}
n_{e}(X) = 1 - \kappa \gamma^{-3/2}_{0} \cos (X/\gamma^{3/2}_{0}),
\end{equation} 

where $s = v_{0}/\omega_{pe}$, $X = x/s$, $\beta = v_{0}/c$, $E \rightarrow E/E_{0}$, $E_{0} = (4 \pi n_{0}m_{0} v^{2}_{0})^{1/2}$, $\Phi = 2e(\phi - \phi_{0})/m_{0} v^{2}_{0}$, $v_{e} \rightarrow v_{e}/v_{0}$ and $n_{e} \rightarrow n_{e}/n_{0}$. 
Here $\phi_{0}$ is an arbitrary additive potential, $\kappa = E_{m}/(4\pi n_{0}m_{0}v^{2}_{0})^{1/2}$ is ratio of maximum field energy density to kinetic energy density and $s\gamma^{3/2}_{0}$ is the wavelength of stationary waves in the linear limit, {\it i.e.}, $\kappa \ll 1$. It is readily seen from Eqs. \eqref{stat:eq6a} - \eqref{stat:eq9a} that in the linear limit fluid variables are harmonic in space. Fig. \ref{linear v0_01 kr_001} and \ref{linear v0_09 kr_001} show the potential, electric field, velocity and density for two different beam velocities $\beta = 0.1$ and $\beta = 0.9$ respectively and for parameter $\kappa = 0.01$. In Fig. \ref{linear v0_01 kr_001} and \ref{linear v0_09 kr_001} continuous curves are obtained from the linear theory and dashed curves are the results of nonlinear theory, which will be discussed in the next section.
\begin{figure}[h!] 
\centering
\subfloat[]{
\includegraphics[width=0.5\linewidth]{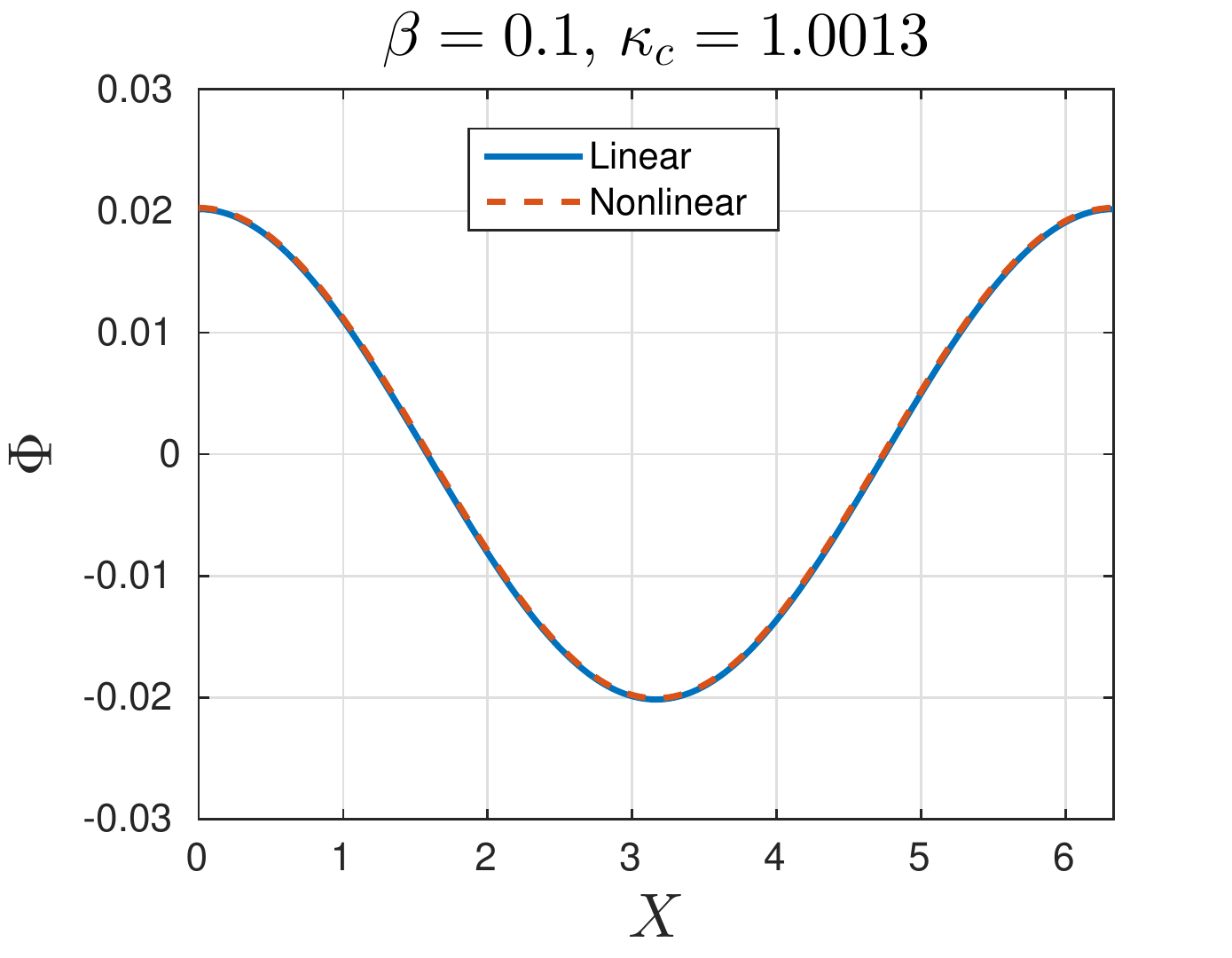}
}
\subfloat[]{
\includegraphics[width=0.5\linewidth]{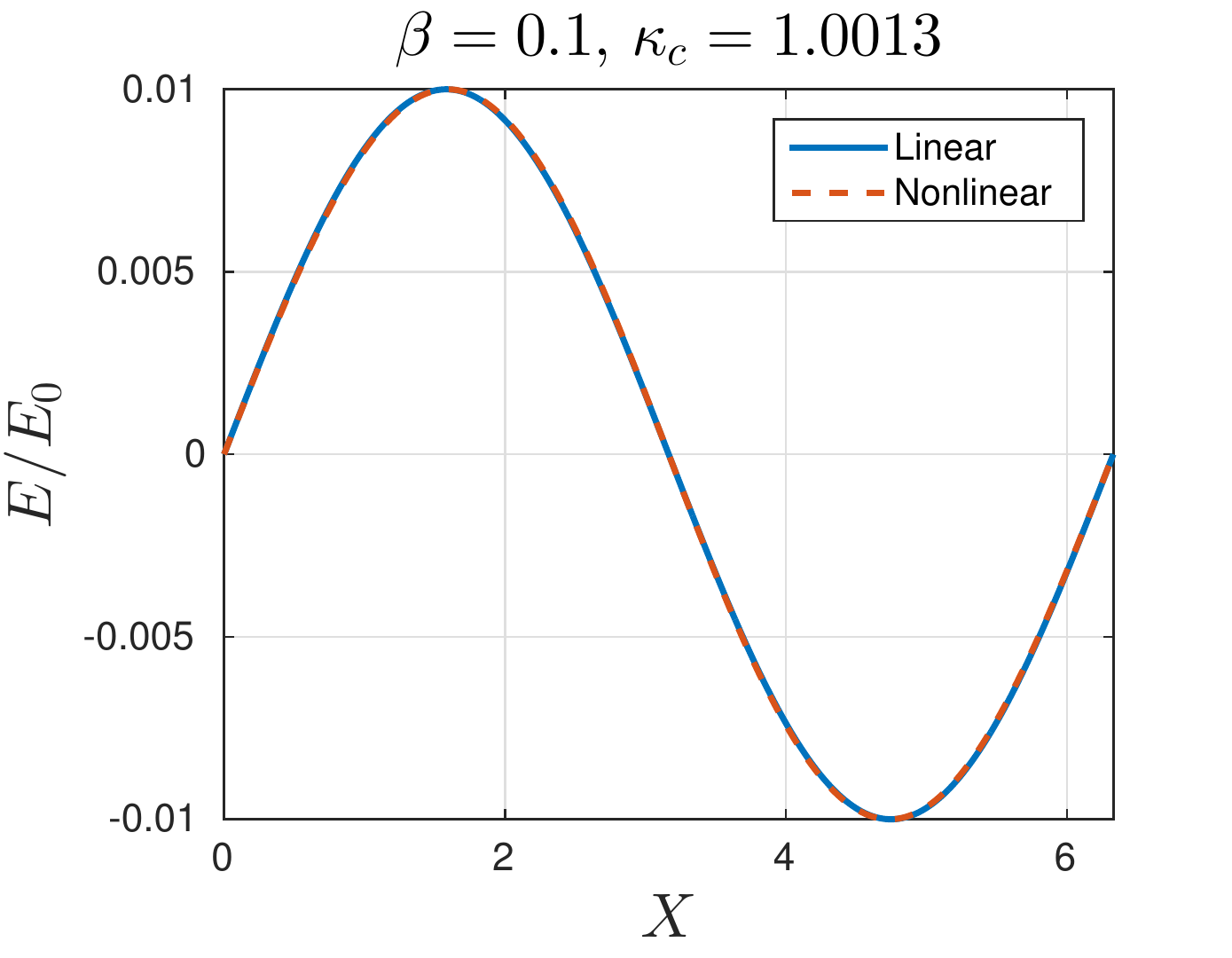}
}\\
\subfloat[]{
\includegraphics[width=0.5\linewidth]{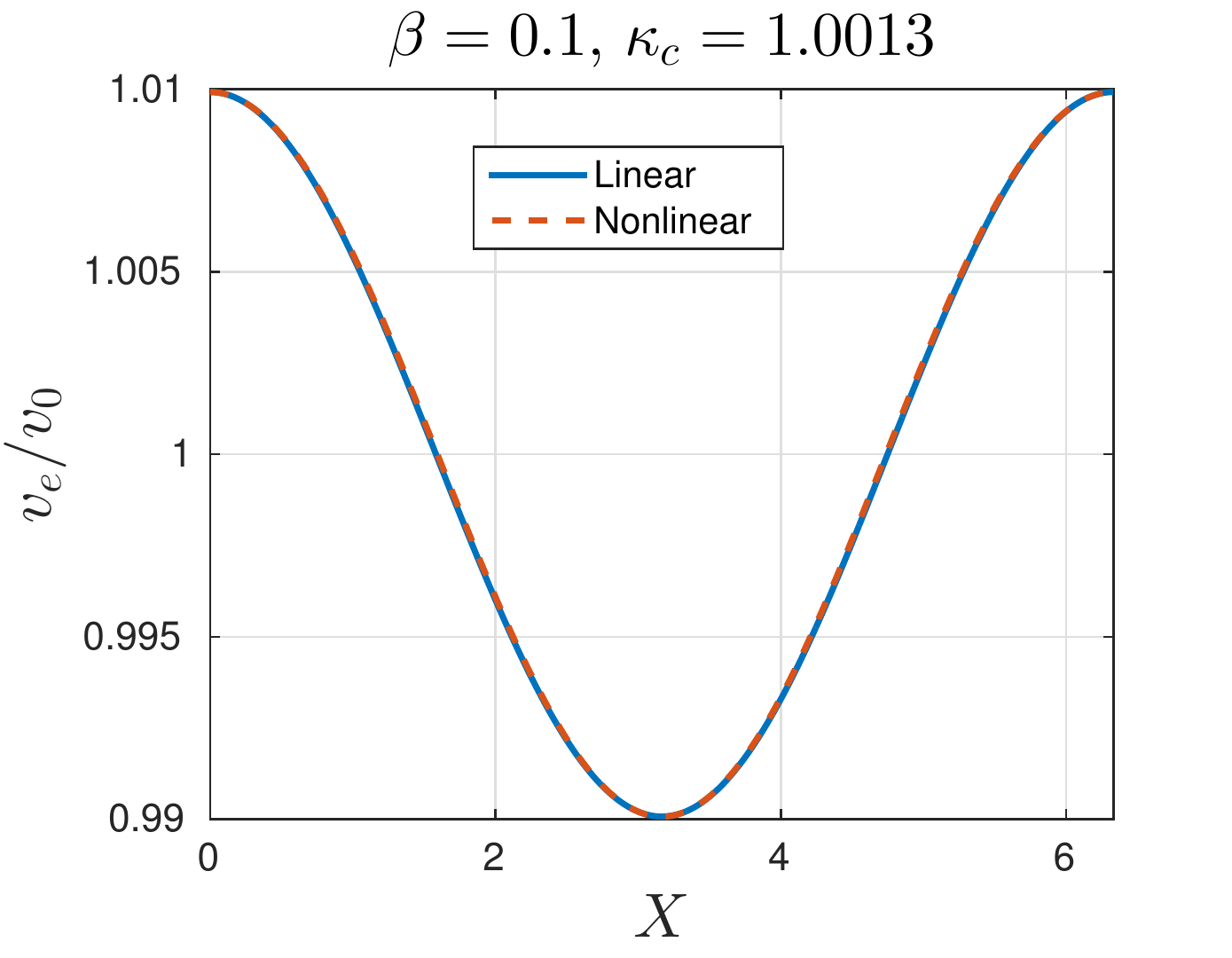}
}
\subfloat[]{
\includegraphics[width=0.5\linewidth]{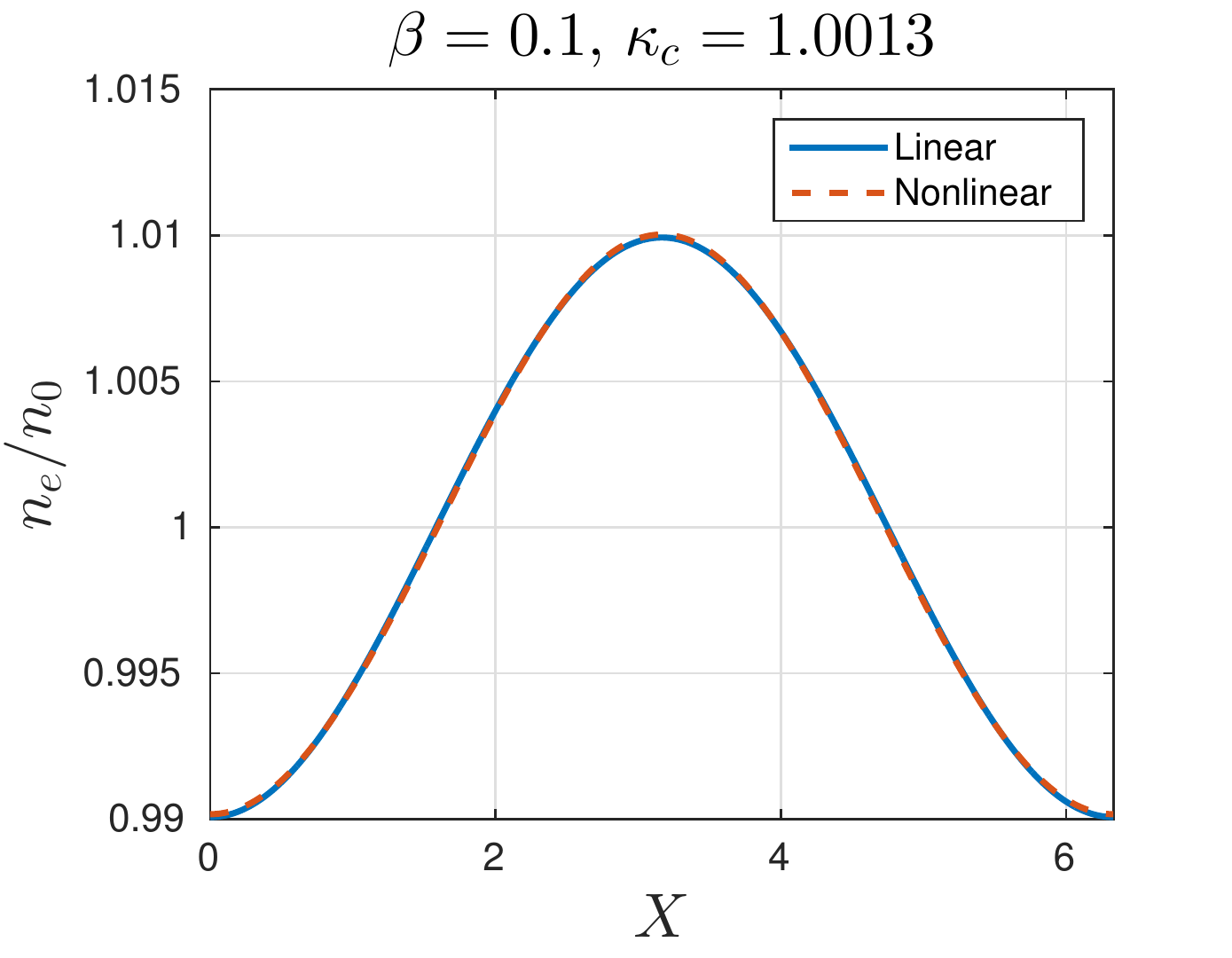}
}
\caption{Fig. shows (a) potential (b) electric field (c) velocity and (d) density for the parameters $\beta = 0.1$ and $\kappa = 0.01$. Here continuous curves are obtained from linear theory and dashed curves are the results of nonlinear theory.}
\label{linear v0_01 kr_001}
\end{figure} 
 \begin{figure}[h!]
\centering
\subfloat[]{
\includegraphics[width=0.5\linewidth]{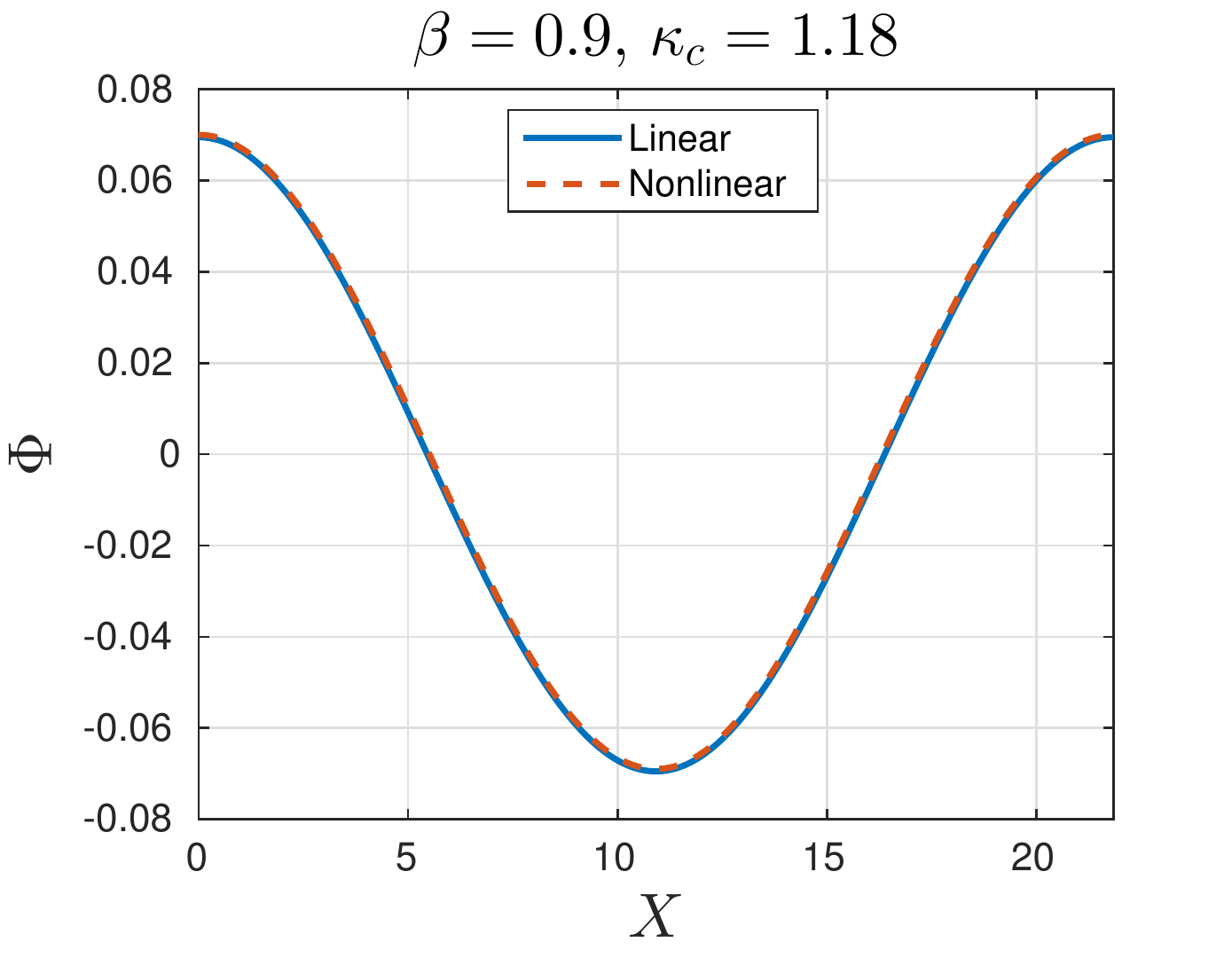}
}
\subfloat[]{
\includegraphics[width=0.5\linewidth]{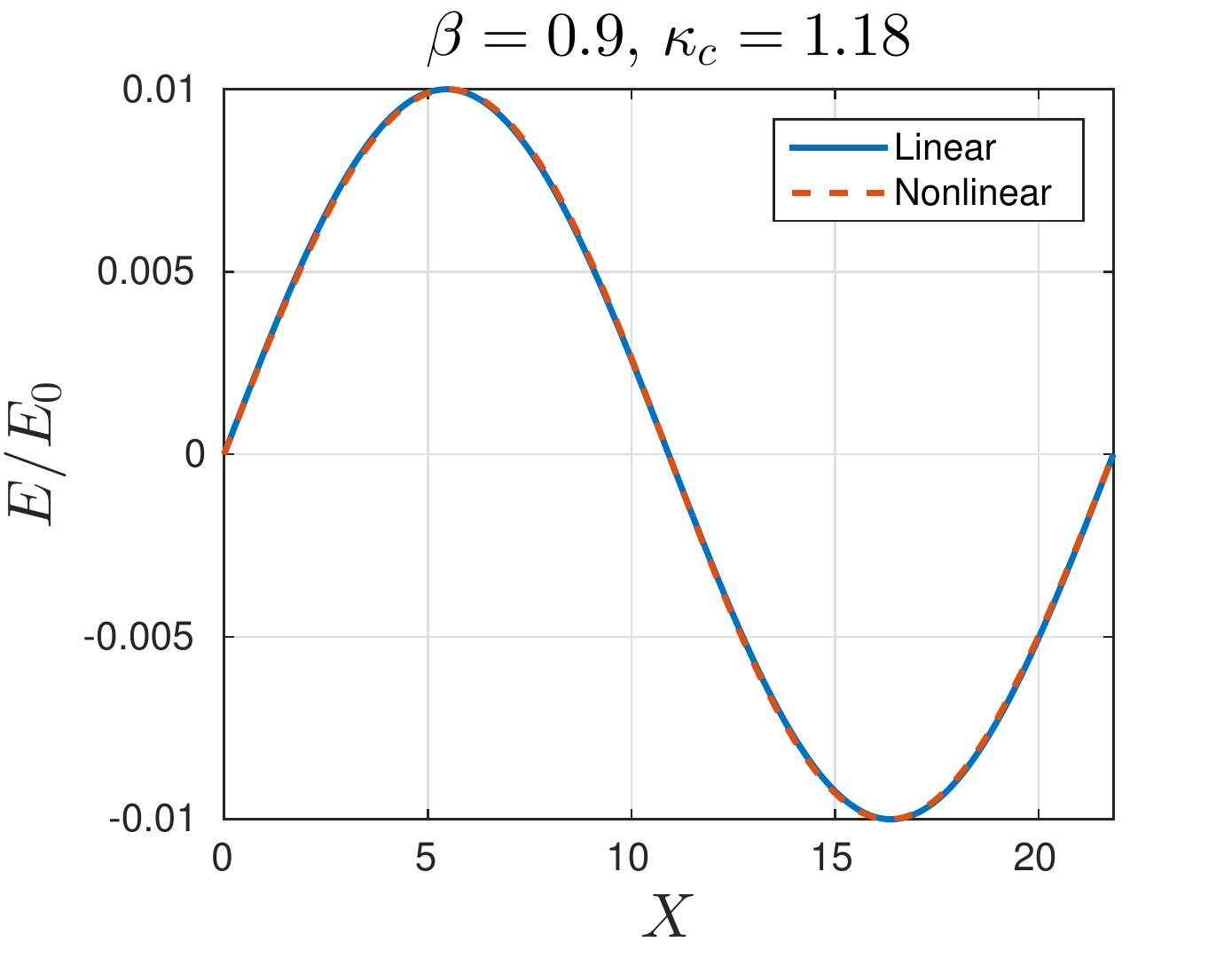}
}\\
\subfloat[]{
\includegraphics[width=0.5\linewidth]{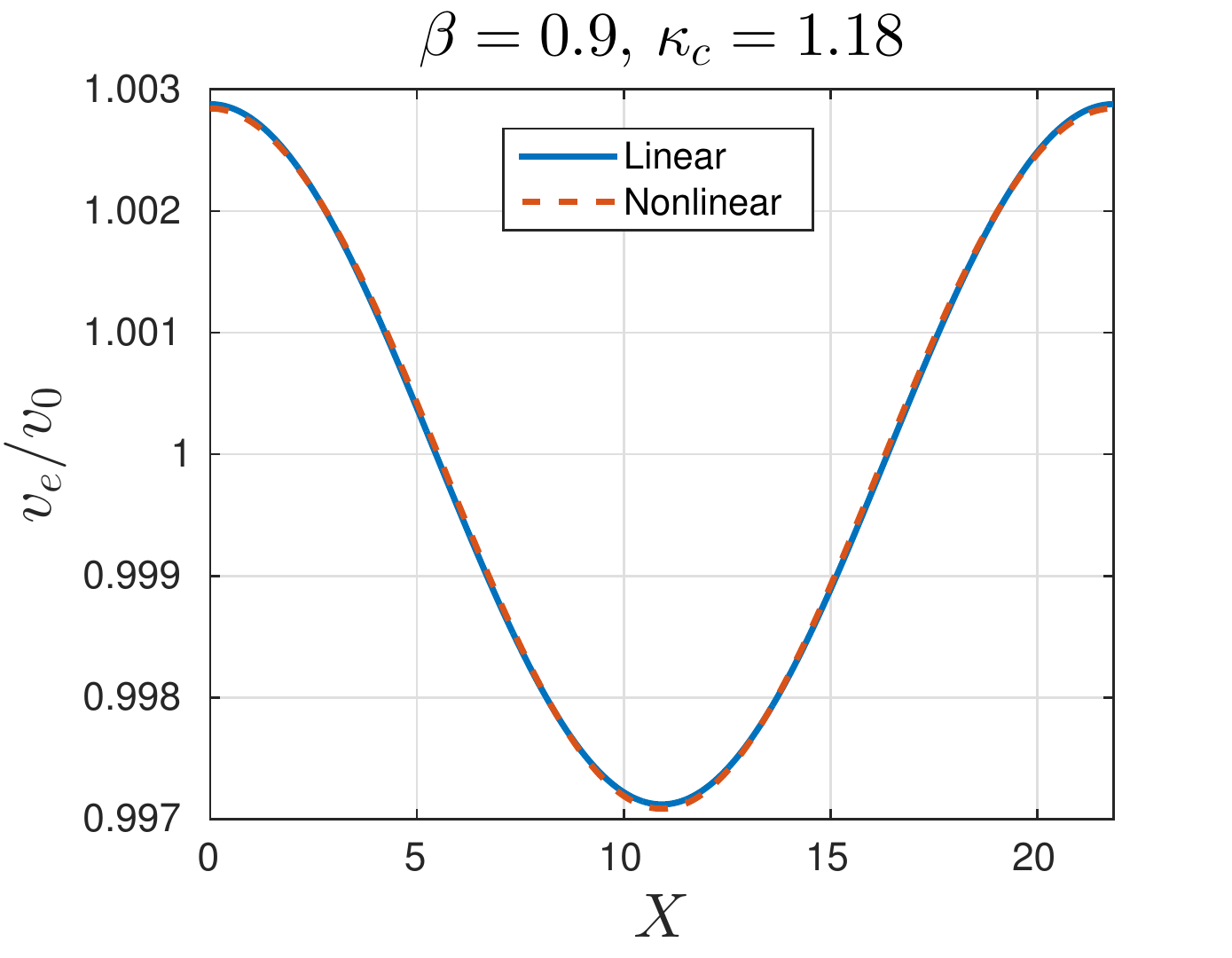}
}
\subfloat[]{
\includegraphics[width=0.5\linewidth]{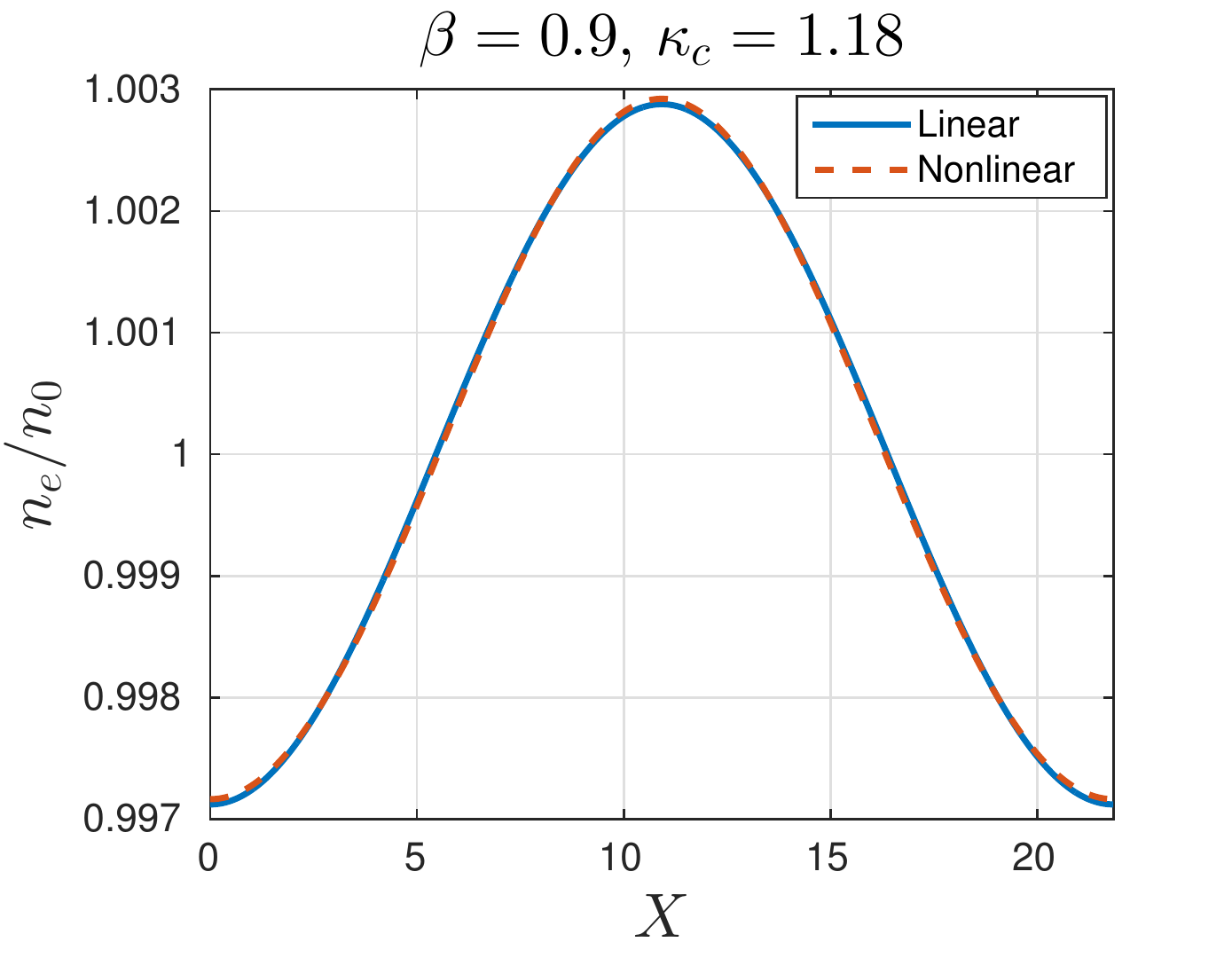}
}
\caption{Fig. shows (a) potential (b) electric field (c) velocity and (d) density for the parameters $\beta = 0.9$ and $\kappa = 0.01$. Here continuous curves are obtained from linear theory and dashed curves are the results of nonlinear theory.}
 \label{linear v0_09 kr_001}
\end{figure} 

\section{Nonlinear Theory} \label{stat:non linear}
%The set of nonlinear stationary relativistic fluid equations are
%\begin{equation}	\label{stat:eq13}
%\frac{\partial n_{e} v_{e} }{\partial x} = 0,
%\end{equation}
%\begin{equation}	\label{stat:eq12}
%v_{e} \frac{\partial p_{e}}{\partial x} = e \frac{\partial \phi}{\partial x},
%\end{equation}
%\begin{equation}	\label{stat:eq14}
%\frac{\partial E}{\partial x} = 4 \pi e (n_{0}-n_{e}).
%\end{equation}
Now integrating Eq.{\eqref{stat:eq2} and assuming that at $\phi = \phi_{0}$; $v = v_{0}$, the relation between electron velocity and electrostatic potential is obtained as \cite{Farokhi2006}
\begin{equation}	\label{stat:eq15}
\frac{m_{0}c^{2}}{\sqrt{1-v^{2}_{e}/c^2}} - \frac{m_{0}c^{2}}{\sqrt{1-v^{2}_{0}/c^2}} = e(\phi(x) - \phi_{0}).
\end{equation}
Using Eqs. \eqref{stat:eq1}, \eqref{stat:eq3} and \eqref{stat:eq15}, the gradient of the electric field as a function of potential can be written as \cite{Farokhi2006}
\begin{equation}	\label{stat:eq16} 
\frac{d^{2} \Phi}{d X^{2}} = - 2 \left( 1 - \beta \frac{ 2 \gamma_{0} + \beta^{2} \Phi}{\sqrt{(2 \gamma_{0} + \beta^{2} \Phi)^{2} - 4}}   \right).
\end{equation}
{Multiplying Eq.\ref{stat:eq16} with $d\Phi/dX$, we obtain}
%or\\
%\begin{equation} \label{stat:eq17}
%\frac{d^{2} \Phi}{d X^{2}} = - \frac{d V_{1}(\Phi)}{d \Phi}, 
%\end{equation}
%  
%\begin{equation}
%\frac{d}{d\Phi} \left\lbrace \frac{1}{2} \left( \frac{d\Phi}{dX} \right)^{2}+ V_{1}(\Phi) \right\rbrace = 0,
%\end{equation}
%or 
%\begin{equation} \label{stat:eq18}
%\frac{1}{2} \left( \frac{d\Phi}{dX} \right)^{2}+ V_{1}(\Phi)  = constant,
%\end{equation}
%which is an energy equation. Here $V_{1}(\Phi)$ is a Sagdeev potential and given by 
%\begin{equation} \label{stat:eq181}
%V_{1}(\Phi) =  2 \left( \Phi - 2 \gamma_{0} \sqrt{1 + \frac{1}{\gamma_{0}}  \Phi + \frac{\beta^{2} }{ 4\gamma^{2}_{0}}\Phi^{2} } \right).
%\end{equation}
%
% Now putting $ d\Phi/dX = -2 \, E $ in the equation \eqref{stat:eq18}, yields
%\begin{equation} \label{stat:eq182}
%E^{2} + \frac{V_{1}(\Phi)}{2} = constant,
%\end{equation} 
%or 
\begin{equation}	\label{stat:eq183}
E^{2} + V(\Phi) = constant,
\end{equation}
{where $d\Phi/dX = -2E$ and $V(\Phi)$ is defined as}
\begin{equation} \label{stat:eq191}
V(\Phi) = 2 \gamma_{0} \left( 1 + \frac{\Phi}{ 2\gamma_{0}} - \sqrt{1 + \frac{1}{\gamma_{0}}  \Phi + \frac{\beta^{2} }{ 4\gamma^{2}_{0}}\Phi^{2} } \right).
\end{equation} 
The $constant$ in Eq.\eqref{stat:eq183} is {derived from $dE^{2}/d\Phi = 0$ at $\Phi = 0$; $V(\Phi)=0$, where $E^{2}$ becomes maximum, {\it i.e.}, $(E/E_{0})^{2} = \kappa^{2}$, then Eq.\eqref{stat:eq183} becomes} \cite{Farokhi2006}
\begin{equation}	\label{stat:eq19}
E^{2} + V(\Phi) = \kappa^{2} .
\end{equation}
Eq.\eqref{stat:eq19} gives a family of curves in the phase space $\Phi - E$ for different values of the parameters $\kappa$ and $\beta$. 

\begin{figure}[h!]
\centering
\subfloat[]{
\includegraphics[width=0.5\linewidth]{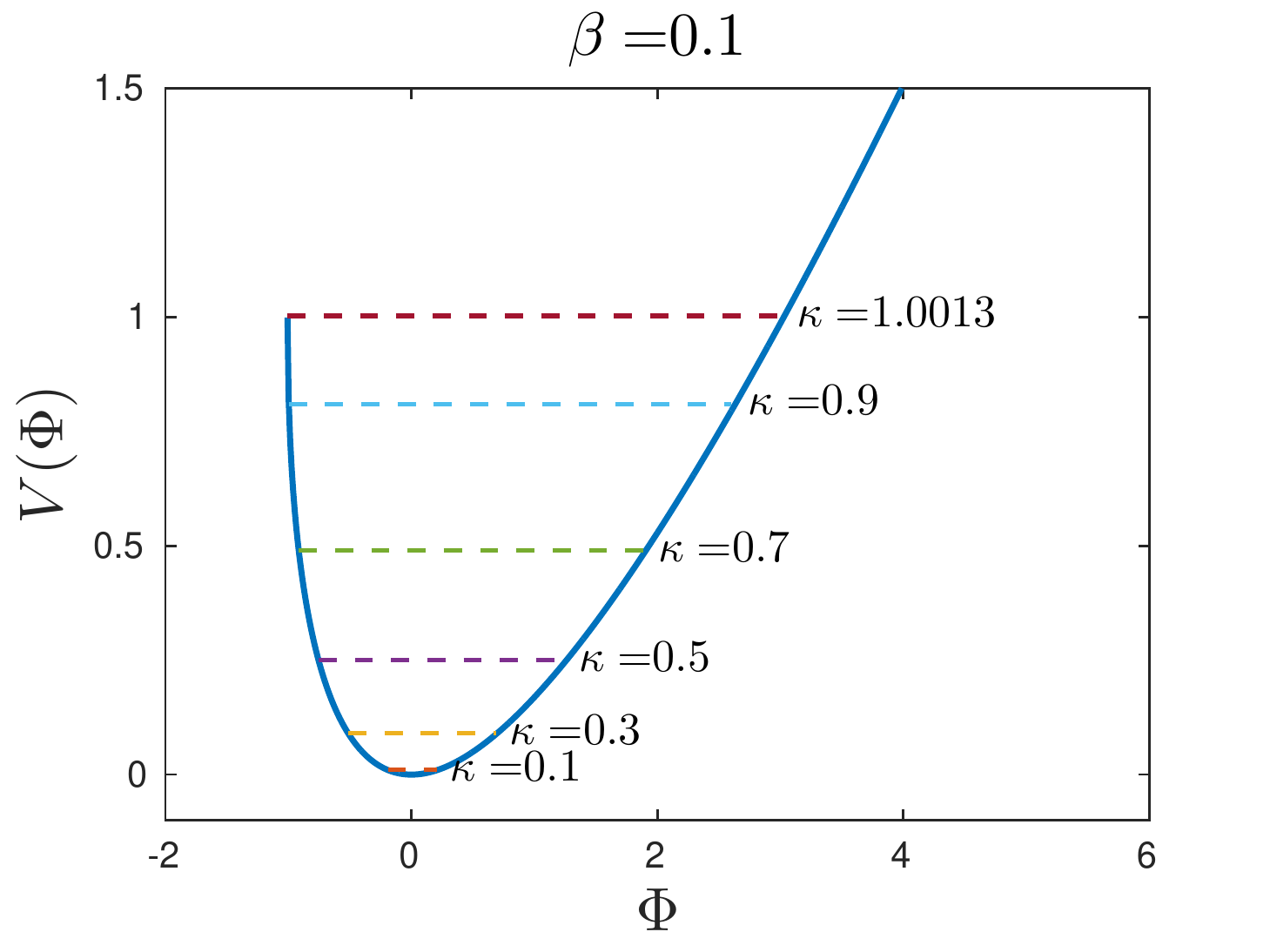}
}
\subfloat[]{
\includegraphics[width=0.5\linewidth]{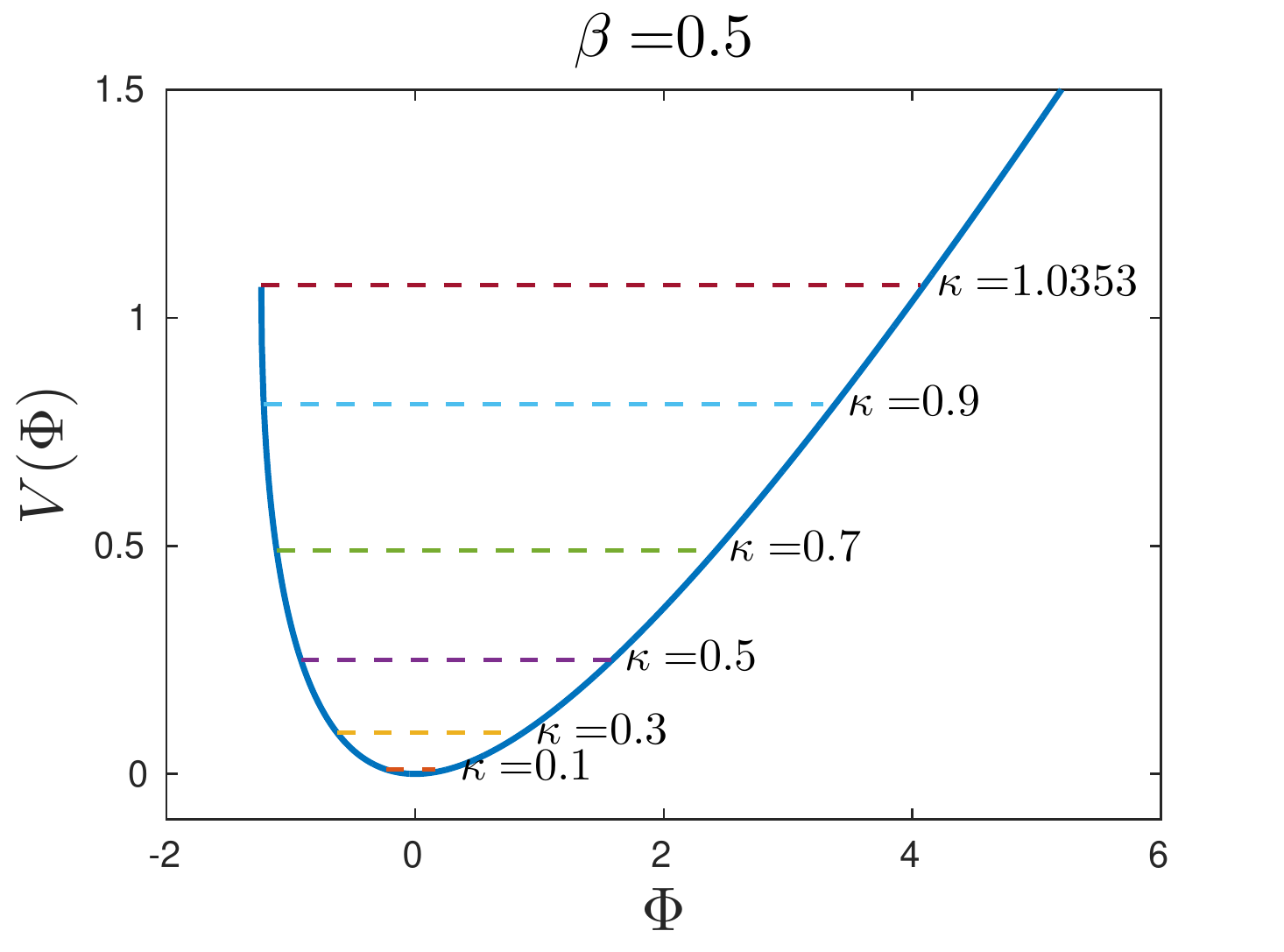}
}\\
\subfloat[]{
\includegraphics[width=0.5\linewidth]{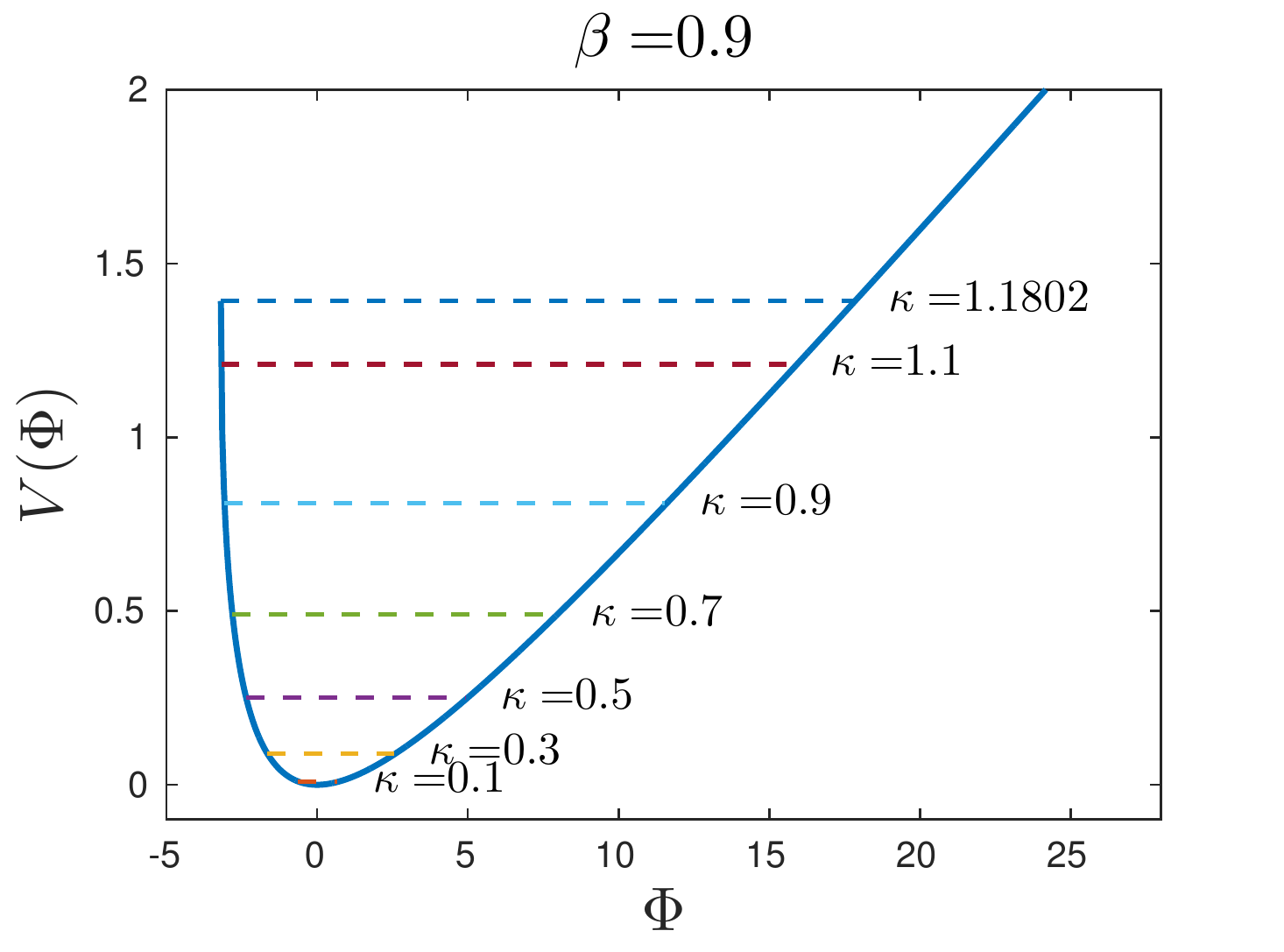}
}
\subfloat[]{
\includegraphics[width=0.5\linewidth]{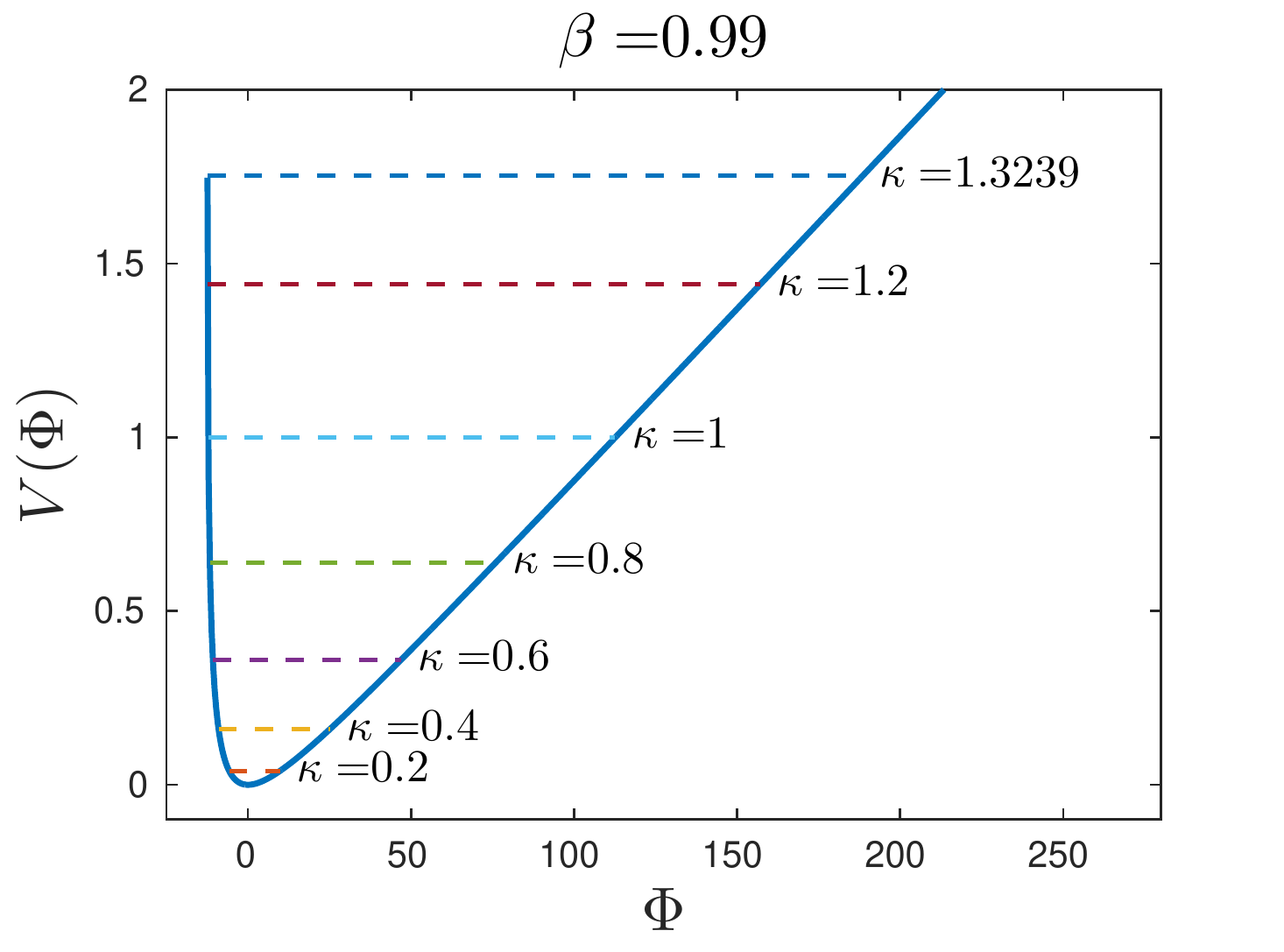}
}
\caption{In this Fig. continuous line shows Sagdeev potential for different beam velocities(a) $\beta = 0.1$ (b) $\beta = 0.5$ (c) $\beta = 0.9$ and (d) $\beta = 0.99$ and dashed line shows level of pseudo-energy for different values of $\kappa$.}
 \label{Sagdeev potential}
\end{figure}
In Fig. \ref{Sagdeev potential} solid blue curve shows variation of Sagdeev potential with the electrostatic potential $\Phi$ for different beam velocities $\beta = 0.1$, $\beta = 0.5$, $\beta = 0.9$ and $\beta = 0.99$. It is noticed here that Sagdeev potential becomes undefined at $\Phi^{c} = -2\gamma^{2}_{0}/(1+\gamma_{0})$ (below this value of potential, the square root term becomes imaginary). {Using Eq.\eqref{stat:eq19} for $E^{2}=0$, $\kappa^{2} = V(\Phi_{c})$, so that $\kappa_{c} = \sqrt{2 \gamma_{0}/(1+ \gamma_{0})}$.} The $\kappa_{c}$ is the critical value of $\kappa$, above which periodic solutions do not exist. The straight lines in Fig. \ref{Sagdeev potential} show different values of $\kappa \lesssim \kappa_{c}$ for which periodic solutions exist; corresponding to these values of $\kappa$ closed orbits are seen in $\Phi - E$ space (Fig. \ref{e-phi phase space}).
\begin{figure} [h!]
\centering
\subfloat[]{
\includegraphics[width=0.5\linewidth]{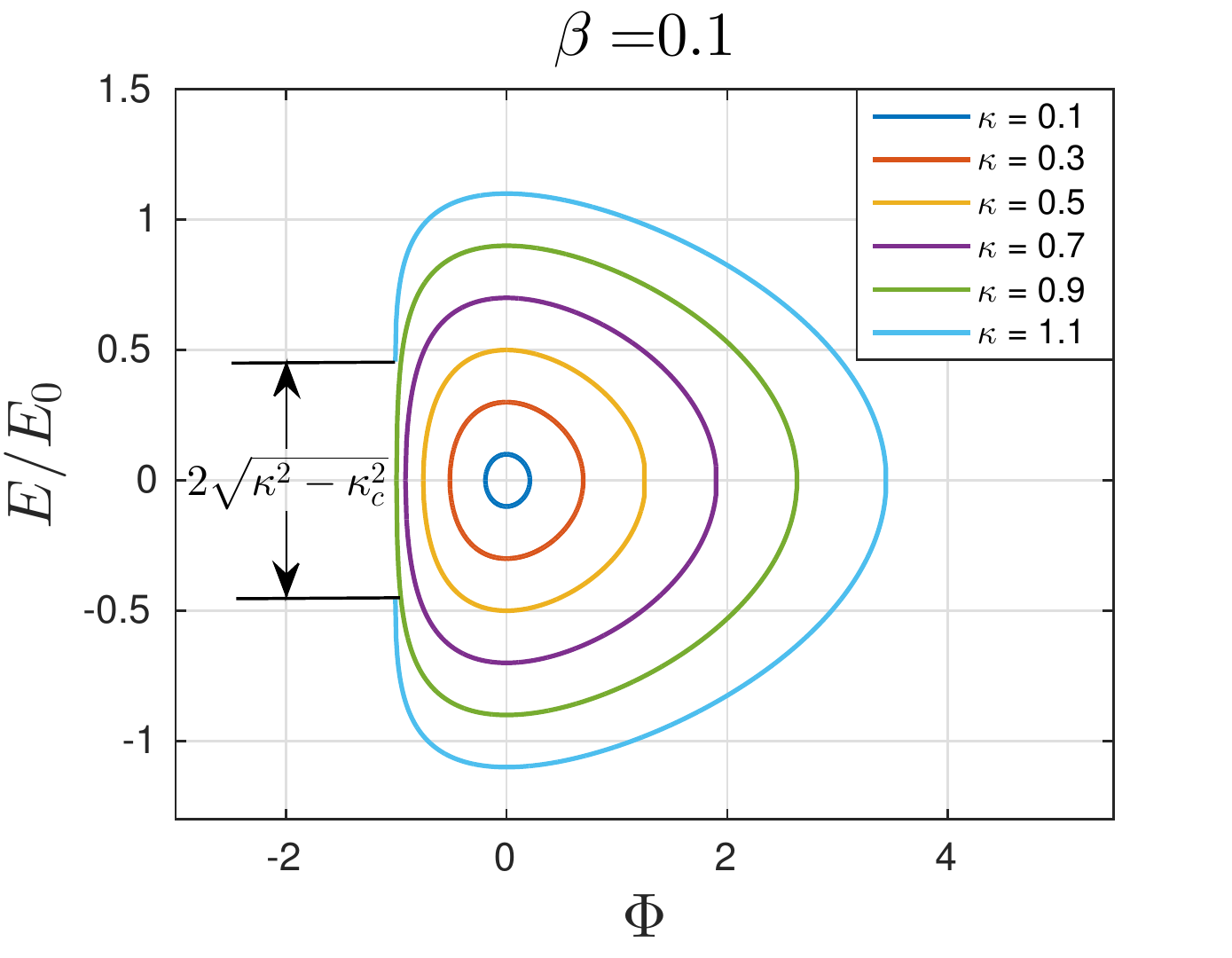}
}
\subfloat[]{
\includegraphics[width=0.5\linewidth]{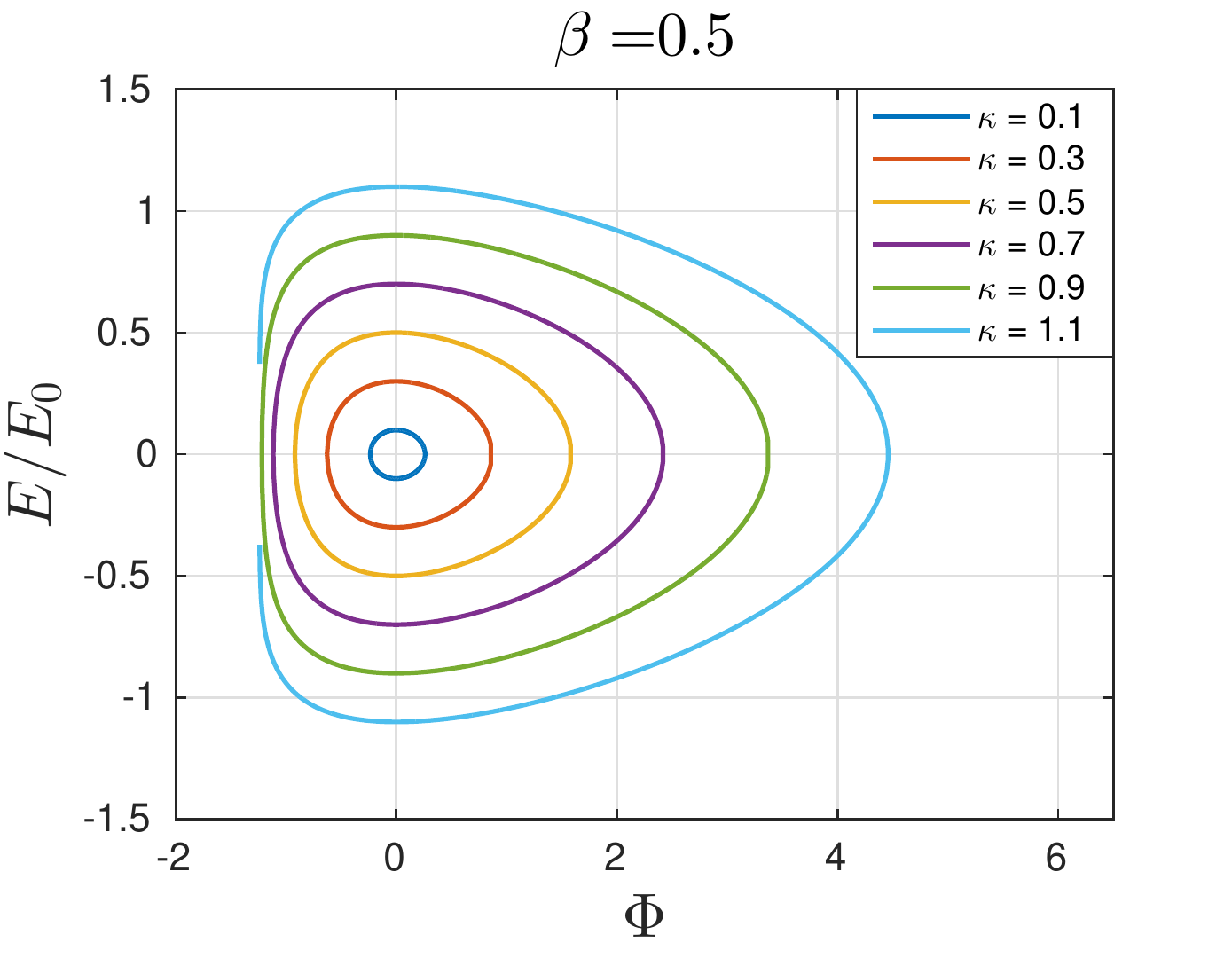}
}\\
\subfloat[]{
\includegraphics[width=0.5\linewidth]{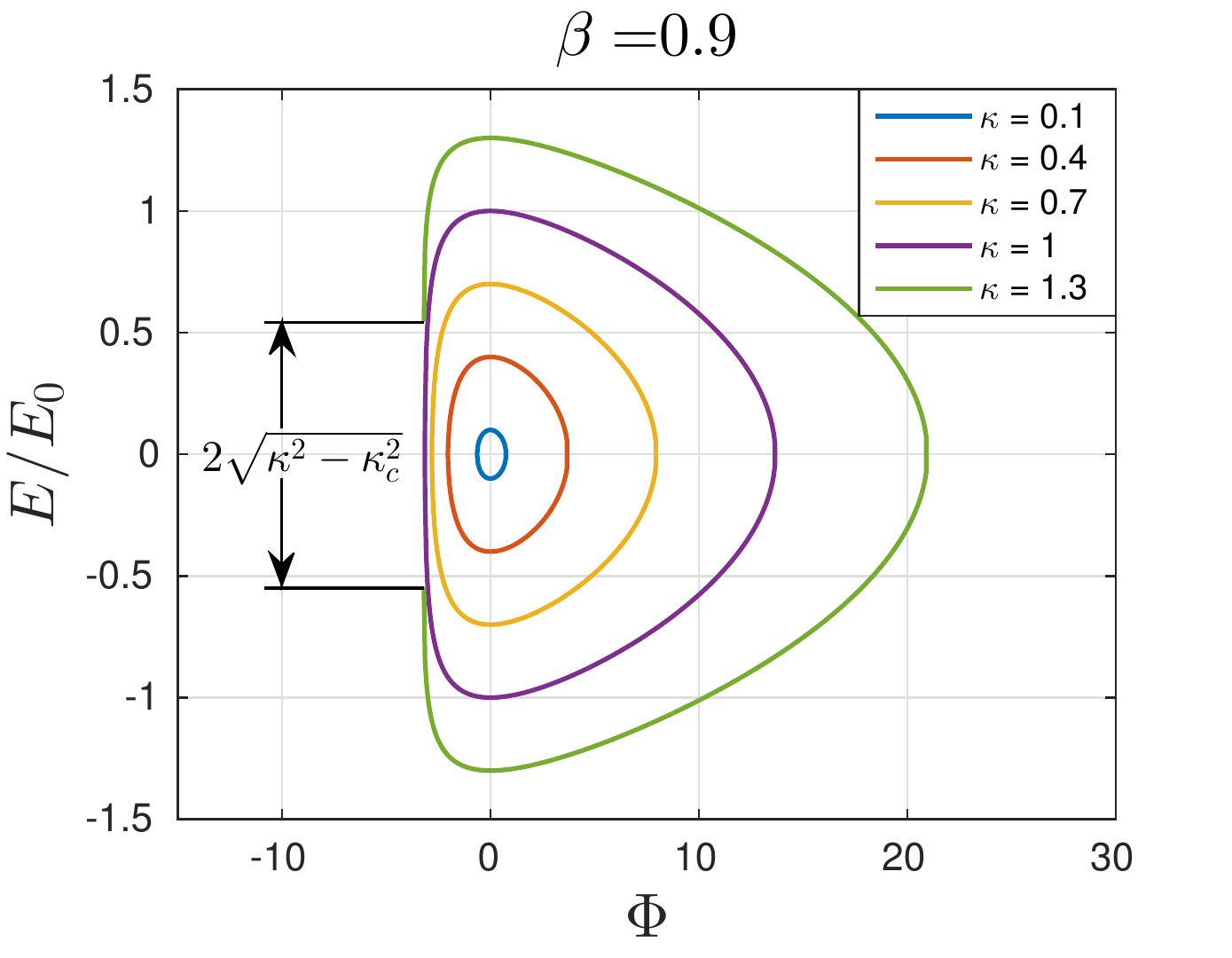}
}
\subfloat[]{
\includegraphics[width=0.5\linewidth]{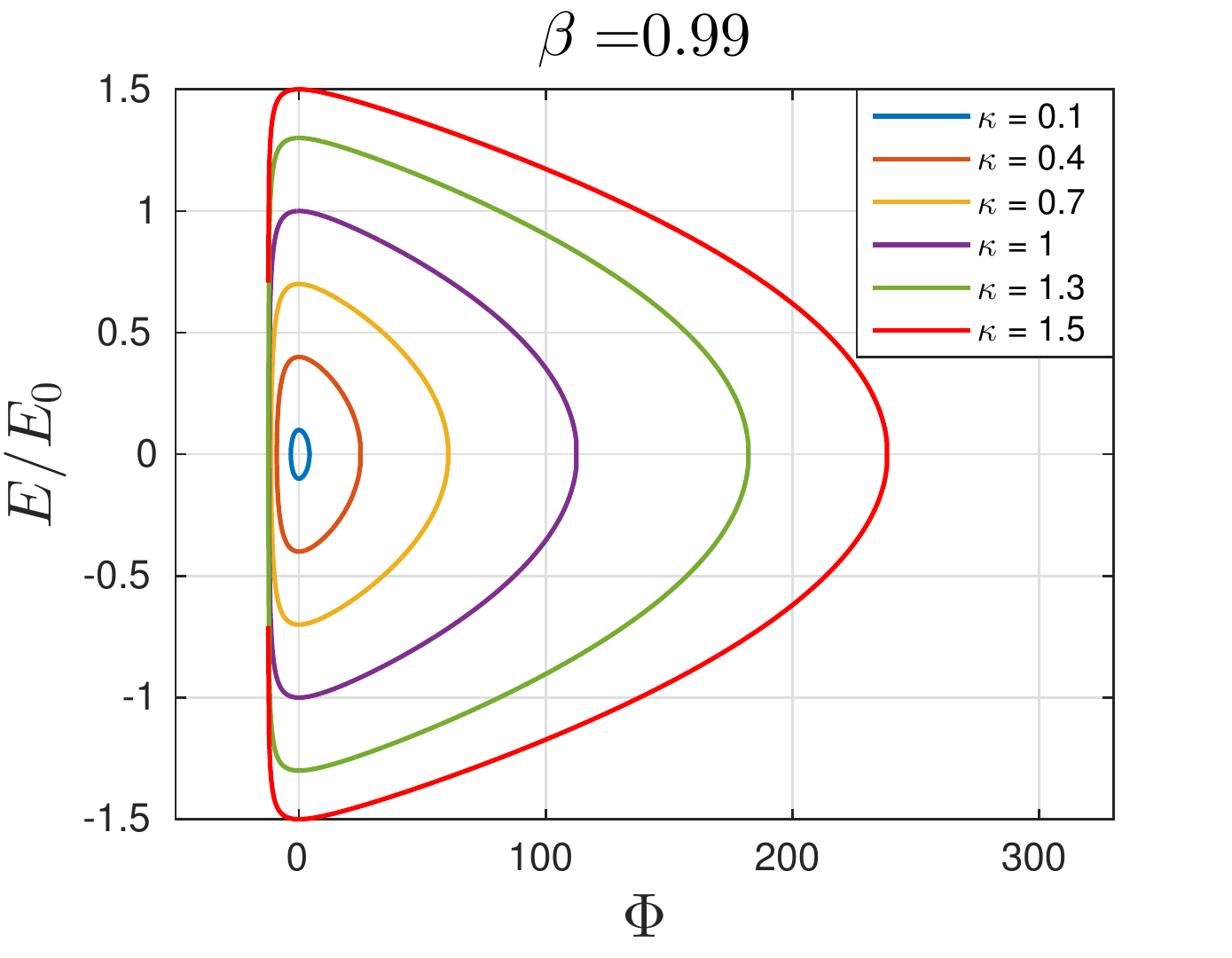}
}
\caption{$\Phi - E$ phase space for different values of $\kappa$ and (a) $\beta = 0.1$, (b) $\beta = 0.5$, (c) $\beta = 0.9$, (d) $\beta = 0.99$.}
 \label{e-phi phase space}
\end{figure}

In Fig. \ref{e-phi phase space} the relation $\Phi - E$ is plotted for different values of the parameters $\kappa$ and $\beta$. It is readily noticed by looking at the Fig. \ref{e-phi phase space} that the variation of $\beta$ modulates the shape of phase space curves as well as changes the range of electrostatic potential $\Phi$. It is also noticed that the plot in phase space becomes discontinuous \cite{Sen_pr_1955} after a critical value of $\kappa$ and this critical value as mentioned above is $\kappa = \kappa_{c} = \sqrt{2 \gamma_{0}/(1+ \gamma_{0})}$. It is found that at the $\kappa = \kappa_{c}$, gradient of electric field becomes infinite, {\it i.e.}, $dE/dX \rightarrow -\infty$,  which is sign of wave breaking \cite{Dawson_prl_1959,Mori_prl_1988} of stationary Langmuir structures in current carrying plasmas.

The range of electrostatic potential $\Phi$ for $0 \leq \kappa \leq \kappa_{c}$ and for $\kappa_{c} \leq \kappa < +\infty$, {are solutions of Eq.\eqref{stat:eq19} for $E = 0$} and are respectively given by Eqs.\eqref{stat:eq20} and \eqref{stat:eq21} below
\begin{multline}	\label{stat:eq20}
%\begin{split}
\kappa \gamma^{2}_{0} \left( \kappa -  \sqrt{\kappa^{2} \beta^{2} + \frac{4}{\gamma_{0}}} \right) \leq \Phi \leq \kappa \gamma^{2}_{0} \left( \kappa +  \sqrt{\kappa^{2} \beta^{2} + \frac{4}{\gamma_{0}}} \right) \hspace{1cm} 0 \leq \kappa \leq \kappa_{c}
%\end{split}
\end{multline}
\begin{multline} \label{stat:eq21}
- \frac{2\gamma^{2}_{0}}{\left(1+\gamma_{0} \right)} \leq \Phi \leq \kappa \gamma^{2}_{0} \left( \kappa  +  \sqrt{\kappa^{2} \beta^{2} + \frac{4}{\gamma_{0}}} \right) \hspace{1cm} \kappa_{c} \leq \kappa < +\infty
\end{multline}

In the range $0 \leq \kappa \leq \kappa_{c}$, phase space curves (Eq.\eqref{stat:eq19}) are continuous and $E$ is found to be oscillating in the range $-\kappa \leq E \leq \kappa$. In the range $\kappa \geq \kappa_{c}$, $E$ becomes discontinuous \cite{Sen_pr_1955} at $\Phi^{c} = -2\gamma^{2}_{0}/(1+\gamma_{0})$ and jumps from $E = \sqrt{\kappa^{2} - \kappa^{2}_{c}}$ to $E = - \sqrt{\kappa^{2} - \kappa^{2}_{c}}$. This implies that $E(X)$ is discontinuous at the positions $X$ satisfying the condition $\Phi(X) = -2\gamma^{2}_{0}/(1+\gamma_{0})$. The critical electrostatic potential at which its gradient ($E(X)$) becomes discontinuous, is not constant as found in the non-relativistic regime ($\Phi^{c} = -1; \gamma_{0} = 1$) \cite{Psimopoulos_pop_1997}, rather, relativity brings the dependency of critical electrostatic potential on the beam velocity through the relation $\Phi^{c} = -2\gamma^{2}_{0}/(1+\gamma_{0})$.

Using $ E = -\frac{1}{2} \frac{d \Phi}{dX}$, and assuming $\Phi = \Phi_{u} =  \kappa \gamma^{2}_{0} \left( \kappa +  \sqrt{\kappa^{2} \beta^{2} + \frac{4}{\gamma_{0}}} \right)$ at $X = 0$, the energy Eq.\eqref{stat:eq19} can be integrated to obtain potential as a function of position as
\begin{equation}	\label{stat:eq23}
X = \frac{1}{2} \int\limits_{\Phi}^{\Phi_u} \frac{d\Phi_1}{ \left[ \kappa^{2} -  \left( 2 \gamma_{0} + \Phi_1 - \sqrt{ \beta^{2}\Phi_{1}^{2} + 4\gamma_{0}\Phi_1 +4\gamma_{0}^{2} } \right) \right]^{1/2} }.
\end{equation}
%
%For simplification, we assume $ (2\gamma_{0}/\beta^{2}) + \Phi = (2/\beta^{2}) \xi$, then integrand of R.H.S. of equation \eqref{stat:eq23} takes the form
%\begin{multline} 	\label{stat:eq24}
%\frac{ d\Phi}{\left( \kappa^{2} -  \left( 2 \gamma_{0} + \Phi - \beta \sqrt{ \left( \frac{2 \gamma_{0}}{\beta^{2}}  + \Phi\right)^{2} - \frac{4}{\beta^{4}} } \right) \right)^{1/2}} = \\ \frac{\sqrt{2} d\xi}{ \beta \left( \frac{\beta^{2}}{2} \left( \kappa^{2} - 2 \gamma_{0} \right) + \gamma_{0} - \xi + \beta ( \xi^2-1)^{1/2} \right)}.
%\end{multline}
%
%For the sake of convenience, we define a new mathematical quantity $\alpha$ as
%\begin{equation}	\label{stat:eq25}
%\alpha = \frac{\beta^{2}}{2} \left( \kappa^{2} - 2 \gamma_{0} \right) + \gamma_{0}
%\end{equation}
%which transforms the equation \eqref{stat:eq24} into
{Using the transformation $\Phi_1 = (2/\beta^{2})(\xi_1 - \gamma_{0})$, we get}
\begin{equation}	\label{stat:eq26}
X = \frac{1}{\beta \sqrt{2}} \int\limits_{\xi}^{\xi_u} \frac{d\xi_{1}}{\left( \alpha - \xi_{1} + \beta \sqrt{ \xi_{1}^2-1} \right)^{1/2}},
\end{equation}
where $\alpha = \frac{\beta^{2}}{2} \left( \kappa^{2} - 2 \gamma_{0} \right) + \gamma_{0}$ and $\xi_{u} = \frac{\beta^{2}}{2} \Phi_{u} + \gamma_{0}$.
{Now a new variable transformation $\chi_1$ is introduced which is defined as}
\begin{equation}	\label{stat:eq27}
\sqrt{\xi_1^{2}-1} = \chi_1^{2} - \xi_1.
\end{equation}
Using the transformation \eqref{stat:eq27}, Eq.\eqref{stat:eq23} becomes
\begin{equation}	\label{stat:eq29}
X = \frac{1}{\beta(1-\beta)^{1/2}} \int\limits^{\chi_{u}}_{\chi} \frac{(\chi^{2}_{1} - 1/\chi^{2}_{1})d\chi_{1}}{((r^{2} - \chi^{2}_{1})(\chi^{2}_{1}-s^{2}))^{1/2}},
\end{equation}
where $r^{2}$ and $s^{2}$ are function of $X$ and defined as 
\begin{eqnarray}	\label{stat:eq30}
r^{2} = \frac{\alpha + \sqrt{\alpha^{2} + \beta^{2} - 1}}{1- \beta},\\
s^{2} = \frac{\alpha - \sqrt{\alpha^{2} + \beta^{2} - 1}}{1- \beta}.
\end{eqnarray}
It must be noted here that substitution of new variable $\chi(X)$, is merely a mathematical manipulation, and does not imply any restriction on the range of the potential. Now Eq. \eqref{stat:eq29} is in standard form and can be reduced easily in the form of elliptic integral upon using new substitution 
%
%To calculate the exact solution of above integral, we substitute
\begin{equation}	\label{stat:eq32}
\sin^{2}\theta_{1} = \frac{r^{2} - \chi_{1}^{2}}{r^{2} - s^{2}}.
\end{equation}
%this implies 
%\begin{equation}	\label{stat:eq33}
%d\chi_{1} = - \frac{(r^{2} - s^{2}) \sin\theta_{1} \cos\theta_{1} }{\sqrt{r^{2} \cos^{2}\theta_{1} +s^{2} \sin^{2}\theta_{1}}} d \theta_{1}.
%\end{equation}
Thus, the exact solution of Eq.\eqref{stat:eq23} can be written as
%\begin{widetext}
%\begin{equation} \label{stat:eq34}
%X r^{3} = \frac{1}{\beta(1-\beta)^{1/2}} \left( \left( \frac{r^{4}(k^{2}-1)+1}{k^{2}-1} \right) E(\theta,k) - \frac{k^{2} \sin2 \theta}{2(k^{2}-1) (1 - k^{2} \sin^{2}\theta)^{1/2} } \right) + \\ c_{1}(\Phi),
%\end{equation}
%\end{widetext}
%where $E(\theta,k)$ is an incomplete elliptic integral of second kind and $c_{1}(\Phi)$ is the constant of integration that can be obtained using the condition that at position $X = 0$, potential is maximum, which is $\Phi_{u} = \kappa \gamma^{2}_{0} \left( \kappa +  \sqrt{\kappa^{2} \beta^{2} + \frac{4}{\gamma_{0}}} \right) $; then the complete solution becomes

\begin{widetext}
\begin{multline}	\label{stat:eq35}
X = \frac{r \gamma_{0}}{\sqrt{1 + \beta}} \Bigg[ 2 \left( E(\theta,k) - E(\theta_{u},k) \right) + \frac{k^{2}(1 - \beta)}{2 \beta} \left( \frac{\sin2 \theta}{\sqrt{(1 - k^{2} \sin^{2}\theta}} -  \frac{\sin2 \theta_{u}}{\sqrt{(1 - k^{2} \sin^{2}\theta_{u}}} \right) \Bigg]
\end{multline}	
\end{widetext}
where $E(\theta,k)$ is an incomplete elliptic integral of second kind and the variables $k$, $\theta_{u}$ and $\theta$ are defined as 
\begin{equation} \label{stat:eq36a}
k^{2} = \frac{r^{2} - s^{2}}{r^{2}} = \frac{2 \sqrt{\alpha^{2} + \beta^{2} - 1}}{\alpha + \sqrt{\alpha^{2} + \beta^{2} - 1}}, 
\end{equation}
\begin{equation} \label{stat:eq36b}
\sin^{2}\theta_{u} = \frac{ 2 r^{2} - ( \beta^{2} \Phi_{u} + 2 \gamma_{0} ) -  \beta \sqrt{\beta^{2}\Phi_{u}^{2} + 4 \gamma_{0} \Phi_{u} + 4 \gamma_{0}^{2} }}{2(r^{2} - s^{2})}
\end{equation}
\begin{equation}	\label{stat:eq36c}
\sin^{2}\theta = \frac{ 2 r^{2} - ( \beta^{2} \Phi + 2 \gamma_{0} ) -  \beta \sqrt{\beta^{2}\Phi^{2} + 4 \gamma_{0} \Phi + 4 \gamma_{0}^{2} }}{2(r^{2} - s^{2})}
\end{equation}
Eq.\eqref{stat:eq35} gives implicit relation between potential and position. The potential $\Phi(X)$ as a function of position $X$ for different values of $\kappa$ and $\beta$ can be obtained by numerical solution of Eq.\eqref{stat:eq35} and \eqref{stat:eq36c}. {\bf From the above equations non-relativistic results of ref. \cite{Psimopoulos_pop_1997} can easily be recovered in the limit  $\beta \rightarrow 0, \gamma_{0} \rightarrow 1$ (see appendix \ref{appendix A})}.

The half wavelength (spatial variation between maxima to minima of the electrostatic potential) of the Langmuir structures can be obtained by putting $\Phi = \Phi_{l}$, the minimum values of $\Phi$ in Eq.\eqref{stat:eq35}.(For the range $\kappa \leq \kappa_{c}$ and $\kappa \geq \kappa_{c} $, $\Phi_{l} = \kappa \gamma^{2}_{0}\left(\kappa - \sqrt{ \kappa^{2} \beta^{2} + 4/\gamma_{0}} \right)$ and $\Phi_{l} = \Phi^{c} = -2\gamma^{2}_{0}/(1+\gamma_{0})$ respectively). In the range $0 \leq \kappa \leq \kappa_{c} $ wavelength turns out to be
\begin{subequations}
\begin{equation}
\lambda = 2 \mu s,
\end{equation}
where
\begin{widetext}
\begin{multline}	\label{stat:eq37}
\mu = \frac{r \gamma_{0}}{\sqrt{1 + \beta}} \Bigg[ 2 \left( E(\theta_{l},k) - E(\theta_{u},k) \right) + \frac{k^{2}(1 - \beta)}{2 \beta} \left( \frac{\sin2 \theta_{l}}{\sqrt{(1 - k^{2} \sin^{2}\theta_{l}}} -  \frac{\sin2 \theta_{u}}{\sqrt{(1 - k^{2} \sin^{2}\theta_{u}}} \right) \Bigg],
\end{multline}
\end{widetext}
\end{subequations}
and for the range $\kappa_{c}  \leq \kappa <  \infty$ it becomes
\begin{subequations}
\begin{equation}
\lambda = 2 \mu_{c} s, 
\end{equation}
where
\begin{multline}	\label{stat:eq38}
\mu_{c} = \frac{r \gamma_{0}}{\sqrt{1 + \beta}} \Bigg[ 2 \left( E(\theta_{c},k) - E(\theta_{u},k) \right) + \frac{k^{2}(1 - \beta)}{2 \beta} \left( \frac{\sin2 \theta_{c}}{\sqrt{(1 - k^{2} \sin^{2}\theta_{c}}} -  \frac{\sin2 \theta_{u}}{\sqrt{(1 - k^{2} \sin^{2}\theta_{u}}} \right) \Bigg],
\end{multline} 
\end{subequations}
and
\begin{equation} \label{stat:eq381}
\sin^{2}\theta_{c} = \frac{ 2 r^{2} - ( \beta^{2} \Phi_{c} + 2 \gamma_{0} ) -  \beta \sqrt{\beta^{2}\Phi_{c}^{2} + 4 \gamma_{0} \Phi_{c} + 4 \gamma_{0}^{2} }}{2(r^{2} - s^{2})},
\end{equation}
Here $\theta_{u}, \, \theta_{l}$ and $\theta_{c}$ are respectively related to $\Phi_{u}, \, \Phi_{l}$ and $\Phi_{c}$ through Eqs.\eqref{stat:eq36b}, \eqref{stat:eq36c} ($\theta \rightarrow \theta_{l}$ and $\Phi \rightarrow \Phi_{l}$) and \eqref{stat:eq381}.
The half wavelengths $\mu(\kappa,\beta)$ and $\mu_{c}(\kappa,\beta)$ are explicit functions of parameters $\kappa$ and $\beta$. Corresponding non-relativistic expression for wavelength can be found in reference \cite{Psimopoulos_pop_1997}. For the non-relativistic case, in the range $0 \leq \kappa \leq 1$, {Psimopoulous} \cite{Psimopoulos_pop_1997} observed that wavelength of the Langmuir structure is constant and independent of $\kappa$, however, in the range $ 1 \leq \kappa < + \infty$, wavelength becomes a function of $\kappa$ and it increases on increasing the value of parameter $\kappa$. In the relativistic regime, it is readily seen that wavelengths (Eq. \eqref{stat:eq37} and \eqref{stat:eq38}) are not only a function of the parameter ($\kappa$) but also has dependence on beam velocity ($\beta$) through the variable $k$; where $k$ is defined by Eq.\eqref{stat:eq36a}.  Figure \ref{wavelength} shows variation of wavelength of the Langmuir structure with the variation of parameter $\kappa$ for two different beam velocities, $i.e.$, $\beta = 0.1$ (\ref{wavelength v0_01}) and $\beta = 0.9$ (\ref{wavelength v0_09}). In fig \ref{wavelength} for the velocity $\beta = 0.1$ (Fig. \ref{wavelength v0_01}), in the range $0 \leq \kappa \leq \kappa_{c}$ (blue color curve), wavelength is almost constant or in other words, in the range $\beta \ll 1$ wavelength of relativistic Langmuir structure turns out to be independent of $\kappa$, a feature which is seen in the non-relativistic case also \cite{Psimopoulos_pop_1997}. However, for the velocity $\beta = 0.9$ (Fig. \ref{wavelength v0_09}), wavelength increases with increasing $\kappa$ as shown in Fig. \ref{wavelength v0_09}, {\it i.e.}, wavelength shows strong dependence on $\kappa$ for large values of $\beta$. Therefore, dependence of wavelength on beam velocity is purely a relativistic effect. In the highly nonlinear range $\kappa_{c} \leq \kappa <  + \infty$, wavelength for all value of $\beta$ increases with increasing $\kappa$ (orange curve in Fig. \ref{wavelength}). The dashed vertical line in Figs. \ref{wavelength v0_01} and \ref{wavelength v0_09} separates the regime $0 \leq \kappa \leq \kappa_{c}$ and $\kappa_{c} \leq \kappa < + \infty$.
\begin{figure}[h!]
\centering
\subfloat[]{ \label{wavelength v0_01}
\includegraphics[width=0.5\linewidth]{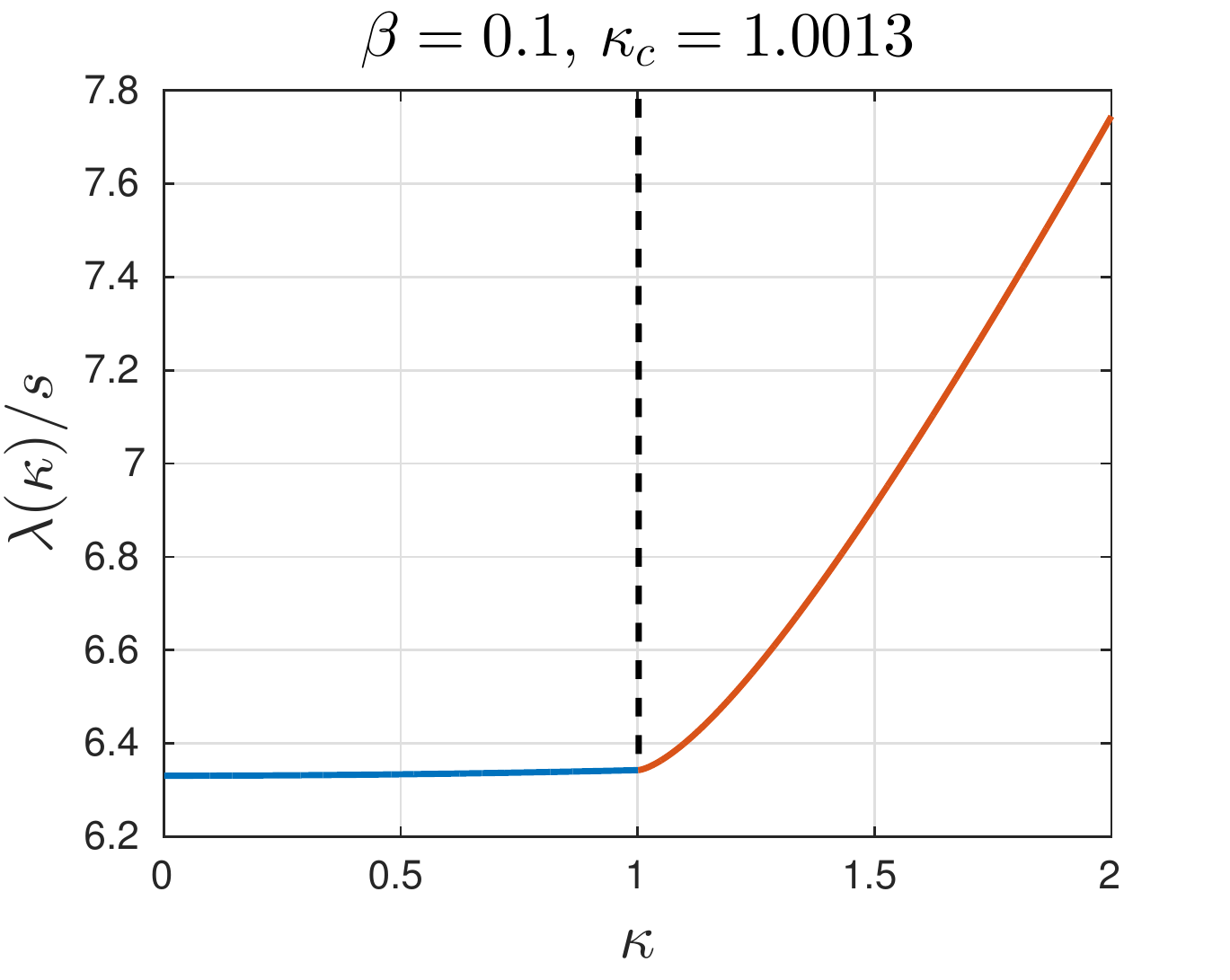}
}
\subfloat[]{ \label{wavelength v0_09}
\includegraphics[width=0.5\linewidth]{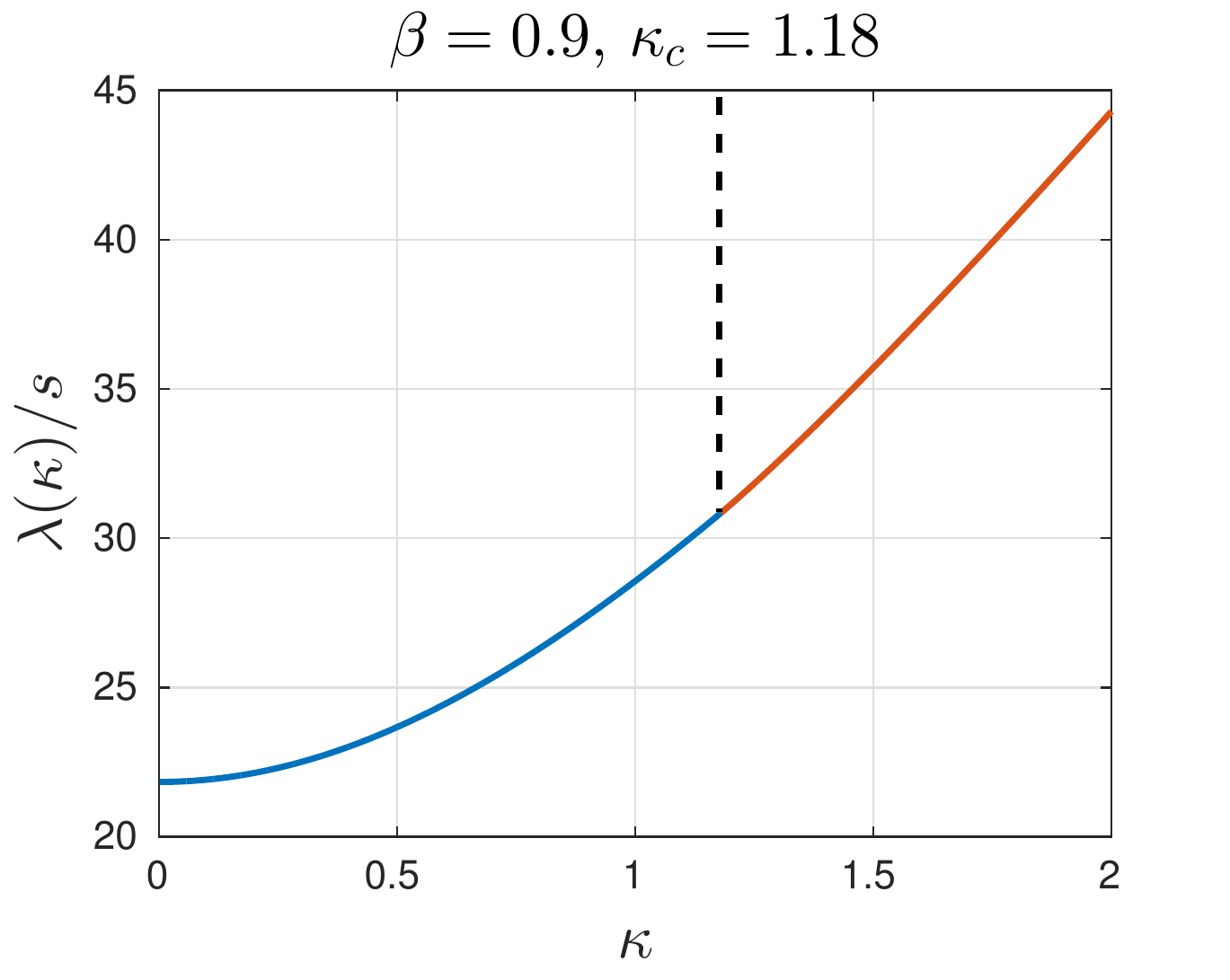}
}
\caption{Variation of wavelength of relativistic Langmuir structure for the velocities (a) $\beta = 0.1$ and (b) $\beta = 0.9$.}
 \label{wavelength}
\end{figure} 
The rate of increase of wavelength with $\kappa$ increases with increasing $\beta$.% Fig. \ref{wavelength comparison} shows wavelength as a function of parameter $\kappa$ for different value of $\beta$= 0.1, 0.9, 0.99 and 0.999.

%\begin{figure} [h!] 
%\centering
%\includegraphics[width=0.5\linewidth]{wavelength_comparison_beta_01_0999.pdf}
%\caption{Variation of wavelength with the parameter $\kappa$ for different values of $\beta$.}
%\label{wavelength comparison}
%\end{figure}

Fig. \ref{potential v0_01} and \ref{potential v0_09} respectively show the potential profile $\Phi(X)$ in the range $\kappa < \kappa_{c}$ and $\kappa > \kappa_{c}$ for two different velocities $\beta = 0.1$ and $\beta = 0.9$. In first case ($\beta = 0.1$), plot (Fig. \ref{potential v0_01 kr_01_09} ) of the potential $\Phi(X)$ is shown for the values of parameter $\kappa = 0.1, 0.3, 0.5, 0.7, 0.9 \, < \kappa_{c} \approx 1.0013$. For these values of $\kappa < \kappa_{c}$ and $\beta = 0.1$, as expected from Fig. \ref{wavelength v0_01} (blue curve), the wavelength is nearly independent of $\kappa$. In second case ($\beta = 0.9$, Fig. \ref{potential v0_09 kr_01_05}), wavelength increases on increasing parameter $\kappa$ (see blue curve in Fig. \ref{wavelength v0_09} ). In the range $\kappa_{c} \geq \kappa$, the minima of the electrostatic potential $\Phi(X)$ is seen at $X = \mu_{c} \approx 3.2$ for the velocity $\beta = 0.1$ ( Fig. \ref{potential v0_01 kr_11}), and for the velocity $\beta \approx 0.9$ ( Fig. \ref{potential v0_09 kr_07}) minima of $\Phi(X)$ is seen at $X = \mu_{c} \approx 16.29$ (These are in agreement with Eq.\eqref{stat:eq38}).

\begin{figure}[h!] 
\centering
\subfloat[]{	\label{potential v0_01 kr_01_09}
\includegraphics[width=0.5\linewidth]{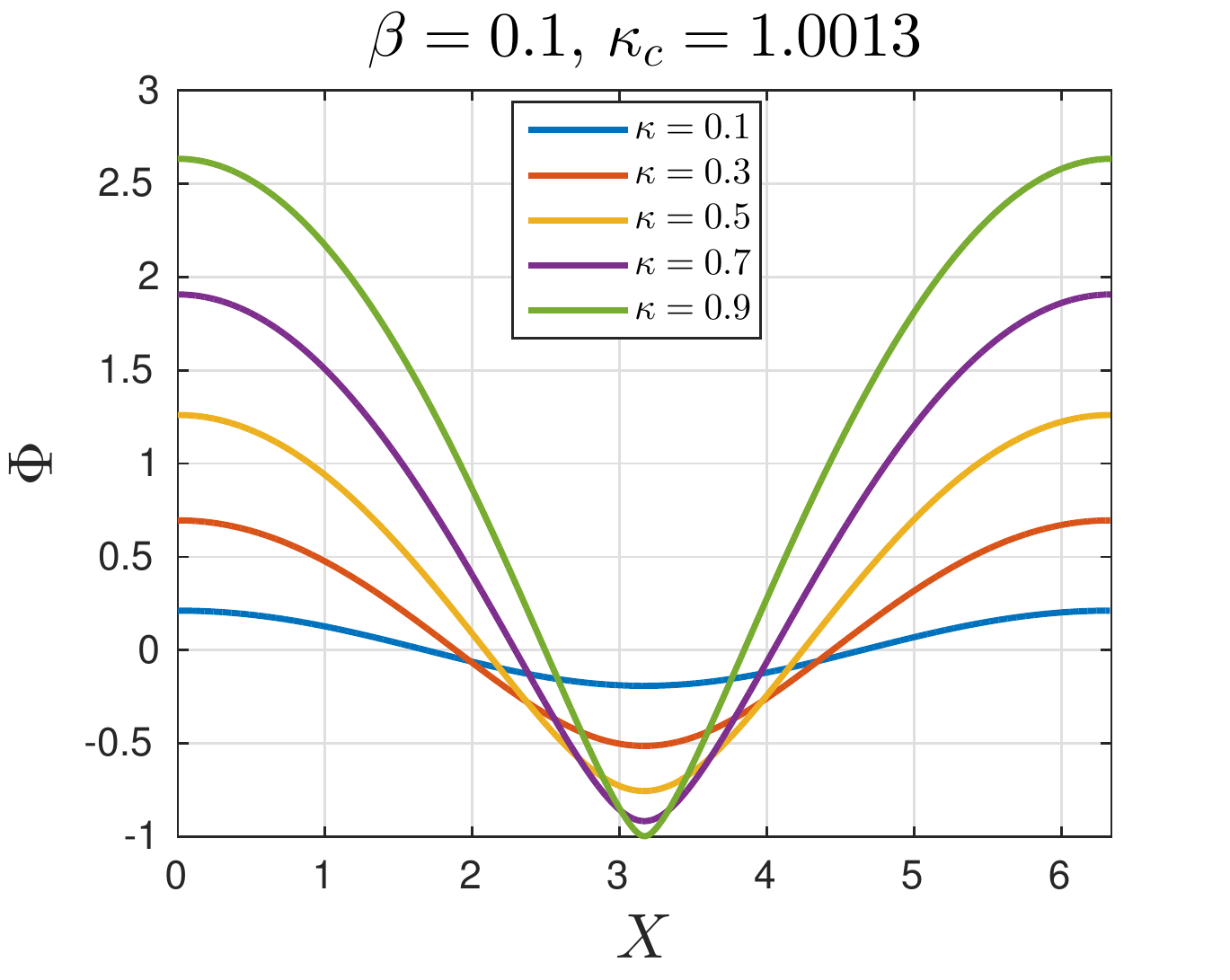}
}
\subfloat[]{ \label{potential v0_09 kr_01_05}
\includegraphics[width=0.5\linewidth]{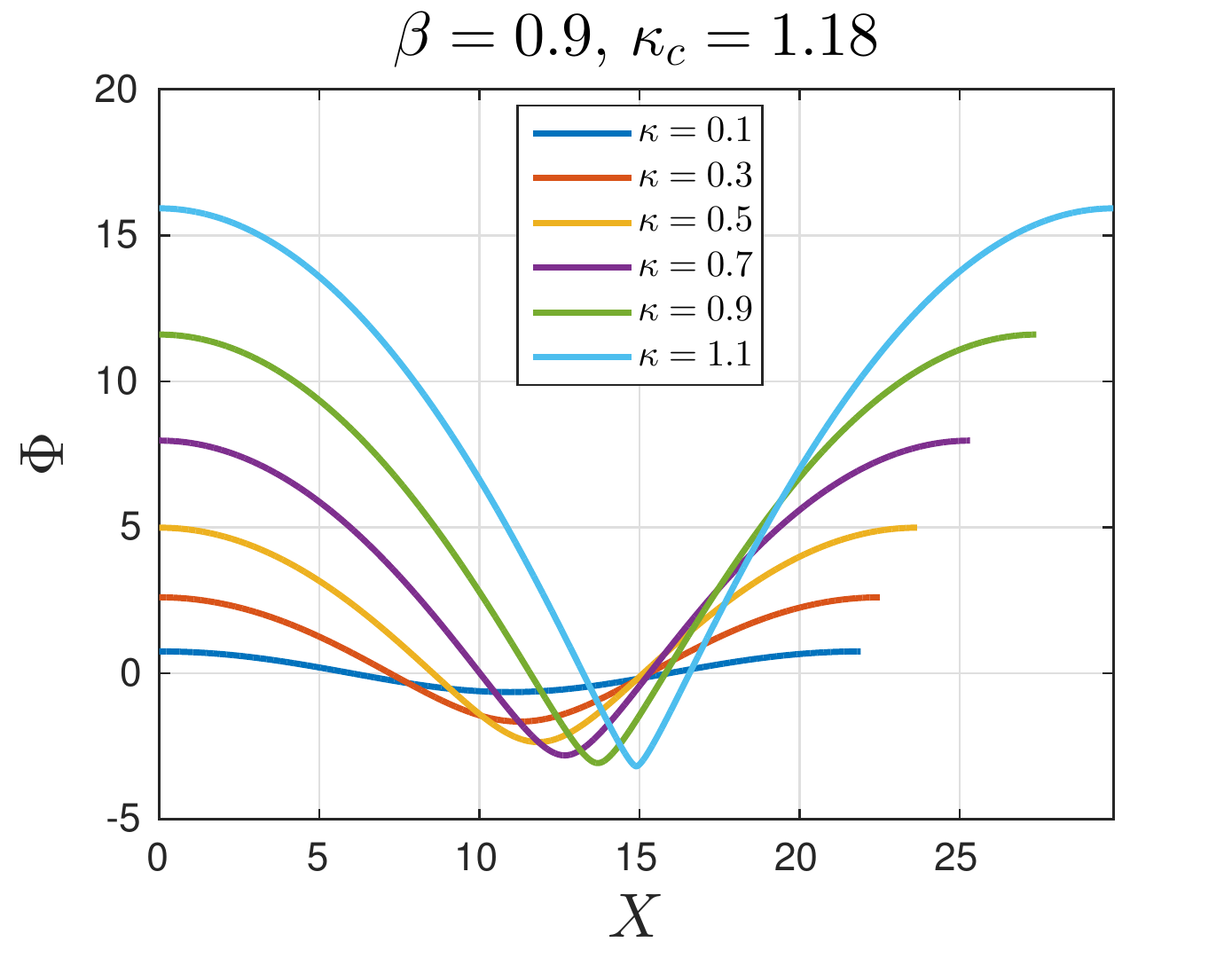}
}
\caption{Plot of electrostatic potential $\Phi(X)$ for (a) $\kappa = 0.1, 0.3, 0.5, 0.7, 0.9$ and $\beta = 0.1$, (b) $\kappa =$ 0.1, 0.3, 0.5, 0.7, 0.9, 1.1 and $\beta = 0.9$.}
\label{potential v0_01}
\end{figure}
\begin{figure}	
\centering
\subfloat[]{	\label{potential v0_01 kr_11}
\includegraphics[width=0.5\linewidth]{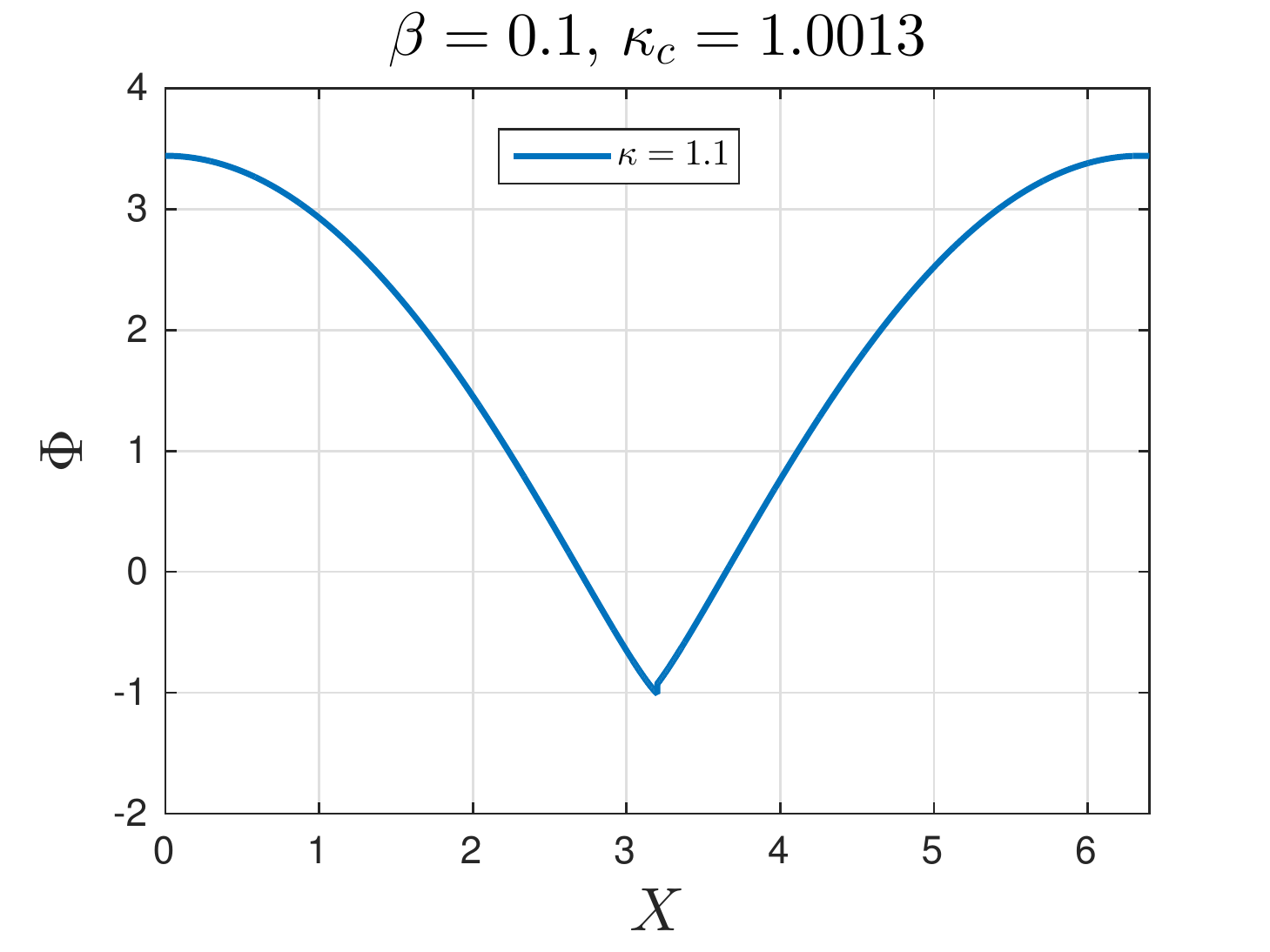}
}
\subfloat[]{	\label{potential v0_09 kr_07}
\includegraphics[width=0.5\linewidth]{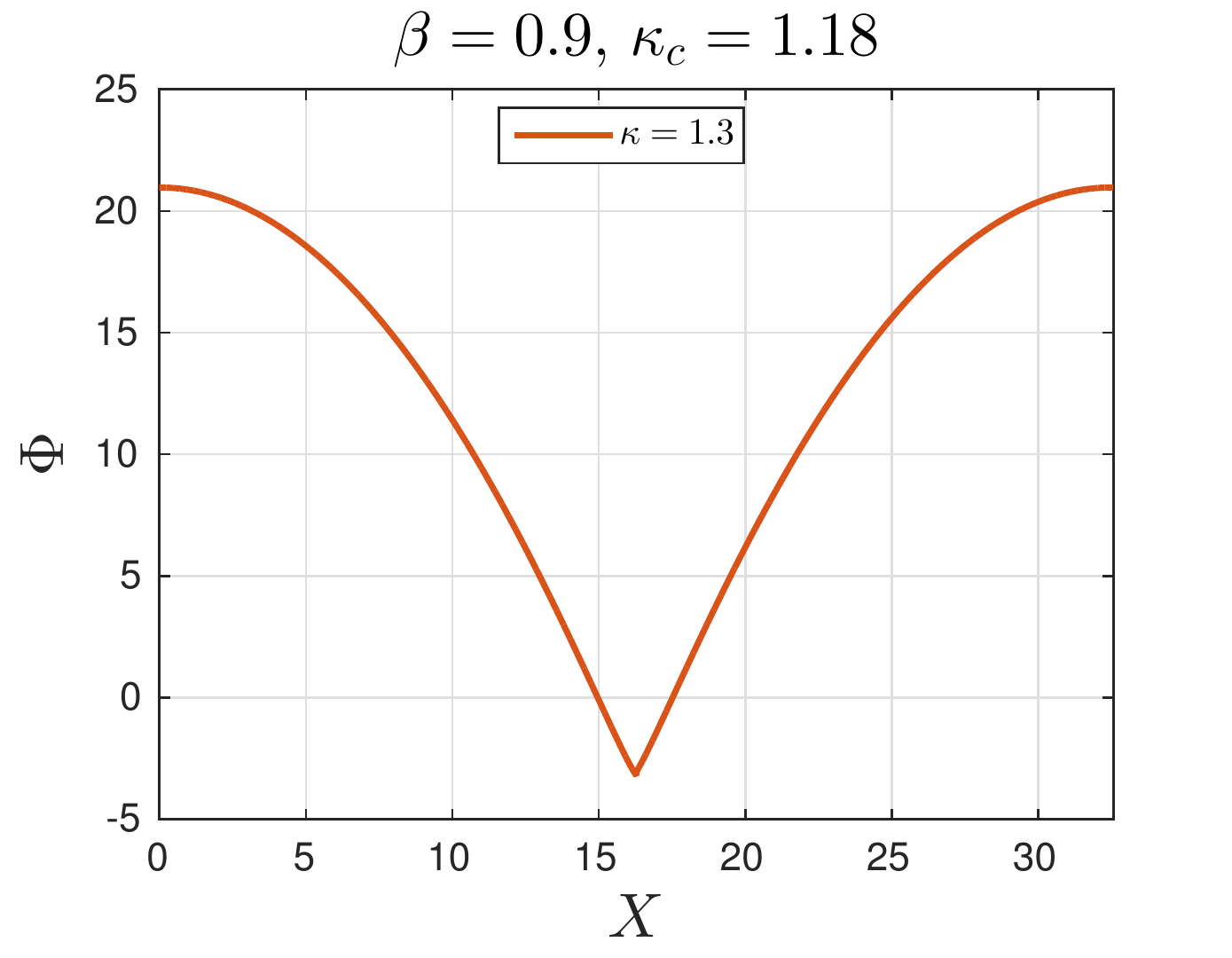}
}
\caption{Plot of electrostatic potential $\Phi(X)$ for the parameters (a) $\kappa = 1.1$, $\beta = 0.1$ and (b) $\kappa = 1.3$, $\beta = 0.9$.}
\label{potential v0_09}
\end{figure}

We now present electric field in terms of position $X$ by solving Eq.\eqref{stat:eq19} and considering two branches depending on the sign of potential $\Phi$ as
\begin{equation*}
\Phi>0 ; \, \, \Phi = \gamma^{2}_{0} (\kappa^{2} - E^{2}) \left( 1 +  \left( \beta^{2} + \frac{4}{\gamma_{0}(\kappa^{2} - E^{2}) }\right)^{1/2} \right),
\end{equation*}
\begin{equation} \label{stat:eq39}
\Phi<0 ; \, \, \Phi = \gamma^{2}_{0} (\kappa^{2} - E^{2}) \left( 1 -  \left( \beta^{2} + \frac{4}{\gamma_{0}(\kappa^{2} - E^{2}) }\right)^{1/2} \right).
\end{equation}
Range of $E$ can be estimated using Eq.\eqref{stat:eq19}: (i) if $0 \leq \kappa \leq \kappa_{c}$, we have $0 \leq |E| \leq \kappa$ for both branches; (ii) if $ \kappa_{c} \leq \kappa < + \infty$ we have $0 \leq |E| \leq \kappa$ for $\Phi > 0$ and $\sqrt{\kappa^{2} - \kappa^{2}_{c}} \leq |E| \leq \kappa$ for $\Phi < 0$. 

Fig. \ref{electric field v0_01 kr_01_09} and \ref{electric field v0_09 kr_01_05} respectively show spatial variation of electric field for two particular case $\beta =$ 0.1 and 0.9. In first case when $\beta =$ 0.1 in Fig. \ref{electric field v0_01 kr_01_09}, gradual steepening of electric field is seen at $X = \mu \approx \pi$ as $\kappa \rightarrow \kappa_{c}$; $\Phi \rightarrow \Phi^{c}$. Similar feature is seen in the second case when $\beta = 0.9$, where gradual steepening of $E$ occurs at $X = \mu$ as shown in Fig. \ref{electric field v0_09 kr_01_05}. When $\kappa > \kappa_{c}$, a discontinuity of $E$ is seen at the position $X = \mu_{c} \approx 3.2$  for $\beta = 0.1$ and at $X = \mu_{c} \approx 16.29$ for $\beta = 0.9$ as shown in Fig. \ref{electric field v0_01 kr_11} and \ref{electric field v0_09 kr_07} respectively.

The gradual steepening of the electric field at $X = \mu$ as $\kappa \rightarrow \kappa_{c}$ can be seen from the expression of spatial derivative of electric field. In the range $0 \leq \kappa <  \kappa_{c}$, we obtain that at $X = \mu$, $E = 0$ and
\begin{equation} \label{stat:eq40}
\frac{dE}{dX} =  \left(1 - \frac{ (2 \gamma_{0} + \beta^{2} \Phi)}{(4 \gamma^{2}_{0}+ 4 \gamma_{0} \Phi + \beta^{2} \Phi^{2} )^{1/2}} \right),
\end{equation}
which is always negative if $\Phi^{c} < \Phi < 0$. A gradual steepening of wave form occurs as $\Phi \rightarrow \Phi^{c}$. At $\Phi = \Phi^{c}$, $\kappa$ becomes $\kappa_{c}$, which implies that at $X = \mu_{c}$, $E = 0$ and $dE/dX = - \infty$. If $\kappa > \kappa_{c}$, $E$ becomes discontinuous at $X = \mu_{c}$ and $E$ jumps from $\sqrt{\kappa^{2} - \kappa^{2}_{c}}$ to $-\sqrt{\kappa^{2} - \kappa^{2}_{c} }$. This jump in electric field implies formation of negatively charged plane at $X = \mu_{c}$. The surface charge density $\rho$ of these planes is defined as
\begin{equation}
\rho =  \Delta E/ 4\pi  = \frac{E_{0}}{2\pi} (\kappa^{2} - \kappa^{2}_{c})^{1/2}.
\end{equation}
It is found that in the limit $v_{0} \rightarrow 0$ and/or $\kappa \rightarrow + \infty$, electron beam is transformed into a crystal of "negatively charged plane" with inter-distance $\lambda_{0} = E_{m}/2\pi n_{0} e$ having surface charge density $\sim E_{m}/2\pi$, which matches with the results found in the non-relativistic regime \cite{Psimopoulos_pop_1997}.

\begin{figure}[h!] 
\centering
\subfloat[]{ \label{electric field v0_01 kr_01_09}
\includegraphics[width=0.5\linewidth]{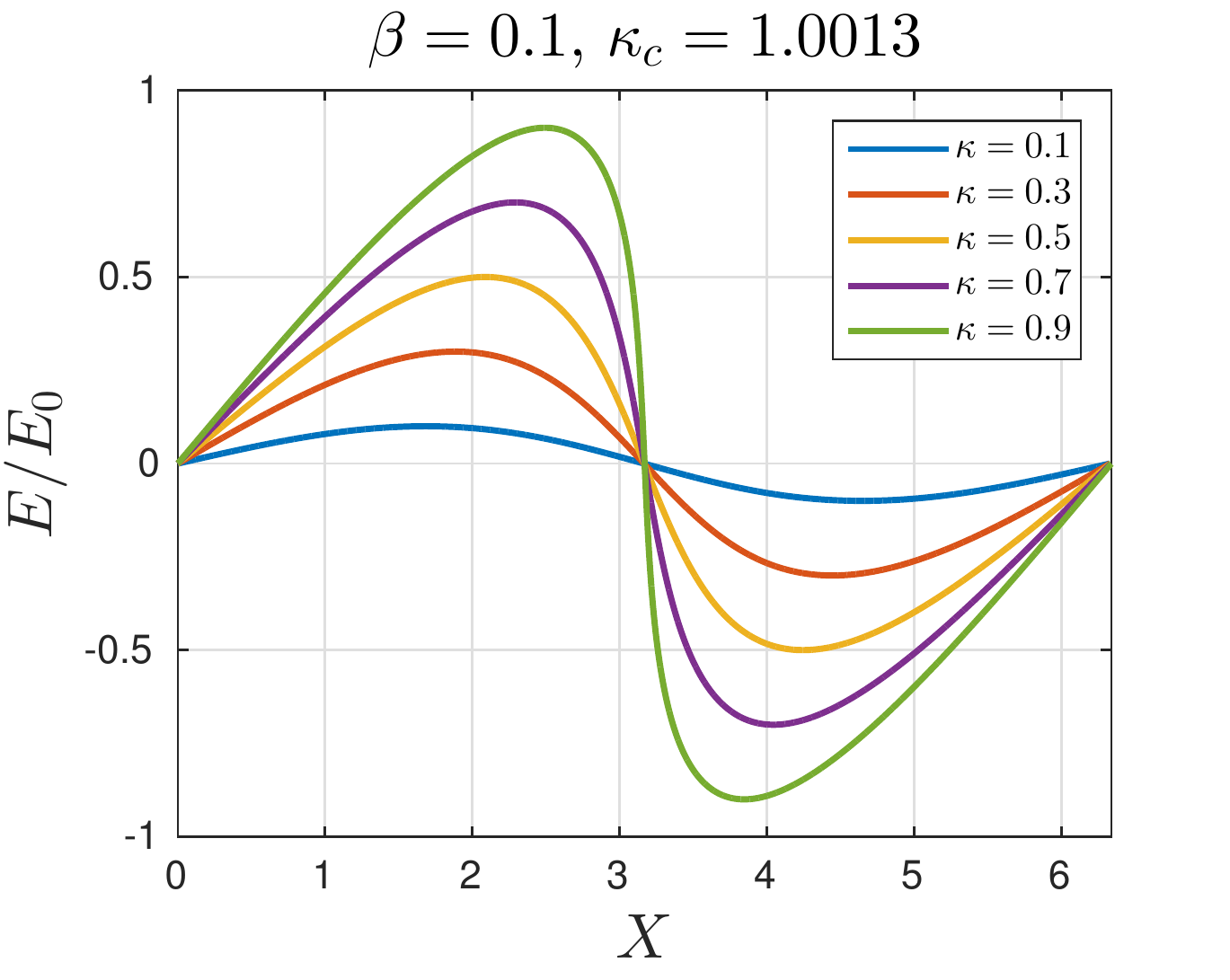}
}
\subfloat[]{	 \label{electric field v0_09 kr_01_05}
\includegraphics[width=0.5\linewidth]{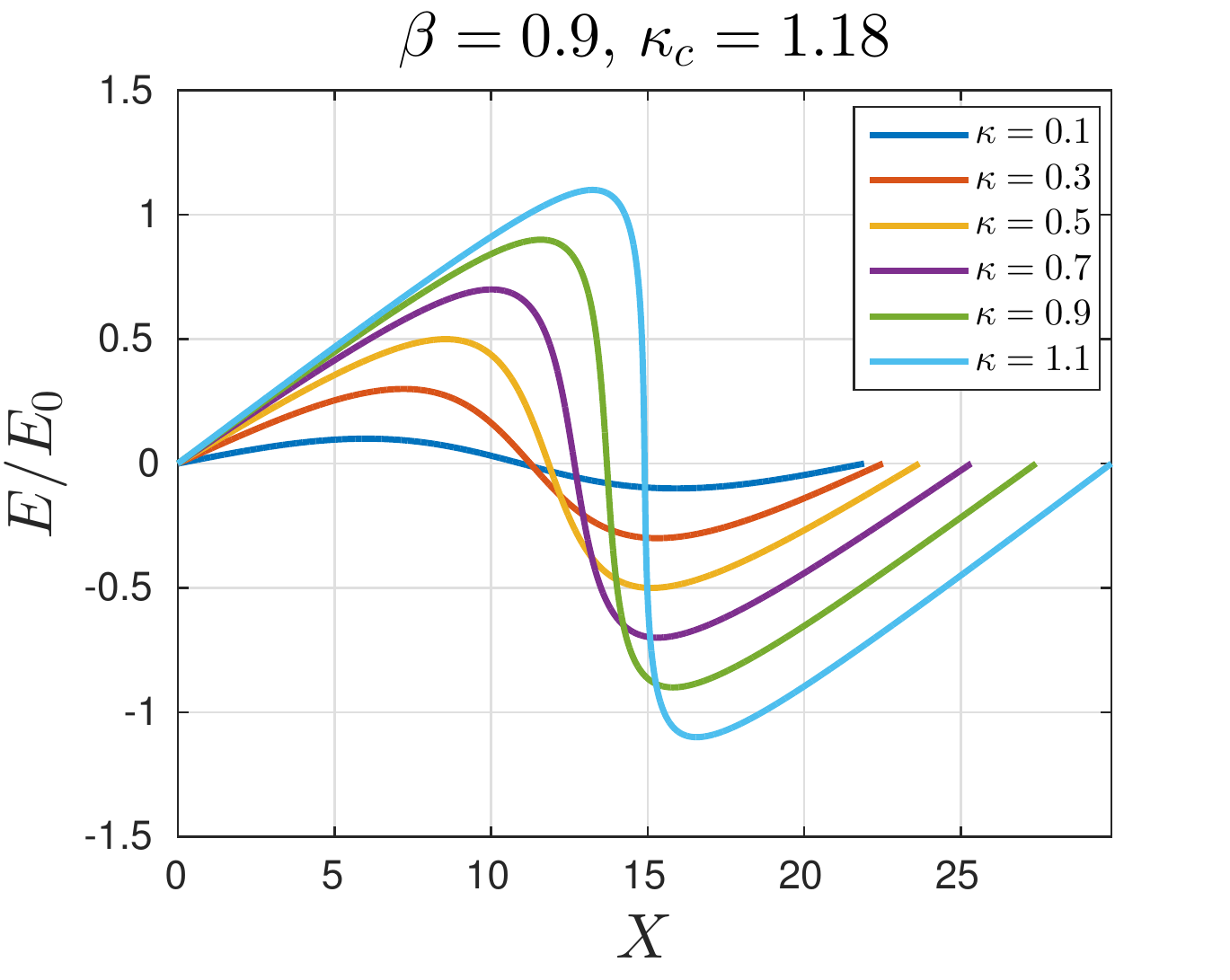}
}
\caption{ Plot of electric field for (a) $\beta = 0.1$; $\kappa = 0.1,\, 0.3,\, 0.5,\, 0.7,\, 0.9$ (b) $\beta = 0.9$; $\kappa = 0.1,\, 0.3,\, 0.5,\, 0.7,\, 0.9, \, 1.1$.}
\label{electric field v0_01}
\end{figure}
\begin{figure}[h!]	
\centering
\subfloat[]{  \label{electric field v0_01 kr_11}
\includegraphics[width=0.5\linewidth]{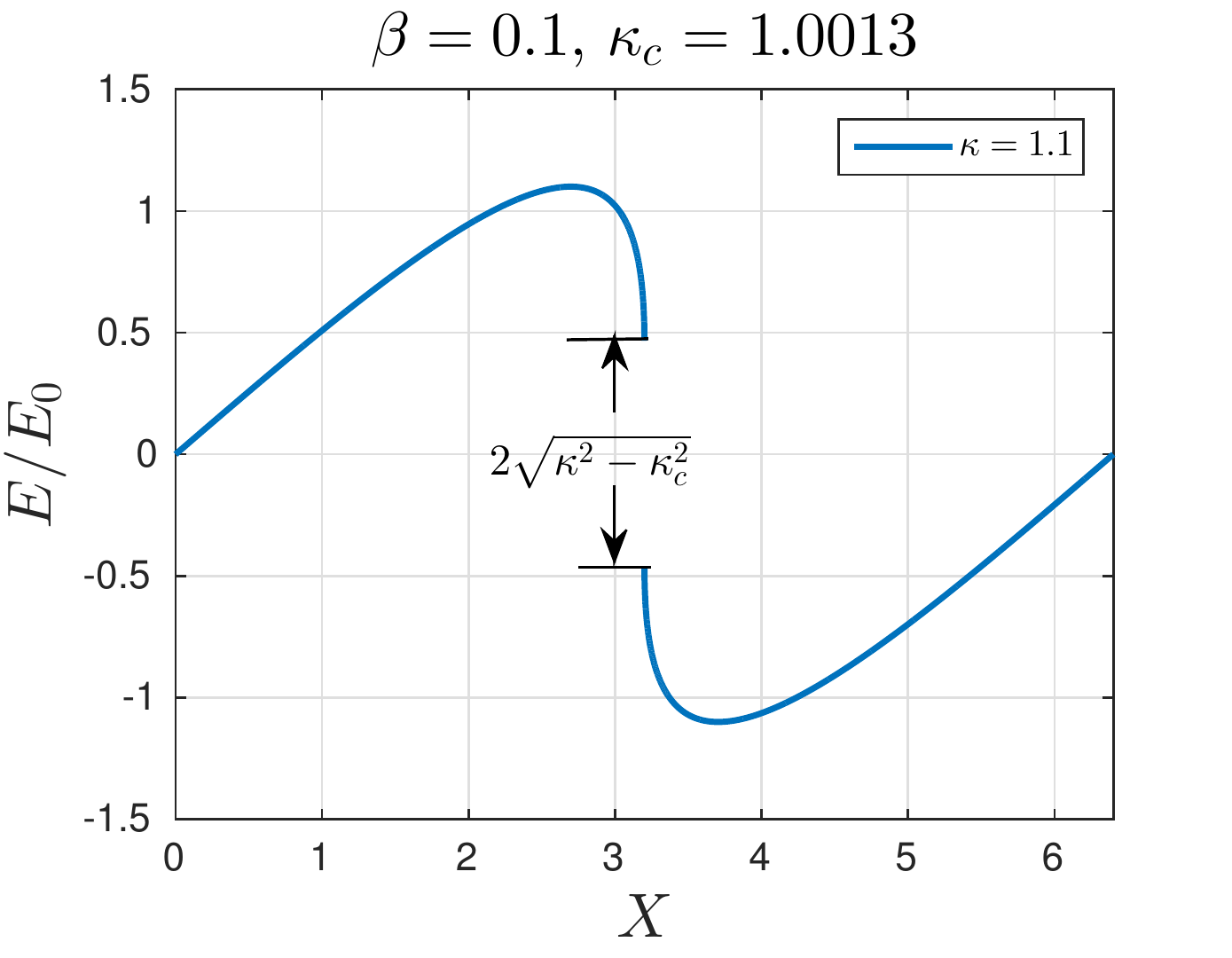}
}
\subfloat[]{ \label{electric field v0_09 kr_07}
\includegraphics[width=0.5\linewidth]{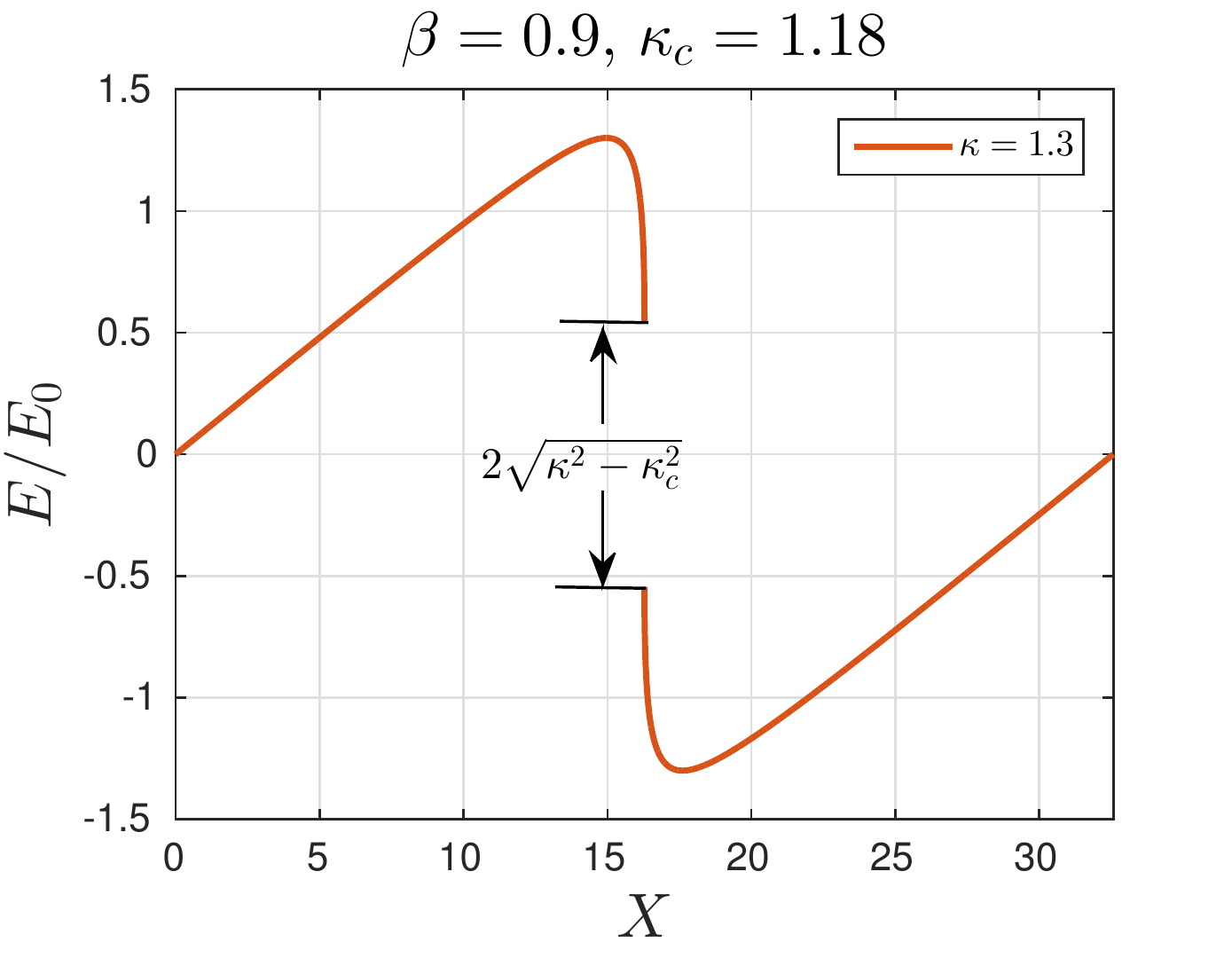}
}
\caption{Plot of electric field, where discontinuity of electric field is seen at (a) $X = \mu_{c} \approx 3.2$ for $\beta = 0.1$; $\kappa = 1.1$ and at (b) $X = \mu_{c} \approx 16.29$ for $\beta = 0.9$; $\kappa = 1.3$.}
\label{electric field v0_09}
\end{figure}

The gradual steepening and discontinuity of the electric field modulates the electron velocity profile. The $E$ vs $v_{e}$ relation can be constructed using Eq.\eqref{stat:eq15} and \eqref{stat:eq19} as
\begin{equation}	\label{stat:eq41}
E^{2} - \kappa^{2} = \frac{2}{\gamma_{0}\beta^{2}} \left( 1 - \frac{\gamma_{0} \left(1 - \beta^{2} v_{e} \right) }{\sqrt{1-v^{2}_{e} \beta^{2}}} \right)
\end{equation}
\begin{figure}[h!]  
\centering
\subfloat[]{
\includegraphics[width=0.5\linewidth]{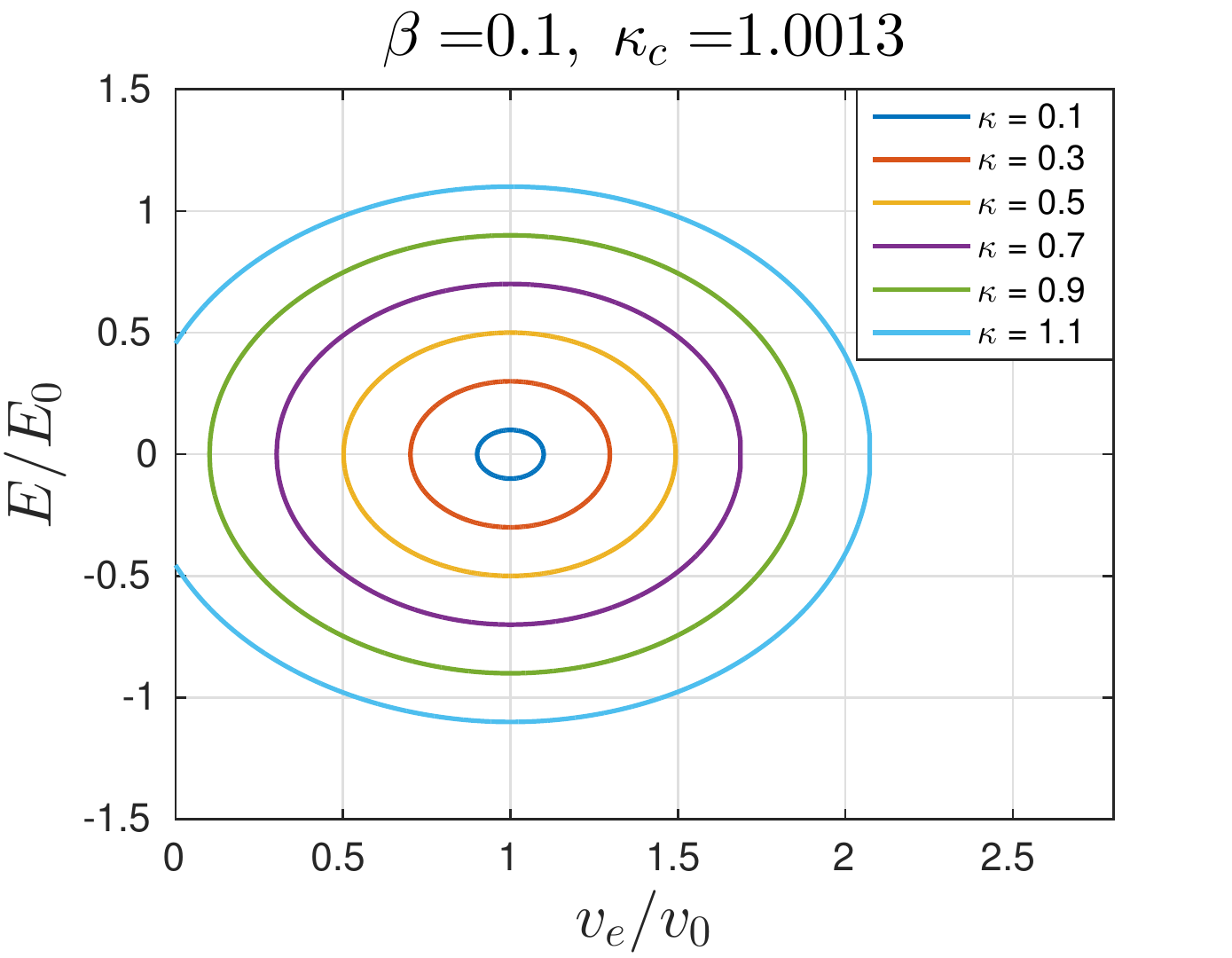}
}
\subfloat[]{
\includegraphics[width=0.5\linewidth]{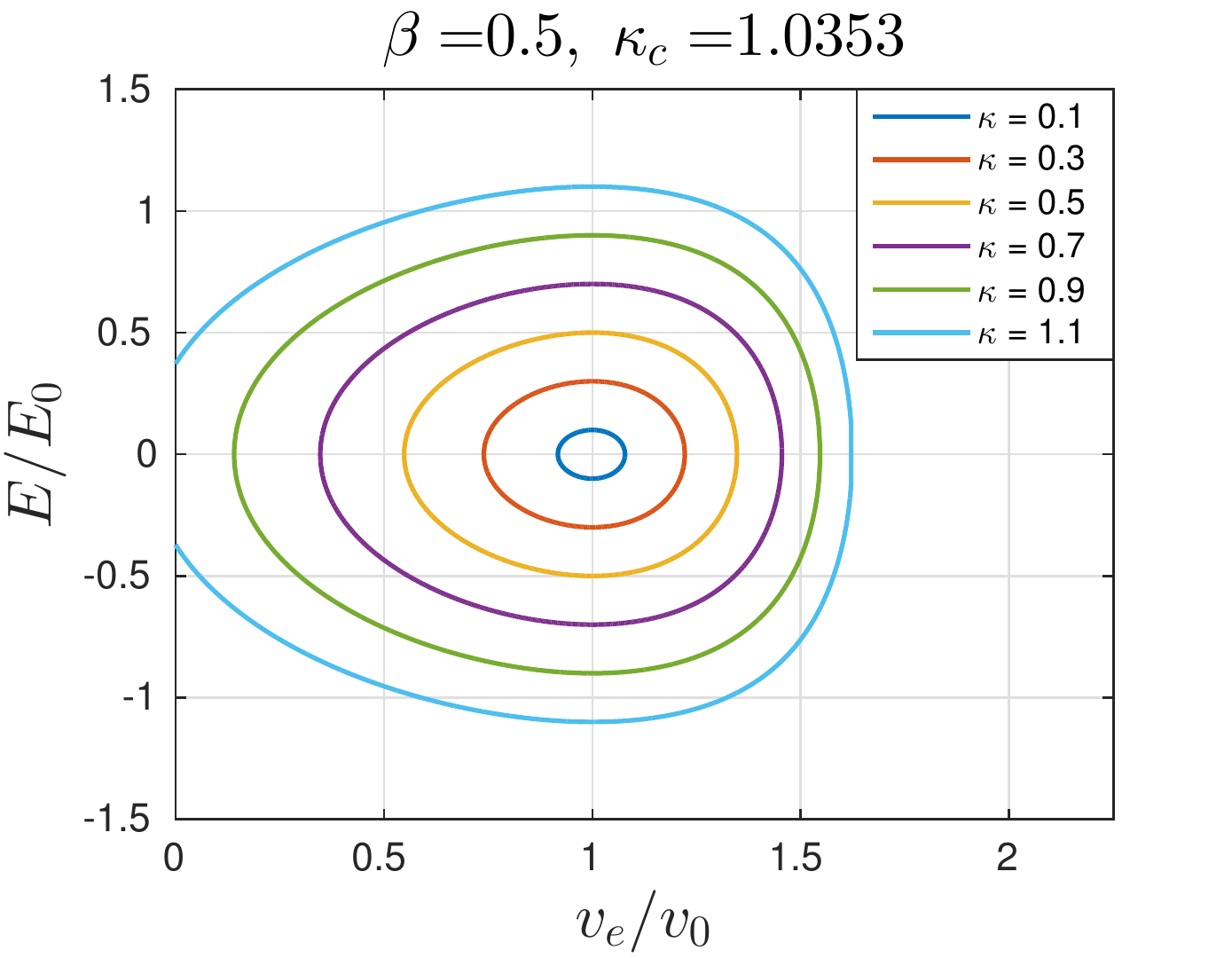}
}\\
\subfloat[]{
\includegraphics[width=0.5\linewidth]{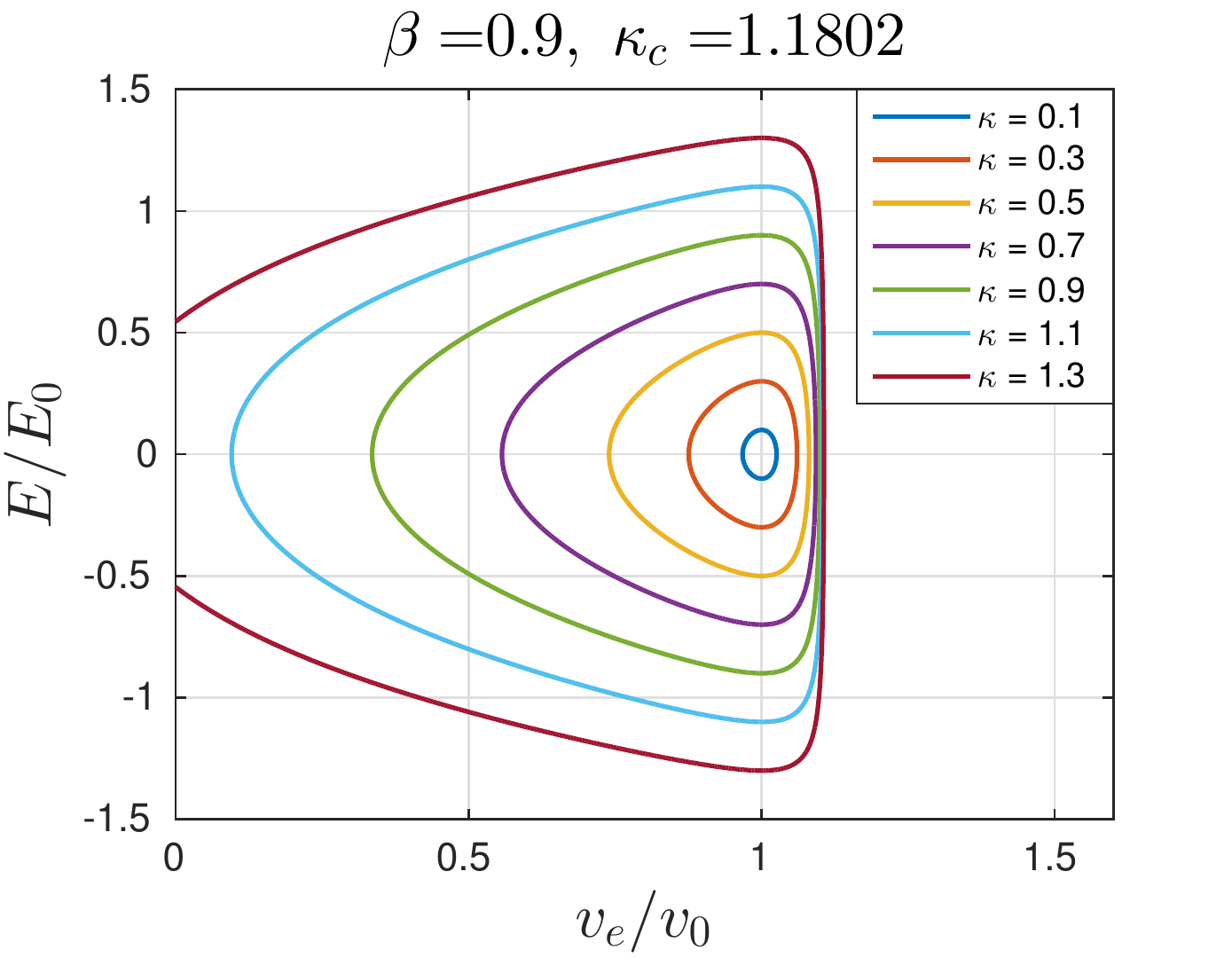}
}
\subfloat[]{
\includegraphics[width=0.5\linewidth]{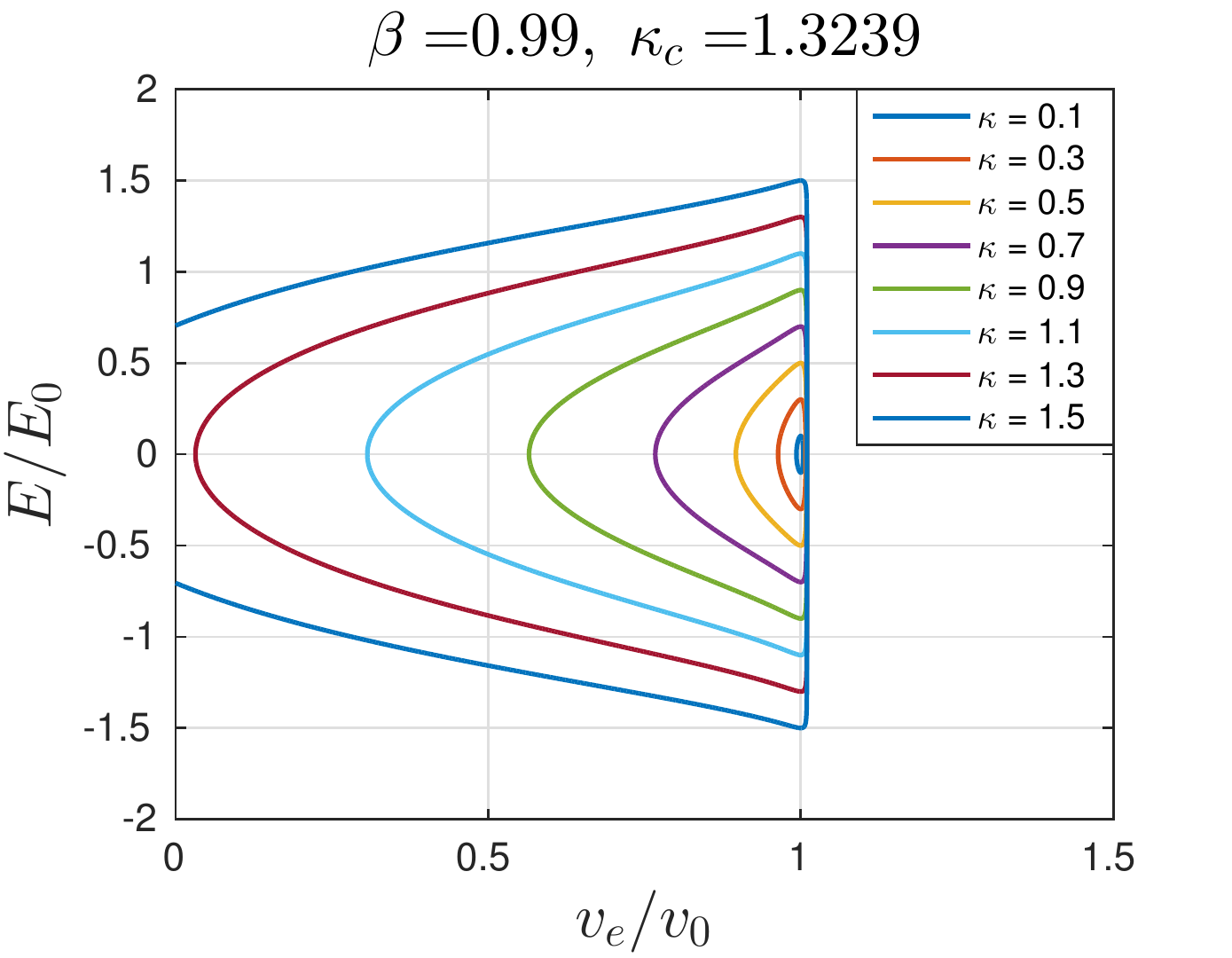}
}
\caption{Plot $E - v_{e}$ phase space for different values of parameter $\kappa$ and for the velocities (a) $\beta = 0.1$, (b) $\beta = 0.5$, (c) $\beta = 0.9$, (d) $\beta = 0.99$.}
\label{e-v phase space}
\end{figure}
Fig. \ref{e-v phase space} shows $E - v_{e}$ phase space for different values of parameter $\kappa$ and $\beta$. It is readily seen that in the range $\kappa > \kappa_{c}$, $E - v_{e}$ phase space becomes discontinuous \cite{Sen_pr_1955} and $E$ jumps from $\sqrt{\kappa^{2} - \kappa^{2}_{c}}$ to $-\sqrt{\kappa^{2} - \kappa^{2}_{c}}$.

The fluid velocity as function of position can be obtained by combining Eq.\eqref{stat:eq15} with Eqs.\eqref{stat:eq35}, \eqref{stat:eq36b} and \eqref{stat:eq36c}. From Eq. \eqref{stat:eq15}, electron fluid velocity can be explicitly written as a function of potential as

\begin{equation} \label{stat:eq42}
v_{e} =  \frac{(4 \gamma^{2}_{0}+ 4 \gamma_{0} \Phi + \beta^{2} \Phi^{2} )^{1/2}}{(2 \gamma_{0} + \beta^{2} \Phi)} ,
\end{equation}
where we have chosen the positive sign only in order to explicitly exclude the existence of trapped electrons. As $\kappa \rightarrow \kappa_{c}$, $\Phi \rightarrow \Phi^{c}$ at the position $X = \mu$, numerator of Eq.\eqref{stat:eq42} tends to zero, thus, a gradual decrement in electron velocity occurs at the position $X = \mu$. If $\kappa \geq \kappa_{c}$ then $\Phi = \Phi^{c}$ at the position $X = \mu_{c}$, this implies that numerator of the Eq. \eqref{stat:eq42} becomes zero at that position, in other words velocity becomes zero. This means electrons stop momentarily at the position $X = \mu_{c}$ and then continue their motion in $+x$ direction. This slowing down (for $\kappa < \kappa_{c}$) and momentarily stopping (for $\kappa > \kappa_{c}$) of the electrons leads to the accumulation of the charge particles at the position $X = \mu_{c}$ which manifests as a density burst; in other words, in order to maintain flux, electron number density increases at the positions where fluid velocity decreases. Fig. \ref{velocity beta_01 kr_01_09} and \ref{velocity beta_09 kr_01_05} show electron velocity for the velocities $\beta =$ 0.1 and 0.9 respectively, and a gradual decrement of electron velocity with increasing $\kappa$ can be seen clearly at the position $X = \mu \approx \pi$ for $\beta = 0.1$ and at $X = \mu_{c}$ for $\beta = 0.9$. Fig. \ref{velocity beta_01 kr_11} and \ref{velocity_beta_09 kr_07} illustrates that in the limit $\kappa \geq \kappa_{c}$, $\Phi = \Phi^{c}$, velocity becomes zero at the position satisfying $X = \mu_{c} \approx 3.2$ for $\beta =  0.1$ and $X = \mu_{c} \approx 16.29$ for $\beta = 0.9$.
\begin{figure} [h!]
\centering
\subfloat[]{ \label{velocity beta_01 kr_01_09}
\includegraphics[width=0.5\linewidth]{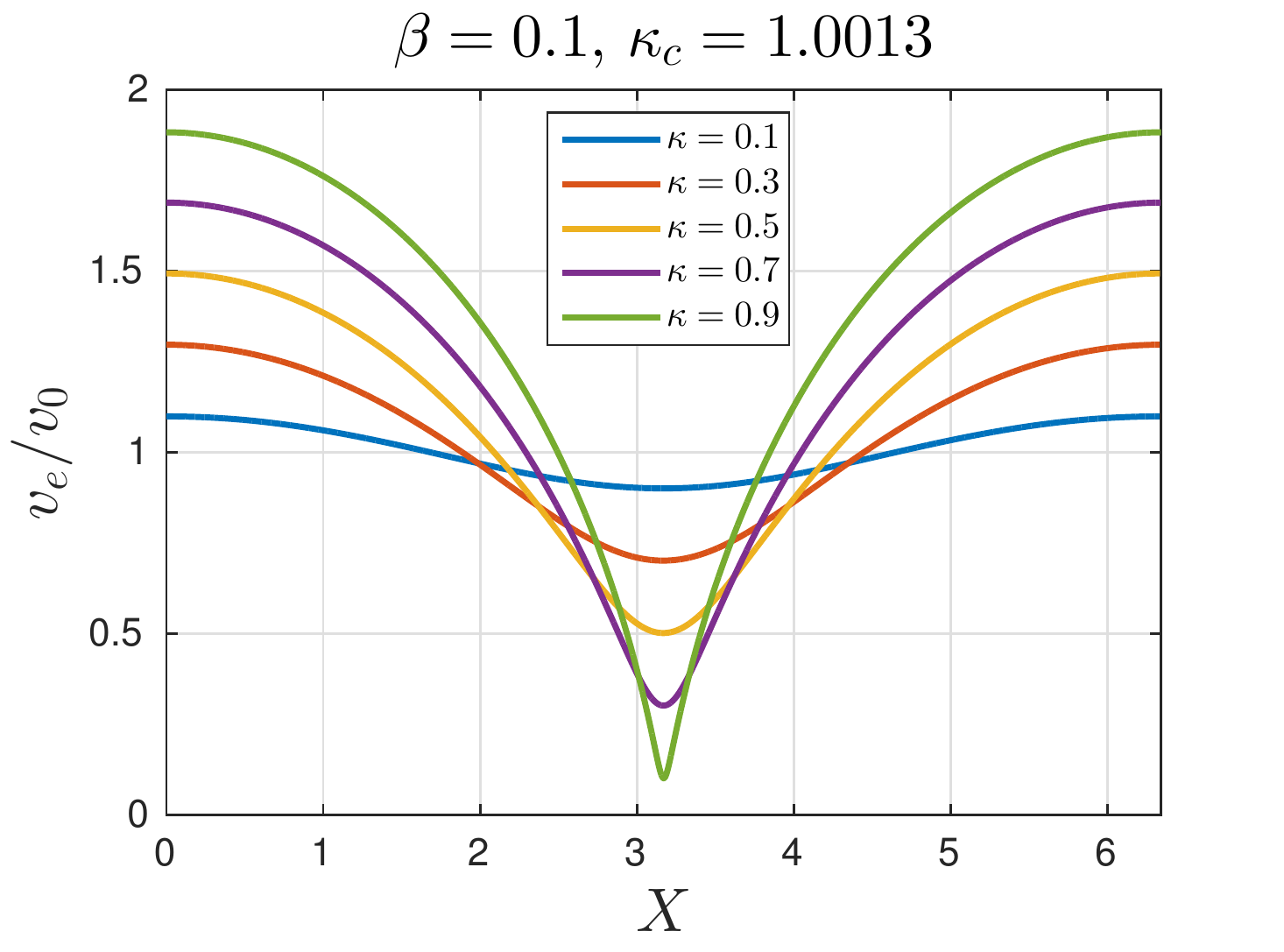}
}
\subfloat[]{	\label{velocity beta_09 kr_01_05}
\includegraphics[width=0.5\linewidth]{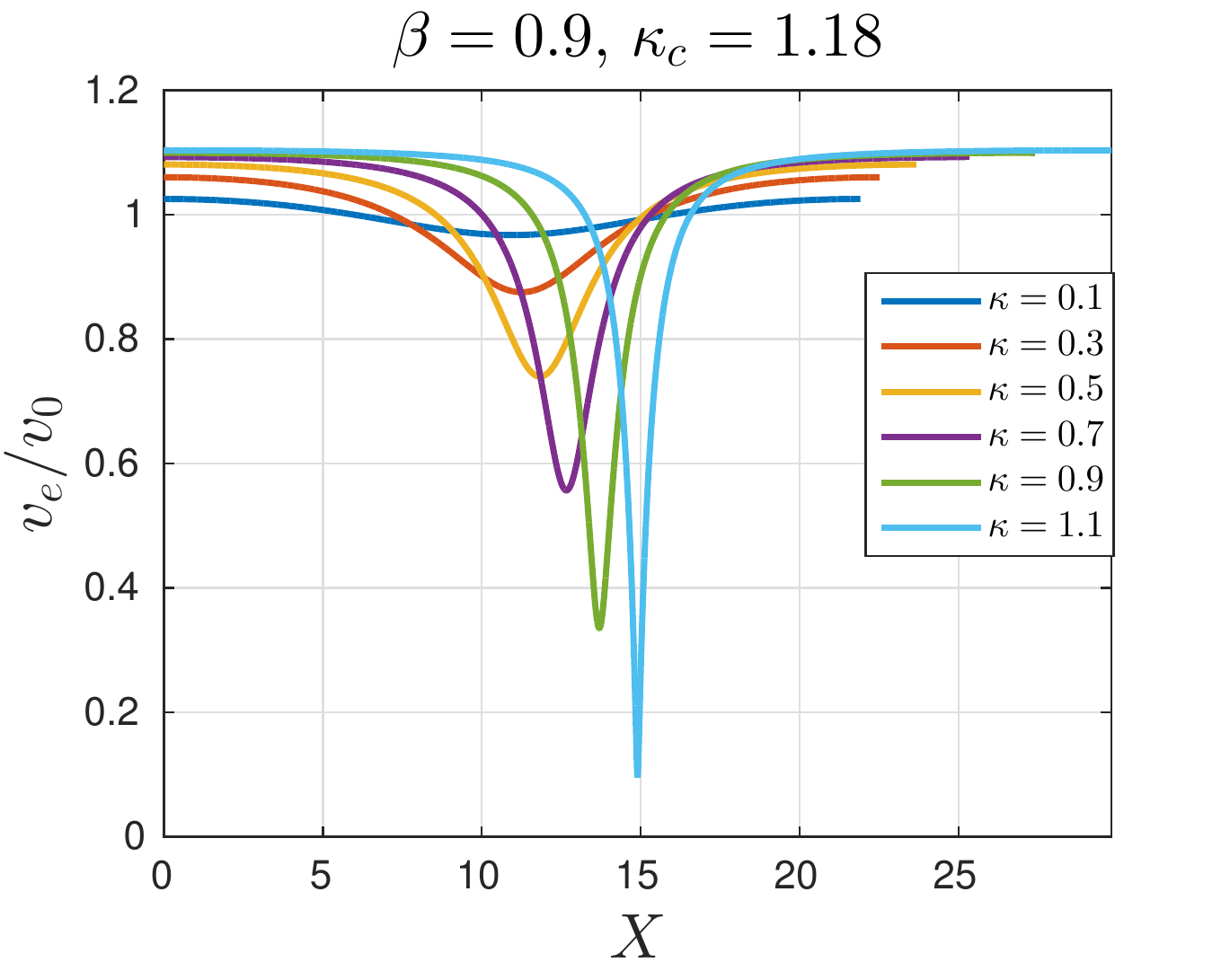}
}
\caption{Plot of electron velocity for the parameters (a) $\beta= 0.1$; $\kappa = 0.1, 0.3, 0.5, 0.7, 0.9$ (b) $\beta = 0.9$; $\kappa = 0.1, 0.3, 0.5, 0.7. 0.9, 1.1$.}
\end{figure}
\begin{figure}
\centering
\subfloat[]{ \label{velocity beta_01 kr_11}
\includegraphics[width=0.5\linewidth]{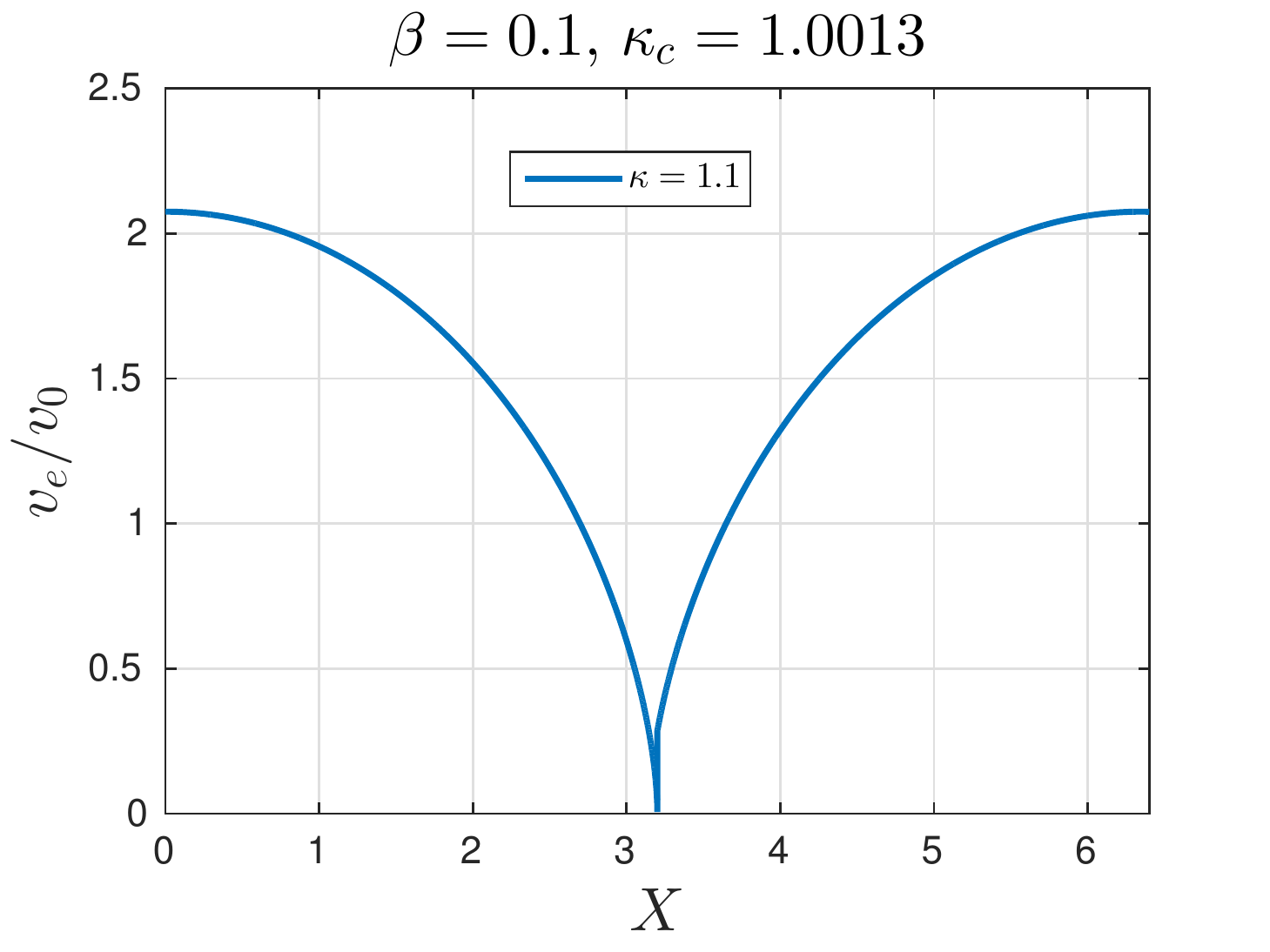}
}
\subfloat[]{	\label{velocity_beta_09 kr_07}
\includegraphics[width=0.5\linewidth]{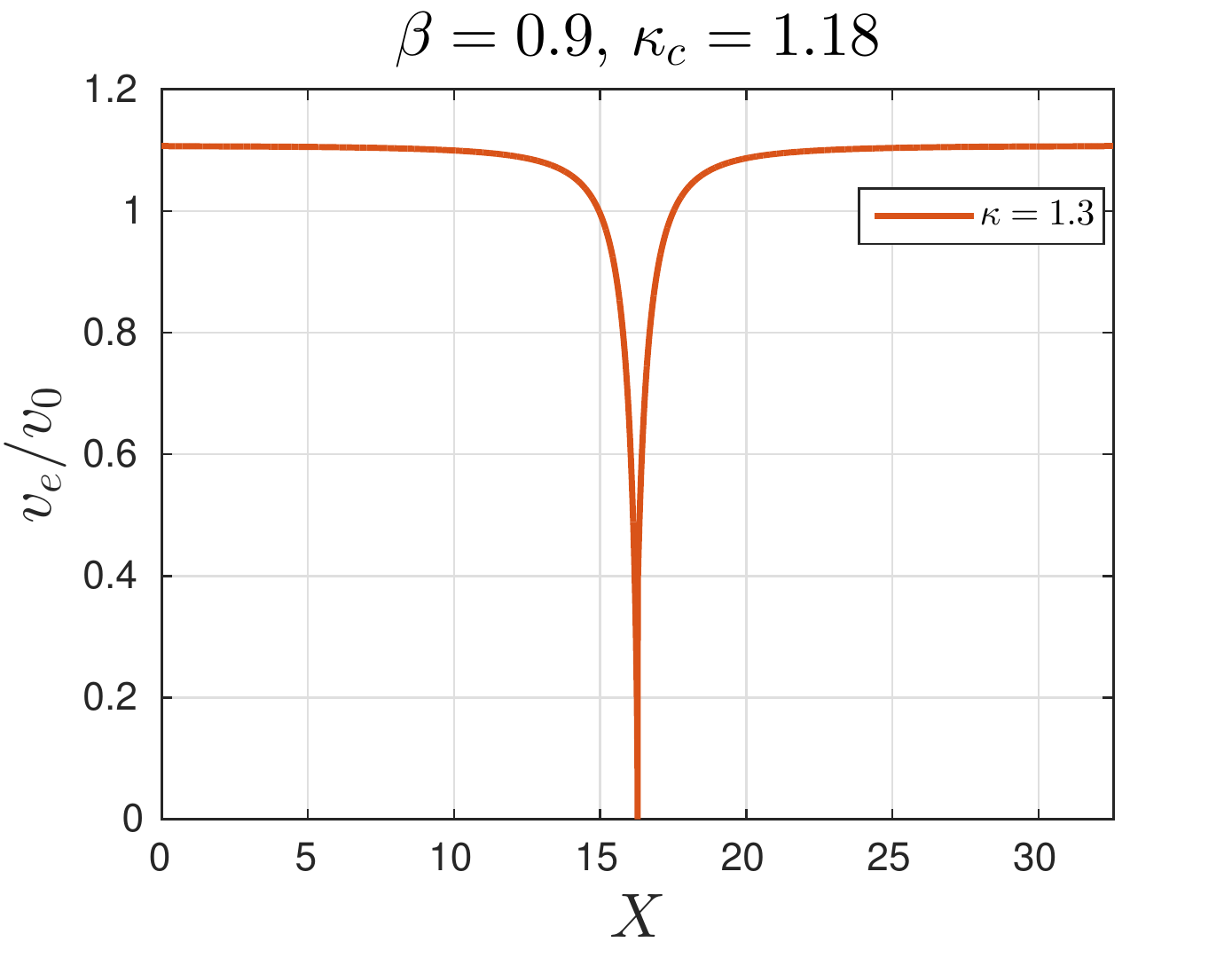}
}
\caption{Plot of electron velocity for the parameters (a) $\beta = 0.1$, $\kappa = 1.1$ and (b) $\beta = 0.9$, $\kappa = 1.3$.}
\end{figure}

The electron density can be written as
\begin{equation} \label{stat:eq43}
n_{e}(X) = 1 -  \frac{\partial E}{\partial X}.
\end{equation}
Using Eq.\eqref{stat:eq40}, the electron density therefore may be written as a function of electrostatic potential as
\begin{equation}	\label{stat:eq44}
n_{e}(X) = \frac{ (2 \gamma_{0} + \beta^{2} \Phi)}{(4 \gamma^{2}_{0}+ 4 \gamma_{0} \Phi + \beta^{2} \Phi^{2} )^{1/2}}
\end{equation}
Eq.\eqref{stat:eq44} along with Eqs.\eqref{stat:eq35}, \eqref{stat:eq36b} and \eqref{stat:eq36c} gives the relation between electron density and spatial position. In the range $0 \leq \kappa < \kappa_{c}$, as $\Phi$ approaches $\Phi^{c}$, steepening of the density can be clearly seen in the Fig. \ref{density v0_01 kr_01_09} for $\beta = 0.1$ at $X \approx \pi$ and in Fig. \ref{density v0_09 kr_01_05} for $\beta = 0.9$ at $X = \mu$. When $\kappa \geq \kappa_{c}$ then $\Phi = \Phi^{c}$ and denominator of the Eq.\eqref{stat:eq44} vanishes. This explosive behavior beyond $\kappa_{c}$ can be clearly seen in Figs. \ref{density v0_01 kr_11} and \ref{density v0_09 kr_07}, where density burst is seen at $X = \mu_{c} \approx 3.2$ for $\beta = 0.1$ (in Fig. \ref{density v0_01 kr_11}) and at $X = \mu_{c} \approx 16.29$ for $\beta = 0.9$ (in Fig. \ref{density v0_09 kr_07}) respectively.
\begin{figure} [h]
\centering
\subfloat[]{ \label{density v0_01 kr_01_09}
\includegraphics[width=0.5\linewidth]{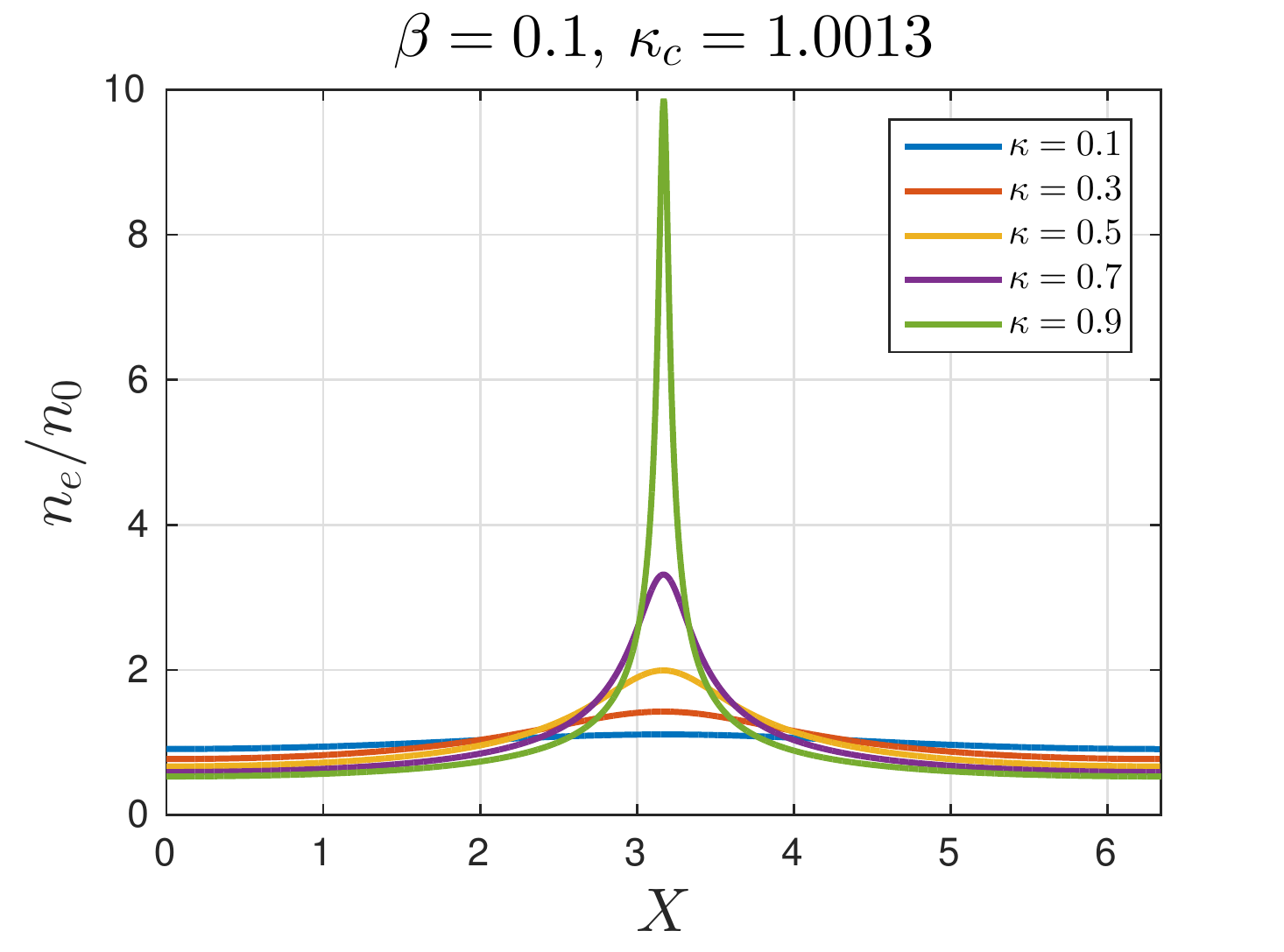}
}
\subfloat[]{ \label{density v0_09 kr_01_05}
\includegraphics[width=0.5\linewidth]{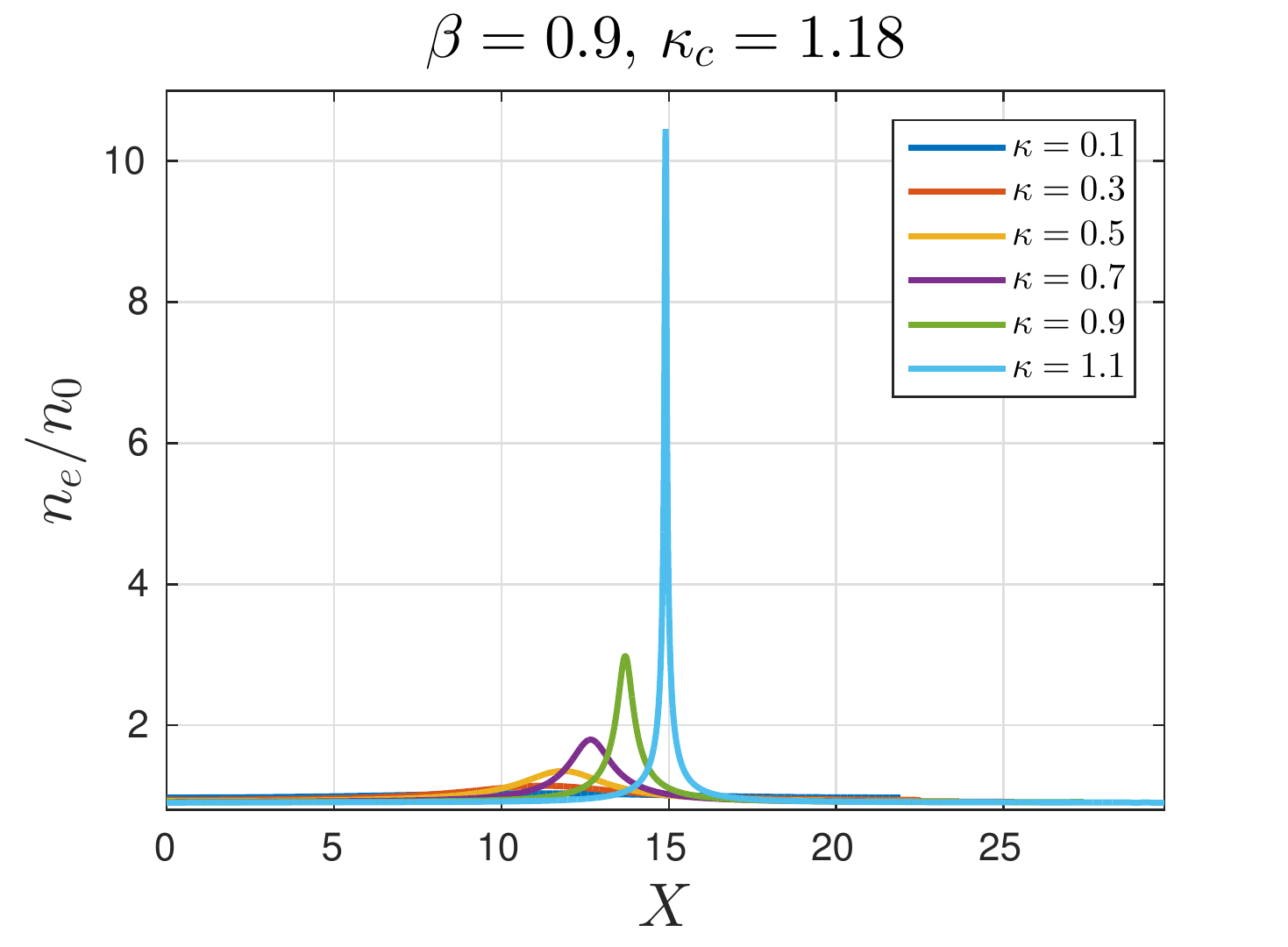}
}
\caption{ Plot of electron density for the parameters (a) $\beta = 0.1$, $\kappa = 0.1, 0.3, 0.5, 0.7, 0.9$ (b) $\beta = 0.9$, $\kappa = 0.1, 0.3, 0.5, 0.7. 0.9, 1.1$.} 
\end{figure}
\begin{figure} [h]	
\centering
\subfloat[]{	\label{density v0_01 kr_11}
\includegraphics[width=0.5\linewidth]{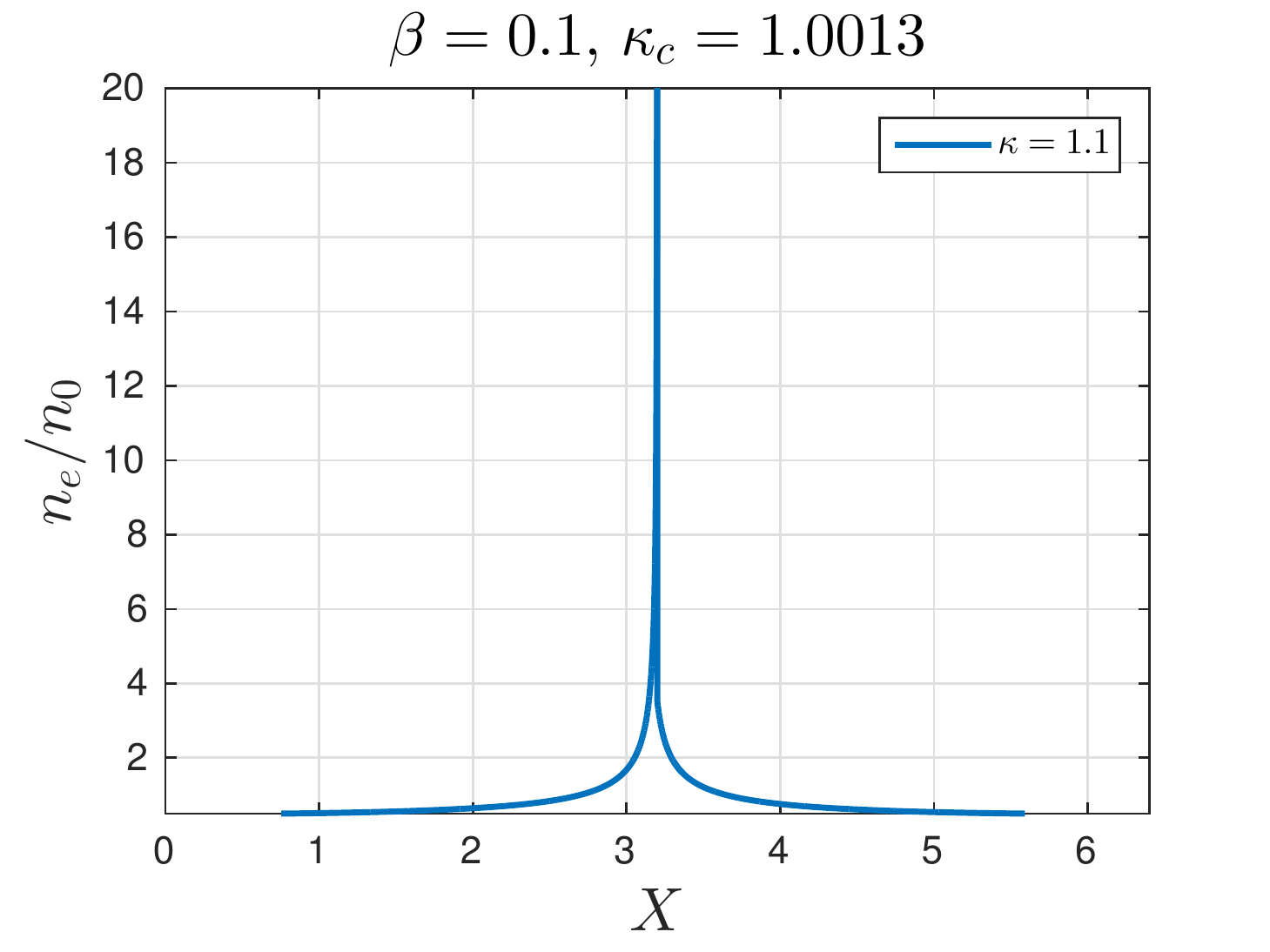}
}
\subfloat[]{	\label{density v0_09 kr_07}
\includegraphics[width=0.5\linewidth]{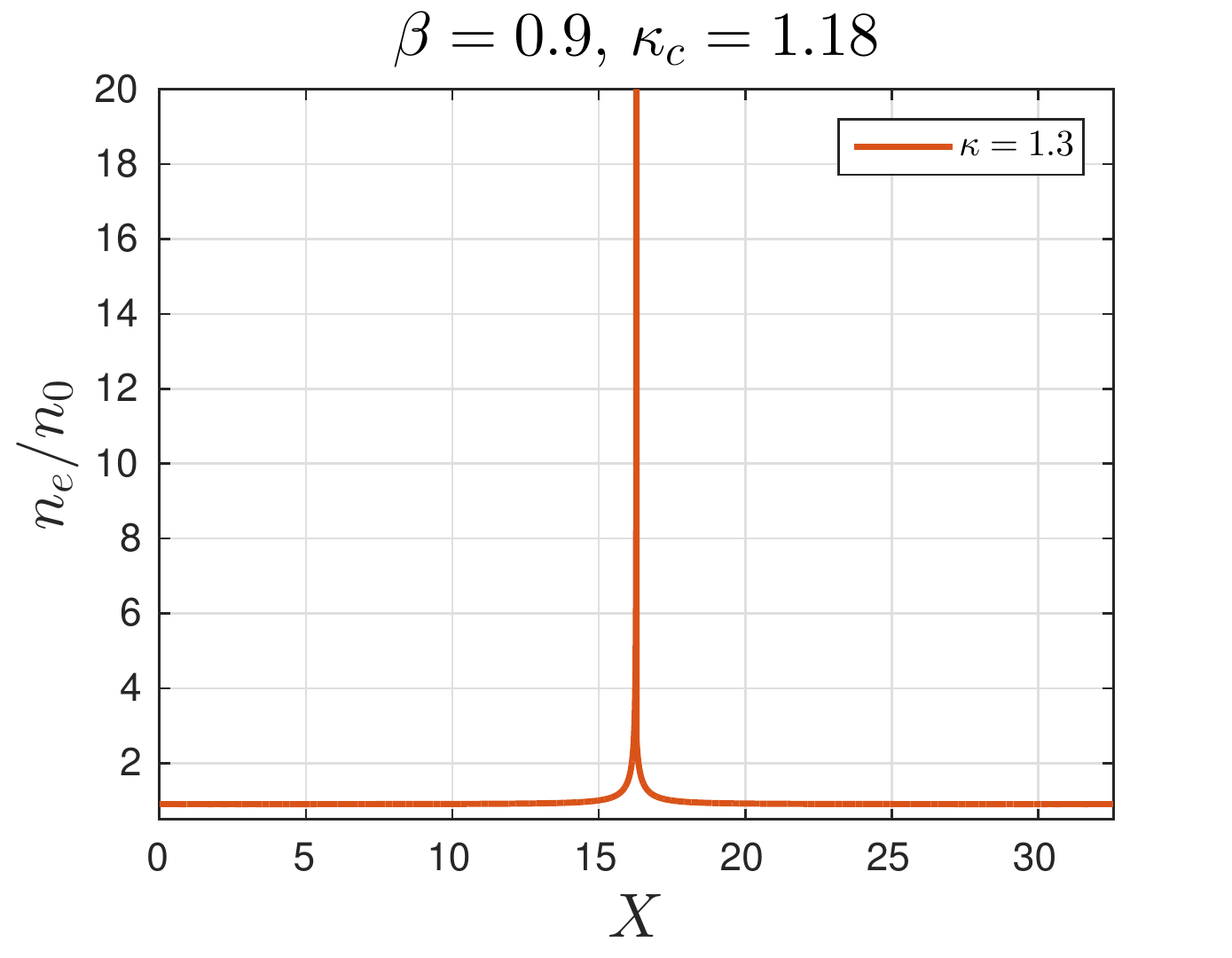}
}
\caption{Plot of electron density for the parameters (a) $\beta = 0.1$, $\kappa = 1.1$ and (b) $\beta = 0.9$, $\kappa = 1.3$. }
\label{density}
\end{figure}

%%%%%%%%%%%%%%%%%%%%%%%%%%%%%%%%%%%%%%%%%%%%%%%%%%%%%%%%%%%%%%%%
\section{Conclusion} \label{stat:conclusion}
An analytical study is carried out for stationary Langmuir structures in relativistic current carrying fluid-Maxwell system. It is observed that profile of nonlinear Langmuir structures is governed by the parameter $\kappa = E_{m}/(4 \pi n_{0} m_{0} v^{2}_{0})^{1/2}$ and $\beta$. Critical value of the parameter $\kappa$ scales with beam velocity $v_{0}$ as $\kappa_{c} = \sqrt{2 \gamma_{0}/(1+ \gamma_{0})}$. Amplitude of  parameter ($\kappa$) embodies the nonlinear effects in the problem. In the linear limit $\kappa \ll \kappa_{c}$, fluid variables vary harmonically in space and results of nonlinear theory coincides with the results of linear theory in this range. In conclusion, Fig. \ref{linear v0_01 kr_001} and \ref{linear v0_09 kr_001} show the fluid variables in the linear limit for the velocity $\beta = 0.1$ and $\beta = 0.9$ respectively, where continuous curves show results obtained from the linear theory and dashed curves show results obtained from the nonlinear theory in the linear limit. Both continuous and dashed curves clearly coincide on each other for both value of $\beta$. As $\kappa \rightarrow \kappa_{c}$ fluid variables gradually begin to shown anharmonic features. In the limit $\kappa \geq \kappa_{c}$, electric field becomes discontinuous \cite{Sen_pr_1955} at certain singular points in space. At the position of electric field discontinuity, electron velocity becomes vanishingly small which results into electron density burst. {Whether these density bursts approach finite values on inclusion of thermal effects, still need to be studied \cite{Infeld_prl_1987, Mori_prl_1988, Trines_pre_2009}.} 
It is found that in the $v_{0} \rightarrow 0$ and/or $\kappa \rightarrow \infty$ range, electron beam is transformed into a crystal of "negatively charged plane" with inter-distance $\lambda_{0} = E_{m}/2\pi n_{0} e$ having surface charge density $\sim E_{m}/2\pi$, which matches with the results found in non-relativistic regime \cite{Psimopoulos_pop_1997}.  Apparently, these 1-D crystals are expected to be unstable since any electron with infinitesimal perturbation in its location will be moved away from the equilibrium position by the electric field. {These solutions which are obtained in the limit $\kappa \geq \kappa_{c} $ cannot exist dynamically. However, the critical value of $\kappa$ has a physical significance. The critical value $\kappa_{c} = \sqrt{2\gamma_{0}/(1+ \gamma_{0})}$
gives the wave breaking \cite{Dawson_prl_1959,Mori_prl_1988} limit of the electron plasma oscillations propagating
on a relativistic electron beam \cite{Chian_pra_1989}.} Study of excitation and stability of these Langmuir structures in the limit $\kappa < \kappa_{c}$, using a PIC/fluid code is left for future studies.

\appendix
\section{Non-relativistic limit of Eq.\eqref{stat:eq35}}
\label{appendix A}
%Using the relation $k^{2} = (r^{2}-s^{2})/r^{2}$, Eq.\eqref{stat:eq35} becomes
We begin with Eq.\eqref{stat:eq35} {\it i.e.}
\begin{equation}
X = \frac{r \gamma_{0}}{\sqrt{1+\beta}}\left[ 2  \left( E(\theta,k)- E(\theta_u,k)\right) +\frac{k^{2}(1-\beta)}{2 \beta}\left( \frac{\sin 2 \theta }{\sqrt{1- k^2 \sin^2 \theta}}- \frac{\sin 2 \theta_u }{\sqrt{1- k^2 \sin^2 \theta_u}} \right) \right].
\label{s22}
\end{equation}
%          Using the relation
%          \begin{eqnarray}
%           \alpha^2 &=& \left[\left(\frac{\kappa^2}{2} -\gamma_{0}\right)\beta^2 +\gamma_{0}\right]^2 \nonumber\\
%                    %
% %                   &=&\left(\frac{\kappa^2}{2} -\gamma_{0}\right)^2 \beta^4 +2\gamma_{0} \left(\frac{\kappa^2}{2} -\gamma_{0}\right)\beta^2+ \gamma^2_{0}\nonumber\\
%                    %
%           \alpha^2 +\beta^2-1 &=&\left(\frac{\kappa^2}{2} -\gamma_{0}\right)^2 \beta^4 +\beta^2\left[1+ 2\gamma_{0} \left(\frac{\kappa^2}{2} -\gamma_{0}\right)\right]+ \gamma^2_{0}-1\nonumber\\
%           %
%          \alpha^2 +\beta^2-1 &=&\left(\frac{\kappa^2}{2} -\gamma_{0}\right)^2 \beta^4 +\beta^2\left[1+ 2\gamma_{0} \left(\frac{\kappa^2}{2} -\gamma_{0}\right) +\gamma^2_{0}\right]          
%          \end{eqnarray}
Using the relation $\alpha = \beta^{2}(\kappa^{2} - 2 \gamma_{0})/2 + \gamma_{0}$, $r^{2}$ can be written as
\begin{eqnarray}
r^2 = \frac{\left(\frac{\kappa^2}{2} -\gamma_{0}\right)\beta^2 + \gamma_{0} + \beta \sqrt{\left(\frac{\kappa^2}{2} -\gamma_{0}\right)^2 \beta^2 +\left[1+ 2\gamma_{0} \left(\frac{\kappa^2}{2} -\gamma_{0}\right) +\gamma^2_{0}\right]}}{1-\beta},       
\end{eqnarray}
In the limit $\beta \rightarrow 0 $, $ \gamma_{0} \rightarrow 1 $  the above term yields
\begin{eqnarray}
           r^2 &\approx& \frac{ \gamma_{0} + \beta \sqrt{1+ \kappa^2 \gamma_{0}-\gamma^2_{0}}}{1-\beta} \nonumber\\
           &\approx& \frac{ 1+ \beta \kappa}{1-\beta}  \nonumber\\
%           \;\;\;\;\;\;\;\;\;\;\; \left( \mbox{ at the limit  $\beta \rightarrow 0$, $ \gamma_{0} \rightarrow 1$ } \right) \nonumber\\
           %
           &\approx& (1+ \beta \kappa)(1+\beta) \nonumber\\
          &\approx&  1+ \beta (1+\kappa).
           \label{u1}          
\end{eqnarray}
          Similarly
          \begin{equation}
           s^2 \approx 1+ \beta(1-\kappa).
           \label{u2}
           \end{equation}
          Therefore 
          \begin{eqnarray}
          k^{2} &=& \frac{r^2-s^2}{r^{2}}  \nonumber\\
                &\approx & \frac{2 \kappa \beta}{1 + \beta(1 + \kappa)} \nonumber\\
                &\approx&  2 \kappa \beta
           \label{u3}
          \end{eqnarray}
In the limit $\beta \rightarrow 0 $, $ \gamma_{0} \rightarrow 1 $  the elliptic integrals yield
$E(\theta,k) \approx \theta $ and $E(\theta_{u},k) \approx \theta_{u} $. Therefore in this limit Eq. (\ref{s22}) becomes
\begin{eqnarray}
X \approx 2 (\theta - \theta_{u} ) +\kappa (\sin 2 \theta - \sin 2 \theta_{u} ) \label{s33}
\end{eqnarray}   

Now 
\begin{eqnarray}
           \sin^2 \theta_{u} &=& \frac{2 r^2 - \beta^2 \phi_u - 2 \gamma_0 - \beta \sqrt{\beta^2 \phi_u^2 + 4 \gamma_{0} \phi_u + 4 \gamma^2_{0}}}{2(r^2-s^2)},\nonumber\\
           &\approx& \frac{2 \{1+ \beta(1+\kappa)\} - 2 \gamma_{0}- 2 \beta \sqrt{ \gamma_{0} \phi_u + \gamma_{0}^{2} }}{2(2 \kappa \beta)},\nonumber\\
            &\approx& \frac{(1+\kappa) - \sqrt{ \phi_u + 1}}{2 \kappa},\nonumber\\
             &\approx& \frac{(1+\kappa) - \sqrt{ \kappa^2+ 2 \kappa + 1}}{2 \kappa} \nonumber\\
             &\approx& 0
\end{eqnarray} 
%           Therefore $\theta_0 =0$ , so Eq. (\ref{s22}) becomes
%\begin{eqnarray}
%           X &=& r\frac{\gamma_{0}}{\sqrt{1+\beta}}\left[ 2  \{E(\theta,k)- E(\theta_0,k)\} +\frac{(1-\beta)}{\beta}\left\{
%           %
%           \frac{k^2 \sin \theta \cos \theta}{\sqrt{1- k^2 \sin^2 \theta}}- \frac{k^2 \sin \theta_0 \cos \theta_0}{\sqrt{1- k^2 \sin^2 \theta_0}}\right\}\right],\nonumber\\
%           %
%           X &\approx& \left[ 2 \theta  +\frac{k^2}{\beta} \sin \theta \cos \theta  \right].
%          \label{s33}
%\end{eqnarray}   
Similarly $\sin^2 \theta$ may be written as 
\begin{equation}
           \sin^2 \theta =\frac{(1+\kappa) - \sqrt{ \phi + 1}}{2 \kappa},
\end{equation}
and
\begin{equation}
           \cos^2 \theta =1- \frac{(1+\kappa) - \sqrt{ \phi + 1}}{2 \kappa}.
\end{equation}
Therefore
\begin{eqnarray}
           \cos^2 \theta \sin^2 \theta &=&\left[ 1- \frac{(1+\kappa) - \sqrt{ \phi + 1}}{2 \kappa}\right]\left[\frac{(1+\kappa) - \sqrt{ \phi + 1}}{2 \kappa}\right],\nonumber\\
           &=&\left[\frac{\kappa^2 - (\sqrt{ 1+\phi}-1)^2}{4 \kappa^2}\right]
           %
           %\sin\theta \cos \theta  &=&\frac{\sqrt{\kappa^2 - (\sqrt{ 1+\phi}-1)^2}}{2 \kappa},
\end{eqnarray} 
   and       %
\begin{eqnarray}
           \sin^2 2\theta &=& 1- \left(\frac{\sqrt{ 1+\phi}-1}{\kappa}\right)^2,\nonumber\\
          \cos^2 2 \theta &=&\left(\frac{\sqrt{ 1+\phi}-1}{\kappa}\right)^2,\nonumber\\
           \cos 2 \theta &=&\left(\frac{\sqrt{ 1+\phi}-1}{\kappa}\right),\nonumber\\
            2 \theta &=& \cos^{-1}\left(\frac{\sqrt{ 1+\phi}-1}{\kappa}\right) = \frac{\pi}{2}-\sin^{-1}\left(\frac{\sqrt{ 1+\phi}-1}{\kappa}\right).
            %
   %       2 \theta  &=& \frac{\pi}{2}-\sin^{-1}\left(\frac{\sqrt{ 1+\phi}-1}{\kappa}\right).
\end{eqnarray}
           Substituting all these values in Eq. (\ref{s33}), we have 
\begin{equation}
%            X &=& \frac{\pi}{2}-\sin^{-1}\left(\frac{\sqrt{ 1+\phi}-1}{\kappa}\right)  +\frac{2 \kappa \beta}{\beta} \frac{\sqrt{\kappa^2 - (\sqrt{ 1+\phi}-1)^2}}{2 \kappa}, \nonumber\\
            %
           X = \frac{\pi}{2}-\sin^{-1}\left(\frac{\sqrt{ 1+\phi}-1}{\kappa}\right)  +\sqrt{\kappa^2 - (\sqrt{ 1+\phi}-1)^2}.
          \label{s44}
\end{equation}   
        which is exactly the non-relativistic result obtained in references \cite{Psimopoulos_pop_1997,Psimopoulos_pop_1997a} .

\begin{acknowledgments}
The authors are thankful to Dr. Mrityunjay Kundu for careful reading of the manuscript.
\end{acknowledgments}
\bibliography{resubmission_rel_langmuir}

\begin{thebibliography}{42}%
\makeatletter
\providecommand \@ifxundefined [1]{%
 \@ifx{#1\undefined}
}%
\providecommand \@ifnum [1]{%
 \ifnum #1\expandafter \@firstoftwo
 \else \expandafter \@secondoftwo
 \fi
}%
\providecommand \@ifx [1]{%
 \ifx #1\expandafter \@firstoftwo
 \else \expandafter \@secondoftwo
 \fi
}%
\providecommand \natexlab [1]{#1}%
\providecommand \enquote  [1]{``#1''}%
\providecommand \bibnamefont  [1]{#1}%
\providecommand \bibfnamefont [1]{#1}%
\providecommand \citenamefont [1]{#1}%
\providecommand \href@noop [0]{\@secondoftwo}%
\providecommand \href [0]{\begingroup \@sanitize@url \@href}%
\providecommand \@href[1]{\@@startlink{#1}\@@href}%
\providecommand \@@href[1]{\endgroup#1\@@endlink}%
\providecommand \@sanitize@url [0]{\catcode `\\12\catcode `\$12\catcode
  `\&12\catcode `\#12\catcode `\^12\catcode `\_12\catcode `\%12\relax}%
\providecommand \@@startlink[1]{}%
\providecommand \@@endlink[0]{}%
\providecommand \url  [0]{\begingroup\@sanitize@url \@url }%
\providecommand \@url [1]{\endgroup\@href {#1}{\urlprefix }}%
\providecommand \urlprefix  [0]{URL }%
\providecommand \Eprint [0]{\href }%
\providecommand \doibase [0]{http://dx.doi.org/}%
\providecommand \selectlanguage [0]{\@gobble}%
\providecommand \bibinfo  [0]{\@secondoftwo}%
\providecommand \bibfield  [0]{\@secondoftwo}%
\providecommand \translation [1]{[#1]}%
\providecommand \BibitemOpen [0]{}%
\providecommand \bibitemStop [0]{}%
\providecommand \bibitemNoStop [0]{.\EOS\space}%
\providecommand \EOS [0]{\spacefactor3000\relax}%
\providecommand \BibitemShut  [1]{\csname bibitem#1\endcsname}%
\let\auto@bib@innerbib\@empty
%</preamble>
\bibitem [{\citenamefont {Bernstein}, \citenamefont {Greene},\ and\
  \citenamefont {Kruskal}(1957)}]{BGK_pr_1957}%
  \BibitemOpen
  \bibfield  {author} {\bibinfo {author} {\bibfnamefont {I.~B.}\ \bibnamefont
  {Bernstein}}, \bibinfo {author} {\bibfnamefont {J.~M.}\ \bibnamefont
  {Greene}}, \ and\ \bibinfo {author} {\bibfnamefont {M.~D.}\ \bibnamefont
  {Kruskal}},\ }\href {\doibase 10.1103/PhysRev.108.546} {\bibfield  {journal}
  {\bibinfo  {journal} {Phys. Rev.}\ }\textbf {\bibinfo {volume} {108}},\
  \bibinfo {pages} {546} (\bibinfo {year} {1957})}\BibitemShut {NoStop}%
\bibitem [{\citenamefont {Schamel}(1986)}]{Schamel_pr_1986}%
  \BibitemOpen
  \bibfield  {author} {\bibinfo {author} {\bibfnamefont {H.}~\bibnamefont
  {Schamel}},\ }\href {\doibase https://doi.org/10.1016/0370-1573(86)90043-8}
  {\bibfield  {journal} {\bibinfo  {journal} {Physics Reports}\ }\textbf
  {\bibinfo {volume} {140}},\ \bibinfo {pages} {161 } (\bibinfo {year}
  {1986})}\BibitemShut {NoStop}%
\bibitem [{\citenamefont {Ng}\ and\ \citenamefont
  {Bhattacharjee}(2005)}]{Ng_prl_2005}%
  \BibitemOpen
  \bibfield  {author} {\bibinfo {author} {\bibfnamefont {C.~S.}\ \bibnamefont
  {Ng}}\ and\ \bibinfo {author} {\bibfnamefont {A.}~\bibnamefont
  {Bhattacharjee}},\ }\href {\doibase 10.1103/PhysRevLett.95.245004} {\bibfield
   {journal} {\bibinfo  {journal} {Phys. Rev. Lett.}\ }\textbf {\bibinfo
  {volume} {95}},\ \bibinfo {pages} {245004} (\bibinfo {year}
  {2005})}\BibitemShut {NoStop}%
\bibitem [{\citenamefont {Eliasson}\ and\ \citenamefont
  {Shukla}(2006)}]{Eliasson_pr_2006}%
  \BibitemOpen
  \bibfield  {author} {\bibinfo {author} {\bibfnamefont {B.}~\bibnamefont
  {Eliasson}}\ and\ \bibinfo {author} {\bibfnamefont {P.}~\bibnamefont
  {Shukla}},\ }\href {\doibase https://doi.org/10.1016/j.physrep.2005.10.003}
  {\bibfield  {journal} {\bibinfo  {journal} {Physics Reports}\ }\textbf
  {\bibinfo {volume} {422}},\ \bibinfo {pages} {225 } (\bibinfo {year}
  {2006})}\BibitemShut {NoStop}%
\bibitem [{\citenamefont {Roberts}\ and\ \citenamefont
  {Berk}(1967)}]{Roberts_prl_1967}%
  \BibitemOpen
  \bibfield  {author} {\bibinfo {author} {\bibfnamefont {K.~V.}\ \bibnamefont
  {Roberts}}\ and\ \bibinfo {author} {\bibfnamefont {H.~L.}\ \bibnamefont
  {Berk}},\ }\href {\doibase 10.1103/PhysRevLett.19.297} {\bibfield  {journal}
  {\bibinfo  {journal} {Phys. Rev. Lett.}\ }\textbf {\bibinfo {volume} {19}},\
  \bibinfo {pages} {297} (\bibinfo {year} {1967})}\BibitemShut {NoStop}%
\bibitem [{\citenamefont {Rajawat}\ and\ \citenamefont
  {Sengupta}(2017)}]{Rajawat_pop_2017}%
  \BibitemOpen
  \bibfield  {author} {\bibinfo {author} {\bibfnamefont {R.~S.}\ \bibnamefont
  {Rajawat}}\ and\ \bibinfo {author} {\bibfnamefont {S.}~\bibnamefont
  {Sengupta}},\ }\href@noop {} {\bibfield  {journal} {\bibinfo  {journal}
  {Physics of Plasmas}\ }\textbf {\bibinfo {volume} {24}},\ \bibinfo {pages}
  {122103} (\bibinfo {year} {2017})}\BibitemShut {NoStop}%
\bibitem [{\citenamefont {Verma}, \citenamefont {Sengupta},\ and\ \citenamefont
  {Kaw}(2012{\natexlab{a}})}]{Verma_pre_2012}%
  \BibitemOpen
  \bibfield  {author} {\bibinfo {author} {\bibfnamefont {P.~S.}\ \bibnamefont
  {Verma}}, \bibinfo {author} {\bibfnamefont {S.}~\bibnamefont {Sengupta}}, \
  and\ \bibinfo {author} {\bibfnamefont {P.}~\bibnamefont {Kaw}},\ }\href
  {\doibase 10.1103/PhysRevE.86.016410} {\bibfield  {journal} {\bibinfo
  {journal} {Phys. Rev. E}\ }\textbf {\bibinfo {volume} {86}},\ \bibinfo
  {pages} {016410} (\bibinfo {year} {2012}{\natexlab{a}})}\BibitemShut
  {NoStop}%
\bibitem [{\citenamefont {Dieckmann}\ and\ \citenamefont
  {Bret}(2009)}]{Dieckmann_aj_2009}%
  \BibitemOpen
  \bibfield  {author} {\bibinfo {author} {\bibfnamefont {M.~E.}\ \bibnamefont
  {Dieckmann}}\ and\ \bibinfo {author} {\bibfnamefont {A.}~\bibnamefont
  {Bret}},\ }\href {http://stacks.iop.org/0004-637X/694/i=1/a=154} {\bibfield
  {journal} {\bibinfo  {journal} {The Astrophysical Journal}\ }\textbf
  {\bibinfo {volume} {694}},\ \bibinfo {pages} {154} (\bibinfo {year}
  {2009})}\BibitemShut {NoStop}%
\bibitem [{\citenamefont {Singh}\ and\ \citenamefont
  {Schunk}(1982{\natexlab{a}})}]{Nsingh_grl_1982}%
  \BibitemOpen
  \bibfield  {author} {\bibinfo {author} {\bibfnamefont {N.}~\bibnamefont
  {Singh}}\ and\ \bibinfo {author} {\bibfnamefont {R.~W.}\ \bibnamefont
  {Schunk}},\ }\href {\doibase 10.1029/GL009i012p01345} {\bibfield  {journal}
  {\bibinfo  {journal} {Geophysical Research Letters}\ }\textbf {\bibinfo
  {volume} {9}},\ \bibinfo {pages} {1345} (\bibinfo {year}
  {1982}{\natexlab{a}})}\BibitemShut {NoStop}%
\bibitem [{\citenamefont {Singh}\ and\ \citenamefont
  {Schunk}(1982{\natexlab{b}})}]{Nsingh_jgr_1982}%
  \BibitemOpen
  \bibfield  {author} {\bibinfo {author} {\bibfnamefont {N.}~\bibnamefont
  {Singh}}\ and\ \bibinfo {author} {\bibfnamefont {R.~W.}\ \bibnamefont
  {Schunk}},\ }\href {\doibase 10.1029/JA087iA05p03561} {\bibfield  {journal}
  {\bibinfo  {journal} {Journal of Geophysical Research: Space Physics}\
  }\textbf {\bibinfo {volume} {87}},\ \bibinfo {pages} {3561} (\bibinfo {year}
  {1982}{\natexlab{b}})}\BibitemShut {NoStop}%
\bibitem [{\citenamefont {Dawson}(1995)}]{Dawson_pop_1995}%
  \BibitemOpen
  \bibfield  {author} {\bibinfo {author} {\bibfnamefont {J.~M.}\ \bibnamefont
  {Dawson}},\ }\href {\doibase 10.1063/1.871304} {\bibfield  {journal}
  {\bibinfo  {journal} {Physics of Plasmas}\ }\textbf {\bibinfo {volume} {2}},\
  \bibinfo {pages} {2189} (\bibinfo {year} {1995})},\ \Eprint
  {http://arxiv.org/abs/https://doi.org/10.1063/1.871304}
  {https://doi.org/10.1063/1.871304} \BibitemShut {NoStop}%
\bibitem [{\citenamefont {Saeki}\ \emph {et~al.}(1979)\citenamefont {Saeki},
  \citenamefont {Michelsen}, \citenamefont {P\'ecseli},\ and\ \citenamefont
  {Rasmussen}}]{Saeki_prl_1979}%
  \BibitemOpen
  \bibfield  {author} {\bibinfo {author} {\bibfnamefont {K.}~\bibnamefont
  {Saeki}}, \bibinfo {author} {\bibfnamefont {P.}~\bibnamefont {Michelsen}},
  \bibinfo {author} {\bibfnamefont {H.~L.}\ \bibnamefont {P\'ecseli}}, \ and\
  \bibinfo {author} {\bibfnamefont {J.~J.}\ \bibnamefont {Rasmussen}},\ }\href
  {\doibase 10.1103/PhysRevLett.42.501} {\bibfield  {journal} {\bibinfo
  {journal} {Phys. Rev. Lett.}\ }\textbf {\bibinfo {volume} {42}},\ \bibinfo
  {pages} {501} (\bibinfo {year} {1979})}\BibitemShut {NoStop}%
\bibitem [{\citenamefont {Fox}\ \emph {et~al.}(2008)\citenamefont {Fox},
  \citenamefont {Porkolab}, \citenamefont {Egedal}, \citenamefont {Katz},\ and\
  \citenamefont {Le}}]{Fox_prl_2008}%
  \BibitemOpen
  \bibfield  {author} {\bibinfo {author} {\bibfnamefont {W.}~\bibnamefont
  {Fox}}, \bibinfo {author} {\bibfnamefont {M.}~\bibnamefont {Porkolab}},
  \bibinfo {author} {\bibfnamefont {J.}~\bibnamefont {Egedal}}, \bibinfo
  {author} {\bibfnamefont {N.}~\bibnamefont {Katz}}, \ and\ \bibinfo {author}
  {\bibfnamefont {A.}~\bibnamefont {Le}},\ }\href {\doibase
  10.1103/PhysRevLett.101.255003} {\bibfield  {journal} {\bibinfo  {journal}
  {Phys. Rev. Lett.}\ }\textbf {\bibinfo {volume} {101}},\ \bibinfo {pages}
  {255003} (\bibinfo {year} {2008})}\BibitemShut {NoStop}%
\bibitem [{\citenamefont {Drake}\ \emph {et~al.}(2003)\citenamefont {Drake},
  \citenamefont {Swisdak}, \citenamefont {Cattell}, \citenamefont {Shay},
  \citenamefont {Rogers},\ and\ \citenamefont {Zeiler}}]{Drake_science_2003}%
  \BibitemOpen
  \bibfield  {author} {\bibinfo {author} {\bibfnamefont {J.~F.}\ \bibnamefont
  {Drake}}, \bibinfo {author} {\bibfnamefont {M.}~\bibnamefont {Swisdak}},
  \bibinfo {author} {\bibfnamefont {C.}~\bibnamefont {Cattell}}, \bibinfo
  {author} {\bibfnamefont {M.~A.}\ \bibnamefont {Shay}}, \bibinfo {author}
  {\bibfnamefont {B.~N.}\ \bibnamefont {Rogers}}, \ and\ \bibinfo {author}
  {\bibfnamefont {A.}~\bibnamefont {Zeiler}},\ }\href {\doibase
  10.1126/science.1080333} {\bibfield  {journal} {\bibinfo  {journal}
  {Science}\ }\textbf {\bibinfo {volume} {299}},\ \bibinfo {pages} {873}
  (\bibinfo {year} {2003})},\ \Eprint
  {http://arxiv.org/abs/http://science.sciencemag.org/content/299/5608/873.full.pdf}
  {http://science.sciencemag.org/content/299/5608/873.full.pdf} \BibitemShut
  {NoStop}%
\bibitem [{\citenamefont {Goldman}\ \emph {et~al.}(2014)\citenamefont
  {Goldman}, \citenamefont {Newman}, \citenamefont {Lapenta}, \citenamefont
  {Andersson}, \citenamefont {Gosling}, \citenamefont {Eriksson}, \citenamefont
  {Markidis}, \citenamefont {Eastwood},\ and\ \citenamefont
  {Ergun}}]{Goldman_prl_2014}%
  \BibitemOpen
  \bibfield  {author} {\bibinfo {author} {\bibfnamefont {M.~V.}\ \bibnamefont
  {Goldman}}, \bibinfo {author} {\bibfnamefont {D.~L.}\ \bibnamefont {Newman}},
  \bibinfo {author} {\bibfnamefont {G.}~\bibnamefont {Lapenta}}, \bibinfo
  {author} {\bibfnamefont {L.}~\bibnamefont {Andersson}}, \bibinfo {author}
  {\bibfnamefont {J.~T.}\ \bibnamefont {Gosling}}, \bibinfo {author}
  {\bibfnamefont {S.}~\bibnamefont {Eriksson}}, \bibinfo {author}
  {\bibfnamefont {S.}~\bibnamefont {Markidis}}, \bibinfo {author}
  {\bibfnamefont {J.~P.}\ \bibnamefont {Eastwood}}, \ and\ \bibinfo {author}
  {\bibfnamefont {R.}~\bibnamefont {Ergun}},\ }\href {\doibase
  10.1103/PhysRevLett.112.145002} {\bibfield  {journal} {\bibinfo  {journal}
  {Phys. Rev. Lett.}\ }\textbf {\bibinfo {volume} {112}},\ \bibinfo {pages}
  {145002} (\bibinfo {year} {2014})}\BibitemShut {NoStop}%
\bibitem [{\citenamefont {Ergun}\ \emph {et~al.}(1998)\citenamefont {Ergun},
  \citenamefont {Carlson}, \citenamefont {McFadden}, \citenamefont {Mozer},
  \citenamefont {Muschietti}, \citenamefont {Roth},\ and\ \citenamefont
  {Strangeway}}]{Ergun_prl_1998}%
  \BibitemOpen
  \bibfield  {author} {\bibinfo {author} {\bibfnamefont {R.~E.}\ \bibnamefont
  {Ergun}}, \bibinfo {author} {\bibfnamefont {C.~W.}\ \bibnamefont {Carlson}},
  \bibinfo {author} {\bibfnamefont {J.~P.}\ \bibnamefont {McFadden}}, \bibinfo
  {author} {\bibfnamefont {F.~S.}\ \bibnamefont {Mozer}}, \bibinfo {author}
  {\bibfnamefont {L.}~\bibnamefont {Muschietti}}, \bibinfo {author}
  {\bibfnamefont {I.}~\bibnamefont {Roth}}, \ and\ \bibinfo {author}
  {\bibfnamefont {R.~J.}\ \bibnamefont {Strangeway}},\ }\href {\doibase
  10.1103/PhysRevLett.81.826} {\bibfield  {journal} {\bibinfo  {journal} {Phys.
  Rev. Lett.}\ }\textbf {\bibinfo {volume} {81}},\ \bibinfo {pages} {826}
  (\bibinfo {year} {1998})}\BibitemShut {NoStop}%
\bibitem [{\citenamefont {Muschietti}\ \emph {et~al.}()\citenamefont
  {Muschietti}, \citenamefont {Ergun}, \citenamefont {Roth},\ and\
  \citenamefont {Carlson}}]{Muschietti_grl_1999}%
  \BibitemOpen
  \bibfield  {author} {\bibinfo {author} {\bibfnamefont {L.}~\bibnamefont
  {Muschietti}}, \bibinfo {author} {\bibfnamefont {R.~E.}\ \bibnamefont
  {Ergun}}, \bibinfo {author} {\bibfnamefont {I.}~\bibnamefont {Roth}}, \ and\
  \bibinfo {author} {\bibfnamefont {C.~W.}\ \bibnamefont {Carlson}},\ }\href
  {\doibase 10.1029/1999GL900207} {\bibfield  {journal} {\bibinfo  {journal}
  {Geophysical Research Letters}\ }\textbf {\bibinfo {volume} {26}},\ \bibinfo
  {pages} {1093}},\ \Eprint
  {http://arxiv.org/abs/https://agupubs.onlinelibrary.wiley.com/doi/pdf/10.1029/1999GL900207}
  {https://agupubs.onlinelibrary.wiley.com/doi/pdf/10.1029/1999GL900207}
  \BibitemShut {NoStop}%
\bibitem [{\citenamefont {Goldman}, \citenamefont {Newman},\ and\ \citenamefont
  {Mangeney}(2007)}]{Goldman_prl_2007}%
  \BibitemOpen
  \bibfield  {author} {\bibinfo {author} {\bibfnamefont {M.~V.}\ \bibnamefont
  {Goldman}}, \bibinfo {author} {\bibfnamefont {D.~L.}\ \bibnamefont {Newman}},
  \ and\ \bibinfo {author} {\bibfnamefont {A.}~\bibnamefont {Mangeney}},\
  }\href {\doibase 10.1103/PhysRevLett.99.145002} {\bibfield  {journal}
  {\bibinfo  {journal} {Phys. Rev. Lett.}\ }\textbf {\bibinfo {volume} {99}},\
  \bibinfo {pages} {145002} (\bibinfo {year} {2007})}\BibitemShut {NoStop}%
\bibitem [{\citenamefont {Holmes}\ \emph {et~al.}()\citenamefont {Holmes},
  \citenamefont {Ergun}, \citenamefont {Newman}, \citenamefont {Ahmadi},
  \citenamefont {Andersson}, \citenamefont {Le~Contel}, \citenamefont
  {Torbert}, \citenamefont {Giles}, \citenamefont {Strangeway},\ and\
  \citenamefont {Burch}}]{Holmes_jgr_2018}%
  \BibitemOpen
  \bibfield  {author} {\bibinfo {author} {\bibfnamefont {J.~C.}\ \bibnamefont
  {Holmes}}, \bibinfo {author} {\bibfnamefont {R.~E.}\ \bibnamefont {Ergun}},
  \bibinfo {author} {\bibfnamefont {D.~L.}\ \bibnamefont {Newman}}, \bibinfo
  {author} {\bibfnamefont {N.}~\bibnamefont {Ahmadi}}, \bibinfo {author}
  {\bibfnamefont {L.}~\bibnamefont {Andersson}}, \bibinfo {author}
  {\bibfnamefont {O.}~\bibnamefont {Le~Contel}}, \bibinfo {author}
  {\bibfnamefont {R.~B.}\ \bibnamefont {Torbert}}, \bibinfo {author}
  {\bibfnamefont {B.~L.}\ \bibnamefont {Giles}}, \bibinfo {author}
  {\bibfnamefont {R.~J.}\ \bibnamefont {Strangeway}}, \ and\ \bibinfo {author}
  {\bibfnamefont {J.~L.}\ \bibnamefont {Burch}},\ }\href {\doibase
  10.1029/2018JA025750} {\bibfield  {journal} {\bibinfo  {journal} {Journal of
  Geophysical Research: Space Physics}\ }\textbf {\bibinfo {volume} {0}},\
  10.1029/2018JA025750},\ \Eprint
  {http://arxiv.org/abs/https://agupubs.onlinelibrary.wiley.com/doi/pdf/10.1029/2018JA025750}
  {https://agupubs.onlinelibrary.wiley.com/doi/pdf/10.1029/2018JA025750}
  \BibitemShut {NoStop}%
\bibitem [{\citenamefont {Shimada}\ and\ \citenamefont
  {Hoshino}(2003)}]{Shimada_pop_2003}%
  \BibitemOpen
  \bibfield  {author} {\bibinfo {author} {\bibfnamefont {N.}~\bibnamefont
  {Shimada}}\ and\ \bibinfo {author} {\bibfnamefont {M.}~\bibnamefont
  {Hoshino}},\ }\href@noop {} {\bibfield  {journal} {\bibinfo  {journal}
  {Physics of Plasmas}\ }\textbf {\bibinfo {volume} {10}} (\bibinfo {year}
  {2003})}\BibitemShut {NoStop}%
\bibitem [{\citenamefont {Shimada}\ and\ \citenamefont
  {Hoshino}(2004)}]{Shimada_pop_2004}%
  \BibitemOpen
  \bibfield  {author} {\bibinfo {author} {\bibfnamefont {N.}~\bibnamefont
  {Shimada}}\ and\ \bibinfo {author} {\bibfnamefont {M.}~\bibnamefont
  {Hoshino}},\ }\href {\doibase 10.1063/1.1652060} {\bibfield  {journal}
  {\bibinfo  {journal} {Physics of Plasmas}\ }\textbf {\bibinfo {volume}
  {11}},\ \bibinfo {pages} {1840} (\bibinfo {year} {2004})},\ \Eprint
  {http://arxiv.org/abs/http://dx.doi.org/10.1063/1.1652060}
  {http://dx.doi.org/10.1063/1.1652060} \BibitemShut {NoStop}%
\bibitem [{\citenamefont {Shukla}\ \emph {et~al.}(1986)\citenamefont {Shukla},
  \citenamefont {Rao}, \citenamefont {Yu},\ and\ \citenamefont
  {Tsintsadze}}]{Shukla_pr_1986}%
  \BibitemOpen
  \bibfield  {author} {\bibinfo {author} {\bibfnamefont {P.}~\bibnamefont
  {Shukla}}, \bibinfo {author} {\bibfnamefont {N.}~\bibnamefont {Rao}},
  \bibinfo {author} {\bibfnamefont {M.}~\bibnamefont {Yu}}, \ and\ \bibinfo
  {author} {\bibfnamefont {N.}~\bibnamefont {Tsintsadze}},\ }\href {\doibase
  https://doi.org/10.1016/0370-1573(86)90157-2} {\bibfield  {journal} {\bibinfo
   {journal} {Physics Reports}\ }\textbf {\bibinfo {volume} {138}},\ \bibinfo
  {pages} {1 } (\bibinfo {year} {1986})}\BibitemShut {NoStop}%
\bibitem [{\citenamefont {Montgomery}\ \emph {et~al.}(2001)\citenamefont
  {Montgomery}, \citenamefont {Focia}, \citenamefont {Rose}, \citenamefont
  {Russell}, \citenamefont {Cobble}, \citenamefont {Fern\'andez},\ and\
  \citenamefont {Johnson}}]{Montgomery_prl_2001}%
  \BibitemOpen
  \bibfield  {author} {\bibinfo {author} {\bibfnamefont {D.~S.}\ \bibnamefont
  {Montgomery}}, \bibinfo {author} {\bibfnamefont {R.~J.}\ \bibnamefont
  {Focia}}, \bibinfo {author} {\bibfnamefont {H.~A.}\ \bibnamefont {Rose}},
  \bibinfo {author} {\bibfnamefont {D.~A.}\ \bibnamefont {Russell}}, \bibinfo
  {author} {\bibfnamefont {J.~A.}\ \bibnamefont {Cobble}}, \bibinfo {author}
  {\bibfnamefont {J.~C.}\ \bibnamefont {Fern\'andez}}, \ and\ \bibinfo {author}
  {\bibfnamefont {R.~P.}\ \bibnamefont {Johnson}},\ }\href {\doibase
  10.1103/PhysRevLett.87.155001} {\bibfield  {journal} {\bibinfo  {journal}
  {Phys. Rev. Lett.}\ }\textbf {\bibinfo {volume} {87}},\ \bibinfo {pages}
  {155001} (\bibinfo {year} {2001})}\BibitemShut {NoStop}%
\bibitem [{\citenamefont {Piran}(1999)}]{Piran_pr_1999}%
  \BibitemOpen
  \bibfield  {author} {\bibinfo {author} {\bibfnamefont {T.}~\bibnamefont
  {Piran}},\ }\href {\doibase https://doi.org/10.1016/S0370-1573(98)00127-6}
  {\bibfield  {journal} {\bibinfo  {journal} {Physics Reports}\ }\textbf
  {\bibinfo {volume} {314}},\ \bibinfo {pages} {575 } (\bibinfo {year}
  {1999})}\BibitemShut {NoStop}%
\bibitem [{\citenamefont {Schamel}(1972)}]{Schamel_pp_1972}%
  \BibitemOpen
  \bibfield  {author} {\bibinfo {author} {\bibfnamefont {H.}~\bibnamefont
  {Schamel}},\ }\href {http://stacks.iop.org/0032-1028/14/i=10/a=002}
  {\bibfield  {journal} {\bibinfo  {journal} {Plasma Physics}\ }\textbf
  {\bibinfo {volume} {14}},\ \bibinfo {pages} {905} (\bibinfo {year}
  {1972})}\BibitemShut {NoStop}%
\bibitem [{\citenamefont {Nicholson}(1983)}]{Nicholson}%
  \BibitemOpen
  \bibfield  {author} {\bibinfo {author} {\bibfnamefont {D.}~\bibnamefont
  {Nicholson}},\ }\href {https://books.google.co.in/books?id=fyRRAAAAMAAJ}
  {\emph {\bibinfo {title} {Introduction to plasma theory}}},\ Wiley series in
  plasma physics\ (\bibinfo  {publisher} {Wiley},\ \bibinfo {year}
  {1983})\BibitemShut {NoStop}%
\bibitem [{\citenamefont {Akhiezer}\ and\ \citenamefont
  {Lyubarskizs}(1951)}]{Akhiezer_dans_1951}%
  \BibitemOpen
  \bibfield  {author} {\bibinfo {author} {\bibfnamefont {A.~I.}\ \bibnamefont
  {Akhiezer}}\ and\ \bibinfo {author} {\bibfnamefont {Y.~G.}\ \bibnamefont
  {Lyubarskizs}},\ }\href@noop {} {\bibfield  {journal} {\bibinfo  {journal}
  {Dokl. Acad. Nauk. SSSR}\ }\textbf {\bibinfo {volume} {80}},\ \bibinfo
  {pages} {193} (\bibinfo {year} {1951})}\BibitemShut {NoStop}%
\bibitem [{\citenamefont {Psimopoulos}\ and\ \citenamefont
  {Tanriverdi}(1997{\natexlab{a}})}]{Psimopoulos_pop_1997}%
  \BibitemOpen
  \bibfield  {author} {\bibinfo {author} {\bibfnamefont {M.}~\bibnamefont
  {Psimopoulos}}\ and\ \bibinfo {author} {\bibfnamefont {S.}~\bibnamefont
  {Tanriverdi}},\ }\href {\doibase 10.1063/1.872136} {\bibfield  {journal}
  {\bibinfo  {journal} {Physics of Plasmas}\ }\textbf {\bibinfo {volume} {4}},\
  \bibinfo {pages} {230} (\bibinfo {year} {1997}{\natexlab{a}})},\ \Eprint
  {http://arxiv.org/abs/http://dx.doi.org/10.1063/1.872136}
  {http://dx.doi.org/10.1063/1.872136} \BibitemShut {NoStop}%
\bibitem [{\citenamefont {Psimopoulos}\ and\ \citenamefont
  {Tanriverdi}(1997{\natexlab{b}})}]{Psimopoulos_pop_1997a}%
  \BibitemOpen
  \bibfield  {author} {\bibinfo {author} {\bibfnamefont {M.}~\bibnamefont
  {Psimopoulos}}\ and\ \bibinfo {author} {\bibfnamefont {S.}~\bibnamefont
  {Tanriverdi}},\ }\href {\doibase 10.1063/1.872622} {\bibfield  {journal}
  {\bibinfo  {journal} {Physics of Plasmas}\ }\textbf {\bibinfo {volume} {4}},\
  \bibinfo {pages} {2778} (\bibinfo {year} {1997}{\natexlab{b}})},\ \Eprint
  {http://arxiv.org/abs/https://doi.org/10.1063/1.872622}
  {https://doi.org/10.1063/1.872622} \BibitemShut {NoStop}%
\bibitem [{\citenamefont {Albritton}\ and\ \citenamefont
  {Rowlands}(1975)}]{Albritton_nf_1975}%
  \BibitemOpen
  \bibfield  {author} {\bibinfo {author} {\bibfnamefont {J.}~\bibnamefont
  {Albritton}}\ and\ \bibinfo {author} {\bibfnamefont {G.}~\bibnamefont
  {Rowlands}},\ }\href@noop {} {\bibfield  {journal} {\bibinfo  {journal}
  {Nucl. Fusion}\ }\textbf {\bibinfo {volume} {15}},\ \bibinfo {pages} {1199}
  (\bibinfo {year} {1975})}\BibitemShut {NoStop}%
\bibitem [{\citenamefont {Davidson}\ and\ \citenamefont
  {Schram}(1968)}]{Davidson_nf_1968}%
  \BibitemOpen
  \bibfield  {author} {\bibinfo {author} {\bibfnamefont {R.}~\bibnamefont
  {Davidson}}\ and\ \bibinfo {author} {\bibfnamefont {P.}~\bibnamefont
  {Schram}},\ }\href {http://stacks.iop.org/0029-5515/8/i=3/a=006} {\bibfield
  {journal} {\bibinfo  {journal} {Nuclear Fusion}\ }\textbf {\bibinfo {volume}
  {8}},\ \bibinfo {pages} {183} (\bibinfo {year} {1968})}\BibitemShut {NoStop}%
\bibitem [{\citenamefont {Akhiezer}\ and\ \citenamefont
  {Polovin}(1956)}]{Akhiezer_spj_1956}%
  \BibitemOpen
  \bibfield  {author} {\bibinfo {author} {\bibfnamefont {A.~I.}\ \bibnamefont
  {Akhiezer}}\ and\ \bibinfo {author} {\bibfnamefont {R.~V.}\ \bibnamefont
  {Polovin}},\ }\href@noop {} {\bibfield  {journal} {\bibinfo  {journal} {Sov.
  Phys. JETP}\ }\textbf {\bibinfo {volume} {3}},\ \bibinfo {pages} {696}
  (\bibinfo {year} {1956})}\BibitemShut {NoStop}%
\bibitem [{\citenamefont {Verma}, \citenamefont {Sengupta},\ and\ \citenamefont
  {Kaw}(2012{\natexlab{b}})}]{Verma_prl_2012}%
  \BibitemOpen
  \bibfield  {author} {\bibinfo {author} {\bibfnamefont {P.~S.}\ \bibnamefont
  {Verma}}, \bibinfo {author} {\bibfnamefont {S.}~\bibnamefont {Sengupta}}, \
  and\ \bibinfo {author} {\bibfnamefont {P.}~\bibnamefont {Kaw}},\ }\href
  {\doibase 10.1103/PhysRevLett.108.125005} {\bibfield  {journal} {\bibinfo
  {journal} {Phys. Rev. Lett.}\ }\textbf {\bibinfo {volume} {108}},\ \bibinfo
  {pages} {125005} (\bibinfo {year} {2012}{\natexlab{b}})}\BibitemShut
  {NoStop}%
\bibitem [{\citenamefont {Verma}, \citenamefont {Sengupta},\ and\ \citenamefont
  {Kaw}(2012{\natexlab{c}})}]{Verma_pop_2012}%
  \BibitemOpen
  \bibfield  {author} {\bibinfo {author} {\bibfnamefont {P.~S.}\ \bibnamefont
  {Verma}}, \bibinfo {author} {\bibfnamefont {S.}~\bibnamefont {Sengupta}}, \
  and\ \bibinfo {author} {\bibfnamefont {P.}~\bibnamefont {Kaw}},\ }\href
  {\doibase 10.1063/1.3693357} {\bibfield  {journal} {\bibinfo  {journal}
  {Physics of Plasmas}\ }\textbf {\bibinfo {volume} {19}},\ \bibinfo {pages}
  {032110} (\bibinfo {year} {2012}{\natexlab{c}})},\ \Eprint
  {http://arxiv.org/abs/http://dx.doi.org/10.1063/1.3693357}
  {http://dx.doi.org/10.1063/1.3693357} \BibitemShut {NoStop}%
\bibitem [{\citenamefont {Infeld}\ and\ \citenamefont
  {Rowlands}(1989)}]{Infeld_prl_1989}%
  \BibitemOpen
  \bibfield  {author} {\bibinfo {author} {\bibfnamefont {E.}~\bibnamefont
  {Infeld}}\ and\ \bibinfo {author} {\bibfnamefont {G.}~\bibnamefont
  {Rowlands}},\ }\href@noop {} {\bibfield  {journal} {\bibinfo  {journal}
  {Phys. Rev. Lett.}\ }\textbf {\bibinfo {volume} {62}},\ \bibinfo {pages}
  {1122} (\bibinfo {year} {1989})}\BibitemShut {NoStop}%
\bibitem [{\citenamefont {Sen}(1955)}]{Sen_pr_1955}%
  \BibitemOpen
  \bibfield  {author} {\bibinfo {author} {\bibfnamefont {H.~K.}\ \bibnamefont
  {Sen}},\ }\href {\doibase 10.1103/PhysRev.97.849} {\bibfield  {journal}
  {\bibinfo  {journal} {Phys. Rev.}\ }\textbf {\bibinfo {volume} {97}},\
  \bibinfo {pages} {849} (\bibinfo {year} {1955})}\BibitemShut {NoStop}%
\bibitem [{\citenamefont {Farokhi}(2006)}]{Farokhi2006}%
  \BibitemOpen
  \bibfield  {author} {\bibinfo {author} {\bibfnamefont {B.}~\bibnamefont
  {Farokhi}},\ }in\ \href@noop {} {\emph {\bibinfo {booktitle} {Theory of
  plasma instabilities: Transport, stability issues and their interaction.
  Proceedings of a technical meeting}}}\ (\bibinfo {year} {2006})\BibitemShut
  {NoStop}%
\bibitem [{\citenamefont {Dawson}(1959)}]{Dawson_prl_1959}%
  \BibitemOpen
  \bibfield  {author} {\bibinfo {author} {\bibfnamefont {J.~M.}\ \bibnamefont
  {Dawson}},\ }\href {\doibase 10.1103/PhysRev.113.383} {\bibfield  {journal}
  {\bibinfo  {journal} {Phys. Rev.}\ }\textbf {\bibinfo {volume} {113}},\
  \bibinfo {pages} {383} (\bibinfo {year} {1959})}\BibitemShut {NoStop}%
\bibitem [{\citenamefont {Katsouleas}\ and\ \citenamefont
  {Mori}(1988)}]{Mori_prl_1988}%
  \BibitemOpen
  \bibfield  {author} {\bibinfo {author} {\bibfnamefont {T.}~\bibnamefont
  {Katsouleas}}\ and\ \bibinfo {author} {\bibfnamefont {W.~B.}\ \bibnamefont
  {Mori}},\ }\href {\doibase 10.1103/PhysRevLett.61.90} {\bibfield  {journal}
  {\bibinfo  {journal} {Phys. Rev. Lett.}\ }\textbf {\bibinfo {volume} {61}},\
  \bibinfo {pages} {90} (\bibinfo {year} {1988})}\BibitemShut {NoStop}%
\bibitem [{\citenamefont {Infeld}\ and\ \citenamefont
  {Rowlands}(1987)}]{Infeld_prl_1987}%
  \BibitemOpen
  \bibfield  {author} {\bibinfo {author} {\bibfnamefont {E.}~\bibnamefont
  {Infeld}}\ and\ \bibinfo {author} {\bibfnamefont {G.}~\bibnamefont
  {Rowlands}},\ }\href {\doibase 10.1103/PhysRevLett.58.2063} {\bibfield
  {journal} {\bibinfo  {journal} {Phys. Rev. Lett.}\ }\textbf {\bibinfo
  {volume} {58}},\ \bibinfo {pages} {2063} (\bibinfo {year}
  {1987})}\BibitemShut {NoStop}%
\bibitem [{\citenamefont {Trines}(2009)}]{Trines_pre_2009}%
  \BibitemOpen
  \bibfield  {author} {\bibinfo {author} {\bibfnamefont {R.~M. G.~M.}\
  \bibnamefont {Trines}},\ }\href {\doibase 10.1103/PhysRevE.79.056406}
  {\bibfield  {journal} {\bibinfo  {journal} {Phys. Rev. E}\ }\textbf {\bibinfo
  {volume} {79}},\ \bibinfo {pages} {056406} (\bibinfo {year}
  {2009})}\BibitemShut {NoStop}%
\bibitem [{\citenamefont {Chian}(1989)}]{Chian_pra_1989}%
  \BibitemOpen
  \bibfield  {author} {\bibinfo {author} {\bibfnamefont {A.~C.~L.}\
  \bibnamefont {Chian}},\ }\href {\doibase 10.1103/PhysRevA.39.2561} {\bibfield
   {journal} {\bibinfo  {journal} {Phys. Rev. A}\ }\textbf {\bibinfo {volume}
  {39}},\ \bibinfo {pages} {2561} (\bibinfo {year} {1989})}\BibitemShut
  {NoStop}%
\end{thebibliography}%
\end{document}